\documentclass[aps,prd,superscriptaddress,nofootinbib,preprint]{revtex4-1}
\pdfoutput=1
\usepackage[utf8]{inputenc}
\usepackage[english]{babel}
\usepackage{amsmath}
\usepackage{subfigure}
\usepackage{amsfonts}
\usepackage{amssymb}
\usepackage{color}
\usepackage{cancel}
\usepackage{graphicx}
\newcommand{\be}{\begin{equation}}
\newcommand{\ee}{\end{equation}}
\newcommand{\bea}{\begin{eqnarray}}
\newcommand{\eea}{\end{eqnarray}}

\newcommand{\nn}{\nonumber}

\catcode`@=12

\begin{document}
  
\preprint{HRI-RECAPP-2019-001, IP/BBSR/2019-1}
	\title{Doubly-charged Higgs Boson at Future Electron-Proton Collider}
\author{P. S. Bhupal Dev}
	\email{bdev@wustl.edu}
	\affiliation{Department of Physics and McDonnell Center for the Space Sciences, Washington University, St.
Louis, MO 63130, USA}
	
	\author{Sarif Khan}
	\email{sarifkhan@hri.res.in}
	 \affiliation{Harish-Chandra Research Institute, Chhatnag Road, 
                          Jhusi, Allahabad 211019, India}
\affiliation{Homi Bhabha National Institute, Training School Complex, Anushakti Nagar, Mumbai 400094,
India}
                          \affiliation{Institut f\"{u}r Theoretische Physik, Georg-August-Universit\"{a}t G\"{o}ttingen, Friedrich-Hund-Platz 1, G\"{o}ttingen, D-37077 Germany}
	\author{Manimala Mitra}
	\email{manimala@iopb.res.in}
	\affiliation{Institute of Physics,  Sachivalaya Marg, Bhubaneswar  
		751005, India}
	\affiliation{Homi Bhabha National Institute, Training School Complex, Anushakti Nagar, Mumbai 400094,
India}
	\author{Santosh Kumar Rai}
	\email{skrai@hri.res.in}
	 \affiliation{Harish-Chandra Research Institute, Chhatnag Road, 
                          Jhusi, Allahabad 211019, India}
	\affiliation{Regional Centre for Accelerator-based Particle Physics,
                Harish-Chandra Research Institute, Chhatnag Road, 
                          Jhusi, Allahabad 211019, India}
	\affiliation{Homi Bhabha National Institute, Training School Complex, Anushakti Nagar, Mumbai 400094,
India}
\begin{abstract} 
We explore the discovery prospect of the doubly-charged component of an $SU(2)_L$-triplet scalar at the future  $e^- p$ collider FCC-eh, 
proposed to operate with an electron beam energy of 60 GeV and a proton beam energy of 50 TeV.  We consider the 
associated production of the doubly-charged Higgs boson along with leptons and jet(s), and its subsequent prompt decay 
to same-sign lepton pair. This occurs for ${\cal O}(1)$ Yukawa coupling of the scalar triplet with charged leptons, which is expected for reasonably small vacuum expectation values of the neutral component of the triplet field that governs the neutrino mass  generation in the type-II seesaw. We present our analysis 
for two different final states, $3l+\ge1j$ and an inclusive $\ge2l+\ge1j$ channel.  Considering its decay to electrons only, we find that the doubly-charged Higgs with mass around a TeV could be observed at $3\sigma$ confidence level with $\mathcal{O}(200)\, \rm{fb}^{-1}$ of integrated luminosity, while masses up to 2 TeV could be probed within a few years of data accumulation.  The signal proposed here becomes essentially 
background-free,  if it is triggered in the $\mu\mu$ mode and a $5\sigma$ discovery is achievable in this channel for a TeV-scale doubly-charged Higgs with an integrated luminosity
as low as $\mathcal{O} (50)\, \rm{fb}^{-1}$. We also highlight the sensitivity of FCC-eh to the Yukawa coupling responsible for production of the doubly-charged Higgs boson as a function of its mass in both the $ee$ and $\mu\mu$ channels.
\end{abstract}
	\maketitle
	\newpage 
	
	\section{Introduction \label{intro}}
	The observations of flavor-changing neutrinos  in neutrino   oscillation experiments have conclusively established that neutrinos have nonzero masses and mixing. Specifically, the solar and atmospheric mass square differences are measured to be $\Delta m^2_{12} \sim 7.6\times 10^{-5}$ $\rm{eV}^2$, and $|\Delta m^2_{13}| \sim 2.5\times 10^{-3}$ $\rm{eV}^2$, and the mixing angles are $\theta_{12} \sim 33^{\circ}, ~\theta_{23} \sim 42^{\circ}$, and $\theta_{13} \sim 8^{\circ}$ \cite{deSalas:2018bym}.  
	These experimental observations can not be explained within the framework of the Standard Model (SM), and hence, some beyond the Standard Model  (BSM) extension is required.  So far,  a number of  BSM models have been proposed to  explain  small neutrino masses~\cite{Mohapatra:2006gs}. 
Among them, one of the most appealing frameworks for light neutrino mass generation is the so-called seesaw mechanism, where the lepton-number-violating (LNV) dimension-5 Weinberg operator $LLHH/\Lambda$ \cite{Weinberg:1979sa, Wilczek:1979hc} (where $L$ and $H$ are respectively the SM lepton and Higgs doublets, and $\Lambda$ is the LNV scale) generates Majorana masses of light neutrinos. There are three tree-level realizations of seesaw~\cite{Ma:1998dn}, namely, type-I \cite{Minkowski:1977sc,Mohapatra:1979ia,Yanagida:1979as,GellMann:1980vs,Glashow:1979nm}, type-II \cite{Konetschny:1977bn, Schechter:1980gr, Magg:1980ut,Cheng:1980qt,Lazarides:1980nt,Mohapatra:1980yp}, and type-III \cite{Foot:1988aq} seesaw, depending on whether we add $SU(2)_L$-singlet fermions or $SU(2)_L$-triplet scalars or $SU(2)_L$-triplet fermions to the SM particle content, respectively. One can also have hybrid combinations, such as type-I+type-II seesaw~\cite{Schechter:1980gr}, which naturally occurs, for instance, in the left-right symmetric extension of the SM~\cite{Mohapatra:1974hk, Mohapatra:1974gc, Senjanovic:1975rk}. 
There also exist several other variations of the minimal seesaw, such as inverse~\cite{Mohapatra:1986aw, Nandi:1985uh, Mohapatra:1986bd}, linear~\cite{Malinsky:2005bi} and extended~\cite{Kang:2006sn, Gavela:2009cd, Dev:2012sg, Law:2013gma} seesaw, as well various radiative neutrino mass generation mechanisms~\cite{Cai:2017jrq}. If the scale of the dimension-5 Weinberg operator for neutrino mass generation is within a few TeV range, the seesaw mechanism can be tested by looking for LNV signatures in current and future colliders~\cite{Atre:2009rg, Deppisch:2015qwa, Golling:2016gvc, Cai:2017mow, deBlas:2018mhx}. 

In this paper, we focus on the LNV signatures of seesaw at future $e^-p$ colliders, such as LHeC~\cite{AbelleiraFernandez:2012cc} and FCC-eh~\cite{Acar:2016rde}. Although their center-of-mass energy is down with respect to the $pp$ option by a factor of $\sqrt{E_p/E_e}\sim 10~(30)$ for LHeC (FCC-eh), the $e^-p$ colliders are capable of exploring a broad range of BSM scenarios and characterizing BSM hints at $ee$ and $pp$ colliders~\cite{Azuelos:2018syu}.  
There have been quite a few studies on the prospects of the $SU(2)_L$-singlet heavy neutrino production at future $e^- p$ colliders~\cite{Liang:2010gm, Blaksley:2011ey,  Duarte:2014zea, Mondal:2015zba,  Lindner:2016lxq, Mondal:2016kof,  Antusch:2016ejd,  Curtin:2017bxr,  Mandal:2018qpg, Das:2018usr, Li:2018wut}.  Here we explore the prospects of the $SU(2)_L$-triplet scalar ($\Delta$), and in particular, the production and decay of the doubly-charged component ($H^{\pm\pm}$) of the triplet at a future $e^{-} p$ collider. For earlier studies of the doubly-charged Higgs production in $e^+p$ collision at HERA, see Refs.~\cite{Accomando:1993ar, Aktas:2006nu, Yue:2007ym}. The triplet Higgs acquires an induced vacuum expectation value (VEV) $v_{\Delta}$ and generates light neutrino masses at tree-level via the type-II seesaw mechanism~\cite{Schechter:1980gr, Magg:1980ut,Cheng:1980qt,Lazarides:1980nt,Mohapatra:1980yp}. We consider  $v_{\Delta}$  to be small, i.e. $v_{\Delta} \sim \mathcal{O}(0.1)$ eV, in the same ballpark  as the upper limit on the sum of neutrino masses from cosmology~\cite{Ade:2015xua}. For this choice of the triplet VEV, the Yukawa couplings of the triplet scalar $\Delta$ with a SM lepton-pair are large: $Y^{\Delta} \sim \mathcal{O}(1)$. This is where the production of a doubly-charged scalar in $e^-p$ collision has a major difference when compared with its production in $pp$ collision. At the LHC, the dominant production modes for the doubly-charged Higgs are pair-production via the $s$-channel $\gamma/Z$ and associated production with a singly-charged Higgs via the $s$-channel $W$ exchange~\cite{Grifols:1989xe, Gunion:1989in, Muhlleitner:2003me, Akeroyd:2005gt, Han:2007bk, Chao:2008mq, Perez:2008ha, delAguila:2008cj, Akeroyd:2009hb, Akeroyd:2010ip, Melfo:2011nx, Alloul:2013raa, Chun:2013vma, delAguila:2013mia, Bambhaniya:2013wza, Dev:2016dja, Mitra:2016wpr, Ghosh:2017pxl, Li:2018jns, Borah:2018yxd,  Antusch:2018bgr, Crivellin:2018ahj, Dev:2018kpa, Du:2018eaw}. The pair-production through $\gamma\gamma$ fusion is also important for higher doubly-charged scalar masses~\cite{Babu:2016rcr}, whereas the single-production through $WW$-fusion~\cite{Huitu:1996su, Maalampi:2002vx, Kanemura:2013vxa, Englert:2013wga, Dutta:2014dba, Kanemura:2014goa, Bambhaniya:2015wna, Sirunyan:2017ret} is suppressed by the smallness of the triplet VEV $v_\Delta$ and is not accessible to the LHC~\cite{Sirunyan:2017ret}. Due to the hadrophobic nature of the triplet scalar, one can never exploit the large Yukawa $Y^{\Delta}$ to enhance the production cross-section of $\Delta$-fields at hadron colliders, although the size and structure of $Y^\Delta$ still play an important role in deciding their dominant decay modes. In contrast, at the $e^-p$ collider, the doubly-charged scalar can be singly produced in association with a lepton ($l$ or $\nu_l$) and a jet (see Fig.~\ref{fig:feyn12}), with the cross-section directly proportional to the square of the Yukawa coupling to the electron flavor, $|Y^{\Delta}_{el}|^2$ (with $l=e,\mu,\tau$). Therefore, as we show in this paper, the Yukawa coupling  $Y^{\Delta}$ can be directly measured at future $e^-p$ collider, as long as it is reasonably large $\sim {\cal O}(1)$ to give an observable signal-to-background ratio. This is largely complementary to the case of future lepton colliders, where the doubly-charged scalar can also be singly/pair-produced via the Yukawa coupling $Y^{\Delta}_{el}$~\cite{Mukhopadhyaya:2005vf, Chen:2008jh, Rodejohann:2010jh, Yue:2010zu, Rodejohann:2010bv, Nomura:2017abh, Agrawal:2018pci, Dev:2018upe, Dev:2018sel}. 

For large diagonal Yukawa couplings, the off-diagonal entries of $Y^{\Delta}$ are  restricted from LFV  processes, such as $\mu \to 3e$, $\mu \to e \gamma$, and $\mu -e$ conversion in nuclei~\cite{Akeroyd:2009nu, Dev:2017ouk,  Dev:2018sel, Dev:2018kpa}.  To evade this conflict, we consider the scenario with large diagonal  $Y^\Delta_{ii}$ (with $i=e,\mu$) and small off-diagonal $Y^\Delta_{ij}$ (with $i\neq j$) that are in agreement with LFV measurements and are also promising for  collider searches of $H^{\pm \pm}$ at $e^{-} p$ collider. However, this  choice can not reproduce the observed neutrino mixing angles, which require a highly non-diagonal PMNS mixing matrix.  The  conflict with the neutrino data can be evaded by considering variations in  the pure type-II seesaw model. A simple possibility is to consider a hybrid seesaw model, that contains both the gauge-singlet, right-handed (RH) neutrinos and the triplet Higgs.   The light neutrino mass is generated by both singlet neutrino (type-I) and the triplet Higgs (type-II) contributions. The phenomenology of the singlet neutrinos at the $e^-p$ collider will be the same as discussed before~\cite{Mondal:2015zba, Antusch:2016ejd}, and our main interest here lies in the doubly-charged Higgs phenomenology. 

The direct search constraints on the doubly-charged scalar from the LHC put a lower bound on $M_{H^{\pm \pm}} \gtrsim 800$ GeV at 95$\%$ C.L~\cite{CMS-PAS-HIG-16-036, Aaboud:2017qph}. For the dominant $H^{\pm \pm}$ production channels considered here, we find that the proposed LHeC with $\sqrt s=1.3$ TeV (or even the high-energy LHeC with $\sqrt s = 1.8$ TeV) cannot probe the doubly-charged Higgs boson beyond its current LHC limit. Therefore, for the collider analysis, we consider the 
future circular collider-based $e^-p$ scenario (FCC-eh) with $\sqrt s= 3.46$ TeV.
We   focus on two different  signals: (i) $3l$ + $\ge 1j$, and (ii) an inclusive search for $H^{--}$ in the $\ge 2l$ + $\ge 1j$ final state, with at least two negatively-charged leptons of same flavor in each case. 
We consider a number of SM backgrounds  that can mimic our signal. By choosing
suitable cuts on the kinematic variables, 
we reduce the background significantly and achieve an observable signal-to-background ratio. We  show that doubly-charged Higgs masses up to 2 TeV can be probed at $3\sigma$ confidence level with 500 fb$^{-1}$ integrated luminosity at FCC-eh. 

The rest of the paper is organized as follows. In Section~\ref{model}, we review the basic relevant features of the triplet Higgs model. In Section~\ref{coll}, we discuss the dominant production channels and the signals for doubly-charged scalar at $e^-p$ collider. Followed by this, we present  a detailed event analysis using  cut-based techniques for the signal and background at FCC-eh. Finally, we present our conclusions  in Section~\ref{conclu}.

\section{Triplet Higgs Model \label{model}}
We consider a hybrid type-I+type-II seesaw model, where the SM particle content is extended by  an additional $SU(2)_L$-triplet scalar with hypercharge $Y=2$: 
\begin{align}
\Delta \ = \ \begin{pmatrix} \frac{\delta^+}{\sqrt 2} & \delta^{++} \\ 
\delta^0 & -\frac{\delta^+}{\sqrt 2}
\end{pmatrix} \, ,
\label{eq:delta}
\end{align}
and the SM gauge-singlet RH neutrinos $N_{R,i}$ (with $i=1,2,3$ being the family index). The complete particles spectrum
is shown in Table \ref{tab:spectrum}, along with the corresponding charges under the SM gauge group. Note that the presence of both type-I and type-II seesaw components can be naturally motivated as originating from a left-right symmetric gauge group, or even higher gauge groups like Pati-Salam or $SO(10)$ at high scale.  	
\begin{table}[t!]
\begin{tabular}{||c||c||c||c||} \hline \hline
\begin{tabular}{c}
    Gauge\\
    Group\\    \hline
    $SU(3)_{c}$\\    \hline
    $SU(2)_{L}$\\     \hline
    $U(1)_{Y}$\\         
\end{tabular}
&
\begin{tabular}{c|c|c}
    \multicolumn{3}{c}{Baryon Fields}\\     \hline
    $Q_{L, i}=(u_{L,i},d_{L,i})^{T}$&$u_{R,i}$&$d_{R,i}$\\    \hline
    $3$ & $3$ & $3$ \\      \hline
    $2$ & $1$ & $1$ \\     \hline
    $1/6$ & $2/3$ & $-1/3$ \\     
\end{tabular}
&
\begin{tabular}{c|c|c}
    \multicolumn{3}{c}{Lepton Fields}\\    \hline
    $L_{i}=(\nu_{L,i},e_{L,i})^{T}$ & ~$e_{R,i}$ & $N_{R,i}$  \\    \hline
    $1$&$1$ & 1 \\    \hline
    $2$&$1$ & 1 \\    \hline
    $-1/2$&$-1$ & 0\\    
\end{tabular}
&
\begin{tabular}{c|c}
    \multicolumn{2}{c}{Scalar Fields}\\    \hline
    $\phi$&$\Delta$\\    \hline
    ~~$1$~~&$1$\\    \hline
    ~~$2$~~&$3$\\    \hline
    ~~$1/2$~~&$1$\\     
\end{tabular}\\ \hline \hline
\end{tabular}
\caption{Particle content and their corresponding
charges under the SM gauge group. The $U(1)_Y$ hypercharges in the last row correspond to $Y/2$. Here $i=1,2,3$ stands for the family index.} 
\label{tab:spectrum}
\end{table}

The complete Lagrangian with the additional triplet scalar and RH neutrino  is given by,
\begin{eqnarray}
\mathcal{L} \ = \ \mathcal{L}_{\rm SM} + \mathcal{L}_{\rm kin}^{N_R}  + {\rm Tr}\left[(D_{\mu} \Delta)^{\dagger} (D^{\mu} \Delta) \right]
- \mathcal{V}(\phi, \Delta) + \mathcal{L}_{\rm Yuk} \, ,
\label{lagrangian}
\end{eqnarray}
where $\mathcal{L}_{\rm SM}$ is the complete Lagrangian for the SM particles and ${\cal L}_{\rm kin}^{N_R}=i\overline{N}_R\gamma^\mu \partial_\mu N_R$ is the kinetic term of the right-handed neutrinos. The
covariant derivative 
for the triplet Higgs takes the following form: 
\begin{eqnarray}
D_{\mu} \Delta \ = \ \partial_{\mu} \Delta
+ \frac{i g}{2} \left[\sigma^{i} W^{i}_{\mu}, \Delta \right]
+ i g^{\prime} B_{\mu} \Delta \, ,
\end{eqnarray}
where $W^{i}_{\mu},\ B_{\mu}$ are the $SU(2)_{L}$ and $U(1)_{Y}$ gauge fields, with the corresponding gauge couplings $g,\ g'$ respectively, and $\sigma^{i}$'s (with $i=1,2,3$) are the Pauli matrices.

The Yukawa term in Eq.~\eqref{lagrangian} takes the following form:
 \begin{eqnarray}
 -\mathcal{L}_{\rm Yuk} \ = \ Y^{\Delta}_{ij} L^T_{i}
 C^{-1}\sigma_2 \Delta L_{j} + Y^D_{ij} \overline{L}_i \widetilde{\phi} N_{R,j} + \frac{1}{2}M_{R,ij}  {N}^T_{R,i}C^{-1} N_{R,j} + {\rm H.c.} \, ,
\label{yukawa-neutrino}
 \end{eqnarray}
where $C$ is the Dirac charge conjugation matrix with respect to the Lorentz group, $\widetilde{\phi}\equiv i\sigma_2\phi^*$ and $Y^D$ is the Dirac Yukawa coupling for neutrinos. The 1st and 3rd term in Eq.~\eqref{yukawa-neutrino} violate  lepton number by two units and give rise to the light neutrino mass at tree-level via type-II and type-I seesaw mechanisms, respectively, after electroweak symmetry breaking. The neutral component of the Higgs doublet acquires  a spontaneous VEV, while the neutral component of the Higgs triplet acquires an induced VEV: 
\begin{eqnarray}
\langle \phi \rangle \ = \ 
\begin{pmatrix}
0  \\
\frac{v}{\sqrt{2}}
\end{pmatrix},\,\,\,\,\,\,
\langle \Delta \rangle \ = \ 
\begin{pmatrix}
0 & 0 \\
\frac{v_{\Delta}}{\sqrt{2}} & 0
\end{pmatrix}\,\,.
\label{phih}
\end{eqnarray}
The Majorana neutrino mass matrix for the light neutrinos have the following form:
\begin{eqnarray}
M_{\nu} \ = \  
M_L - M^T_D M^{-1}_R M_D \,, 
\label{neumass}
\end{eqnarray}
where $ M_L=\frac{v_{\Delta}}{\sqrt{2}} Y^{\Delta}$ and $M_D= \frac{v}{\sqrt{2}}Y^D$.  In the limit $Y^D \to 0$, the light neutrino mass matrix becomes type-II dominant: $M_\nu \simeq   
\frac{v_{\Delta}}{\sqrt{2}} Y^{\Delta}$. Since the triplet VEV $v_\Delta$ governs the light neutrino mass in type-II seesaw, it is expected to be much smaller than the electroweak VEV $v$, i.e. $v_\Delta\ll v$. In fact, the electroweak precision data, and in particular, the $\rho$-parameter constraint requires that $v_\Delta \lesssim 2$ GeV~\cite{Dev:2018kpa}. For the  collider study of  doubly-charged Higgs production  and decay (see Section~\ref{coll}), we consider  low VEV of the triplet
Higgs, $v_{\Delta} \sim 0.1 $ eV, that generates large Yukawa $Y^{\Delta} \sim \mathcal{O}(1)$. In particular, we consider  large 
diagonal elements of $Y^{\Delta}_{ii}$ (with $i=e,\mu$), and small off-diagonal Yukawa $Y^{\Delta}_{ij}$ (with $i \neq j$) in order to avoid the bounds from LFV processes mediated by the doubly-charged scalars~\cite{Akeroyd:2009nu, Barry:2013xxa, Bambhaniya:2015ipg, Dev:2017ouk, Dev:2018kpa}. Therefore, we need the type-I seesaw contribution in Eq.~\eqref{neumass} to reproduce the observed pattern of the PMNS mixing matrix in the light neutrino sector. Apart from fitting the neutrino mass, the Dirac Yukawa couplings $Y^D$ play no other role in our subsequent discussion.


The scalar potential in Eq.~\eqref{lagrangian} has the following form: 
\begin{eqnarray}
\mathcal{V}(\phi, \Delta) & \ = \ & -\mu_\phi^2 \left(\phi^{\dagger} \phi \right)
+ \frac{\lambda}{4} \left(\phi^{\dagger} \phi\right)^2
- \mu^2_{\Delta} {\rm Tr}\left[\Delta^{\dagger} \Delta\right]
+ \left[\mu \phi^T i \sigma_2 \Delta^{\dagger} \phi  + {\rm H.c.} \right] \nn \\
&&+ \lambda_1 \left(\phi^{\dagger} \phi\right) {\rm Tr}\left[\Delta^{\dagger} \Delta\right]
+\lambda_2 \left( {\rm Tr}\left[\Delta^{\dagger} \Delta\right] \right)^2 
+ \lambda_3 {\rm Tr}\left[\left(\Delta^{\dagger} \Delta \right)^2 \right]
+ \lambda_4 (\phi^{\dagger} \Delta \Delta^{\dagger} \phi)\,.
\label{potential}
\end{eqnarray} 
In the above potential, one can introduce another term
$\lambda_5 (\phi^{\dagger} \Delta^{\dagger} \Delta \phi)$, but this quartic
coupling $\lambda_5$ can be easily absorbed by redefining the couplings 
$\lambda_1$ and $\lambda_4$, by virtue of the relation  
$\phi^{\dagger} \Delta \Delta^{\dagger} \phi
 + \phi^{\dagger} \Delta^{\dagger} \Delta \phi = 
 \phi^{\dagger} \phi {\rm Tr}\left[\Delta^{\dagger} \Delta\right]$,
 which is valid because $\Delta$ is a traceless $2 \times 2$ matrix [cf.~Eq.~\eqref{eq:delta}]. 


We expand the doublet and triplet scalar fields around their VEVs to obtain 10 real-valued field components:   
\begin{eqnarray}
 \phi \ = \ 
\begin{pmatrix}
\phi^{+}  \\
\frac{v + h_0 + i \xi_1}{\sqrt{2}}
\end{pmatrix},\,\,\,\,\,\,
 \Delta \ = \ 
\begin{pmatrix}
\frac{\delta^{+}}{\sqrt{2}} & \delta^{++} \\
\frac{v_{\Delta} + \delta_0 + i \xi_2}{\sqrt{2}} & -\frac{\delta^{+}}{\sqrt{2}}
\end{pmatrix}\,\,.
\label{phi-delta}
\end{eqnarray}
We consider $\mu^2_\phi > 0$ and $\mu^2_{\Delta} > 0$ in Eq.~\eqref{potential}. We also assume the parameter $\mu$ in the cubic term of  Eq.~\eqref{potential} to be real. Upon minimization of the scalar potential with respect to the VEVs, we get the 
following relations: 
\begin{eqnarray}
\mu^2_{\Delta} &\ = \ & \frac{2 \mu v^2 - \sqrt{2} (\lambda_1 + \lambda_2)v^2 v_{\Delta}
- 2\sqrt{2} (\lambda_2 + \lambda_3)v^3_{\Delta}}{2\sqrt{2}\, v_{\Delta}} \,, \label{eq:mudelta} \\
\mu^2_\phi & \ = \ & \frac{\lambda v^2}{4} - \sqrt{2} \mu v_{\Delta}
+ \frac{\lambda_1 + \lambda_4}{2} v^2_{\Delta} \, ,
\end{eqnarray}
and the mass terms for the scalar fields. 
The mass term of the doubly-charged Higgs  has the following form:  
\begin{eqnarray}
M^2_{H^{\pm\pm}} \ = \ \frac{\sqrt{2} \mu v^2
- \lambda_4 v^2 v_{\Delta} - 2 \lambda_3 v^3_{\Delta}}{2 v_{\Delta}} \, ,
\label{eq:mass_dch}
\end{eqnarray} 
where we have defined the mass basis of the doubly-charged Higgs as $\delta^{\pm\pm}=H^{\pm \pm}$.


The mass matrix of the singly-charged Higgs in the basis ($\phi^{\pm},\delta^{\pm}$) takes the
following form:
\begin{eqnarray}
 M^2_{\pm} \ = \  \left(\sqrt{2} \mu - \frac{\lambda_4 v_{\Delta}}{2}\right)
\begin{pmatrix}
v_{\Delta} & -\frac{v}{\sqrt{2}} \\
-\frac{v}{\sqrt{2}} & \frac{v^2}{2 v_{\Delta}}
\end{pmatrix}\,\,.
\label{phih}
\end{eqnarray}
We define the mass basis for the singly-charged scalars as ($G^{\pm},H^{\pm}$), which are related to the Lagrangian fields by an orthogonal matrix parameterized by the mixing angle $\beta^{\prime}$ with $\tan \beta^{\prime} = \frac{\sqrt{2}\, v_{\Delta}}{v}$ in the following way: 
\begin{eqnarray}
 \begin{pmatrix}
G^{\pm} \\
H^{\pm}
\end{pmatrix}
 \ =
 \ \begin{pmatrix}
\cos \beta^{\prime} & \sin \beta^{\prime} \\
-\sin \beta^{\prime} & \cos \beta^{\prime}
\end{pmatrix}\,\, 
\begin{pmatrix}
\phi^{\pm} \\
\delta^{\pm}
\end{pmatrix}\,.
\label{charged-higgs-mass-matrix}
\end{eqnarray}
After diagonalization, we get the
following eigenvalue for the physical singly-charged scalar: 
\begin{eqnarray}
M^2_{H^{\pm}} &\ = \ & \frac{(v^2 + 2 v^2_{\Delta}) (2\sqrt{2}\mu
- \lambda_4 v_{\Delta})}{4 v_{\Delta}}\, , 
\label{eq:mass_sch}
\end{eqnarray}
whereas $M^{2}_{G^{\pm}} = 0$, corresponding to the Goldstone bosons $G^{\pm}$ responsible
for the mass of the singly-charged gauge bosons $W^{\pm}$. In the limit $v_\Delta \ll v$, using Eq.~\eqref{eq:mudelta}, we can relate the doubly and singly-charged scalar masses in Eqs.~\eqref{eq:mass_dch} and \eqref{eq:mass_sch} as follows: 
\begin{align}
M^2_{H^{\pm\pm}} \ \simeq \ M^2_{H^\pm} - \frac{1}{2}\lambda_4 v^2 \, .
\end{align}  
We assume the quartic coupling $\lambda_4\ll 1$, so that the cascade decay of $H^{\pm\pm}\to H^\pm W^{\pm(*)}$ is kinematically suppressed, as compared to the dilepton decay mode $H^{\pm\pm}\to l^\pm l^\pm$. The diboson channel $H^{\pm\pm}\to W^\pm W^\pm$ is always suppressed, as long as $v_\Delta\lesssim 0.1$ MeV~\cite{Perez:2008ha}. 


Among the four neutral scalars, two of them are CP even and the remaining two 
are CP odd, as can be seen from Eq.~(\ref{phi-delta}). Assuming CP conservation, we can write the  CP-even and CP-odd mass matrices as follows: 
\begin{eqnarray}
M^2_{\rm even} \ & = & \ \begin{pmatrix}
\frac{\lambda v^2}{2} & -\sqrt{2} \mu v+
(\lambda_1 + \lambda_4)v_{\Delta}v\\
-\sqrt{2} \mu v+
(\lambda_1 + \lambda_4)v_{\Delta}v  & 
\frac{\sqrt{2} \mu v^2 + 4 (\lambda_2 + \lambda_3)v^3_{\Delta}}{2 v_{\Delta}} 
\end{pmatrix} \ \equiv \ 
\begin{pmatrix}
A & B \\
B & C
\end{pmatrix}, \label{eq:neutral1} \\
M^2_{\rm odd} \ & = & \ \sqrt{2} \mu
\begin{pmatrix}
2 v_{\Delta} & -v \\
-v & \frac{v^2}{2 v_{\Delta}}
\end{pmatrix} \, . 
\label{eq:neutral2}
\end{eqnarray}
We define the mass eigenbasis both for CP-even and CP-odd scalars, which are 
related to the Lagrangian fields in the following manner: 
\begin{eqnarray}
\begin{pmatrix}
h \\
H^0
\end{pmatrix}
\ =
 \ \begin{pmatrix}
\cos \alpha & \sin \alpha \\
-\sin \alpha & \cos \alpha
\end{pmatrix}
\begin{pmatrix}
h_0 \\
\delta_0
\end{pmatrix} \,\,\, {\rm and}\,\,\,
\begin{pmatrix}
G^0 \\
A^0
\end{pmatrix}
\ = \ 
\begin{pmatrix}
\cos \beta & \sin \beta \\
-\sin \beta & \cos \beta
\end{pmatrix}
\begin{pmatrix}
\xi_1 \\
\xi_2
\end{pmatrix}\, .
\end{eqnarray}
Here $h$ represents the SM-like Higgs and $G^0$ represents the
Goldstone boson associated with the SM $Z$-boson. 
The mixing angles related to the neutral fields depend on the VEVs and the quartic
couplings in the following manner: 
\begin{eqnarray}
\tan 2 \alpha \ = \  \frac{2 B}{A - C} \,, \quad  {\rm and}\, \quad \tan \beta \ = \ \frac{2 v_{\Delta}}{v}
= \sqrt{2} \tan \beta^{\prime}\,.
\end{eqnarray} 
The eigenvalues of the mass matrices \eqref{eq:neutral1} and \eqref{eq:neutral2} are given by 
\begin{eqnarray}
M^2_{h} &\ =\ & \frac{1}{2} \left[A + C - \sqrt{(A-C)^2 + 4B^2}\right] \, , \nn \\
M^2_{H^0} & \ = \ & \frac{1}{2} \left[A + C + \sqrt{(A-C)^2 + 4B^2}\right] \, , \nn \\
M^2_{A^0} & \ = \ & \frac{\mu (v^2 + 4 v^2_{\Delta})}{\sqrt{2} v_{\Delta}} \, , \nn \\
M^2_{G^0} & \ = \ & 0 \,.
\end{eqnarray}
We can easily write down the quartic couplings and the $\mu$-parameter appearing in Eq.~(\ref{potential})
in terms of the above-mentioned physical masses: 
\begin{eqnarray}
\lambda_1 & \ = \ & - \frac{2}{v^2 + 4 v^2_{\Delta}} M^2_{A^0}
+ \frac{4}{v^2 + 2 v^2_{\Delta}} M^2_{H^{\pm}} + \frac{\sin 2\alpha}{2 v v_{\Delta}}
(M^2_{h} - M^2_{H^0}) \, , \nn \\
\lambda_2 & \ = \ & \frac{1}{v^2_{\Delta}} \left[\frac{\sin^2 {\alpha}\  M^2_{h}
+ \cos^2 {\alpha} \ M^2_{H^0}}{2} + \frac{v^2 M^2_{A^0}}{2 (v^2 + 4 v^2_{\Delta})} 
- \frac{2 v^2 M^2_{H^{\pm}}}{v^2 + 2 v^2_{\Delta}} + M^2_{H^{\pm\pm}}\right] \, , \nn \\
\lambda_3 & \ = \ & \frac{1}{v^2_{\Delta}} \left[\frac{-v^2 M^2_{A^0}}{v^2 + 4 v^2_{\Delta}} 
+ \frac{2 v^2 M^2_{H^{\pm}}}{v^2 + 2 v^2_{\Delta}} - M^2_{H^{\pm\pm}}\right] \, , \nn \\
\lambda_4 & \ = \ & \frac{4 M^2_{A^0}}{v^2 + 4 v^2_{\Delta}}
- \frac{4 M_{H^{\pm}}}{v^2 + 2 v^2_{\Delta}} \, , \nn \\
\lambda & \ = \ & \frac{2}{v^2} \left[\cos^2 {\alpha}\ M^2_{h} + \sin^2 {\alpha}\ M^2_{H^0} \right] \nn \\
\mu & \ = \ & \frac{\sqrt{2} v_{\Delta} M^2_{A^0}}{v^2 + 4 v^2_{\Delta}}\,.
\end{eqnarray}
There are miscellaneous constraints on the quartic couplings from electroweak
precision measurements, absence of tachyonic modes, boundedness of the potential, unitarity, vacuum stability and naturalness considerations~\cite{Arhrib:2011uy, Chun:2012jw, Dev:2013ff, Kobakhidze:2013pya, Chabab:2015nel, Haba:2016zbu, Das:2016bir, Xu:2016klg, Dev:2017ouk, Ghosh:2017pxl}, all of which have been taken into account in choosing the benchmark points in our subsequent collider analysis.  
  
\section{Signatures at $e^-p$ Collider\label{coll}}
In this section, we analyze  the discovery prospects of a doubly-charged Higgs in the future $e^- p$ collider. We consider the FCC-eh design with beam energies $ 60$ GeV (for electron) and 
$ 50$ TeV (for proton), so the center-of-mass energy is $\sqrt s\simeq \sqrt{4E_eE_p}=3.46$ TeV and a TeV-scale doubly-charged Higgs can be produced on-shell, through  associated production,  along with a charged lepton or neutrino and an  energetic jet.  We consider the following two production modes (see Fig.\,\ref{fig:feyn12}):
\begin{figure}[t!]
	\centering
	\includegraphics[width=0.65\textwidth]{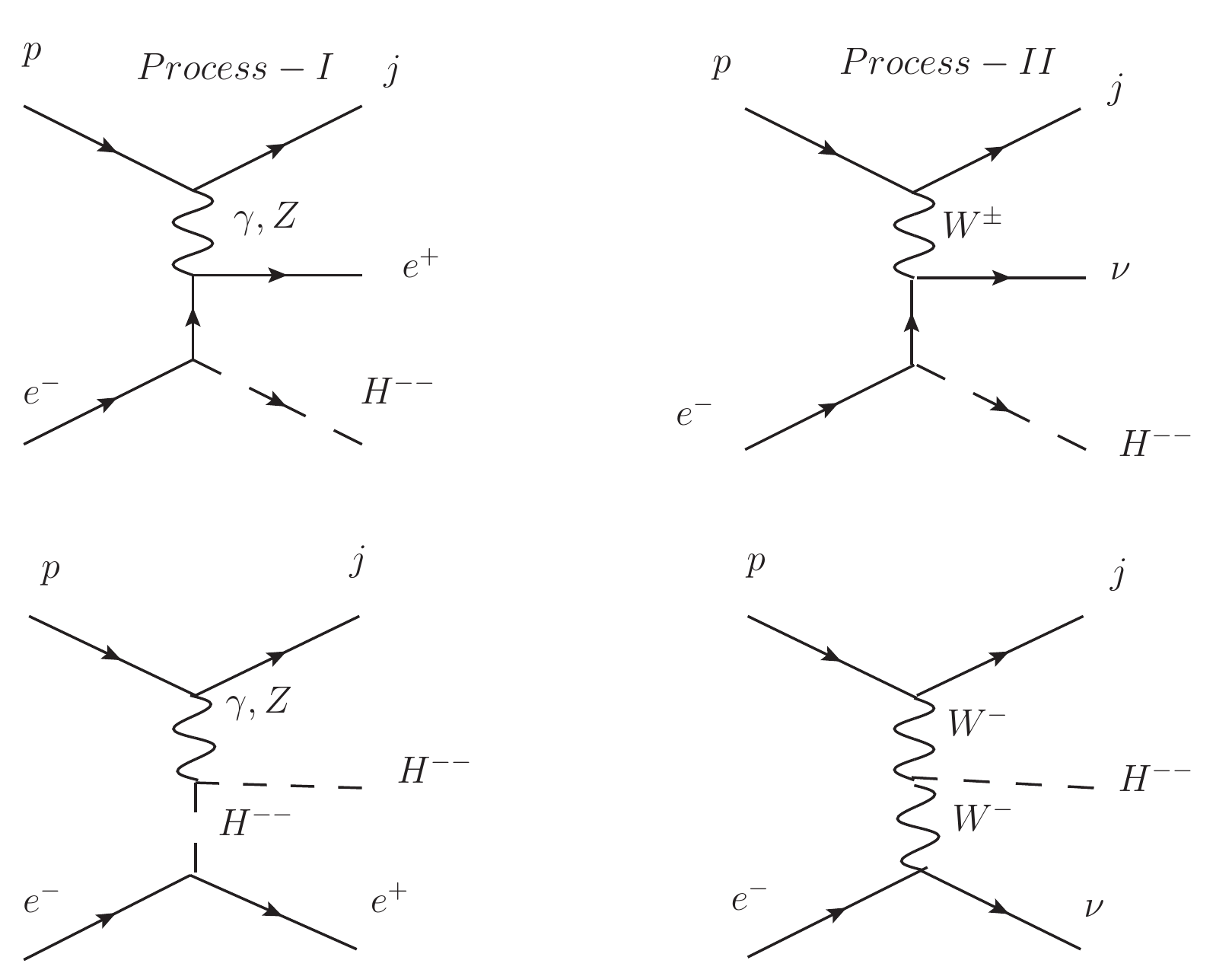}
	\caption{Feynman diagrams for the single production of doubly-charged Higgs boson
	at $e^- \, p$ collider. An additional diagram exists for {\it Process-I} with the $t$-channel photon exchange 
	appearing before the doubly-charged Higgs is radiated, which is not shown here but included in our analysis.}
	\label{fig:feyn12}
\end{figure}
\begin{itemize}
\item
{\bf Process\,-\,I} : $e^{-} p \ \rightarrow \ e^{+}\,j\,H^{--}$, \, $H^{--} \rightarrow l^{-}\,l^{-}$\, ,

\item
{\bf Process\,-\,II} : $e^{-} p \ \rightarrow \ j\,\nu\,H^{--}$, \, $H^{--} \rightarrow l^{-}\,l^{-}$\,,

\end{itemize}
where $l = e, \mu$. The diagrams in Fig.\,\ref{fig:feyn12} clearly show the dominant production mode, depending on the choice of 
the triplet VEV. While the first three diagrams would be relevant for low values of the triplet VEV as it would lead to a larger 
Yukawa $Y^{\Delta}_{ee}$  coupling, they become less relevant when the VEV is chosen larger, in which case the vector-boson-fusion (VBF) diagram becomes dominant.\footnote{Thus, in the small VEV limit, our analysis equally applies to both $SU(2)_L$ and $SU(2)_R$ scalar triplets, such as in the left-right symmetric model. A polarized electron beam could, in principle, distinguish between the two contributions, by analyzing the angular distributions of the final-state leptons, for instance. We have only considered the unpolarized beam in this analysis and the effect of polarization is left for future study.}  In {\it Process-I}, the contribution of $Z$-exchange is much smaller, as compared to the photon exchange. As for the photon-mediated processes, depending on the photon virtuality, the proton may remain intact (elastic process) or dissociates (inelastic process).   As discussed below, a selection cut of 40 GeV on the transverse  momentum of the final state jet is imposed in our analysis. This implies large photon virtuality (compared to the mass of the hadronic final state), and hence, the inelastic process, where the interaction involves a quark inside the proton, gives the dominant contribution.

It is worthwhile to note that  the production of a doubly-charged 
scalar at FCC-eh has a major difference when compared with its production at the LHC. At the LHC, one can never exploit 
the large Yukawa $Y^{\Delta}_{ee}$, although it still plays a very effective role in deciding the dominant decay modes for the 
doubly-charged Higgs. 
Notably, large $Y^{\Delta}$  at the LHC leads to the heavier-flavor charged lepton (tau) to be pair-produced more effectively from the $H^{\pm\pm}$ decay when the neutrino masses exhibit ``normal hierarchy", whereas for ``inverted hierarchy" the electron and muon flavors will be pair-produced more copiously than the tau flavor. On the other hand, for $Y^\Delta\lesssim 10^{-7}$, either the decay of $H^{\pm\pm}\to l^\pm l^\pm$ is displaced~\cite{Dev:2018kpa, Antusch:2018svb}, or the diboson decay modes, such as $H^{\pm\pm}\to W^\pm W^\pm$, take over the dilepton mode~\cite{Kanemura:2013vxa}, depending on the $H^{\pm\pm}$ mass. So  $Y^{\Delta}_{ee}$ still 
has a secondary role in the phenomenology of doubly-charged Higgs at hadron colliders, but cannot be {\it directly} probed due to its hadrophobic nature.
In this work we explore this region of parameter space where the triplet VEV is small, $v_\Delta\sim {\cal O}(0.1)$ eV, and  $Y^{\Delta}_{ee}$ can be directly probed, as it plays the all important role in the doubly-charged scalar phenomenology at the FCC-eh. 

It is also worth noting that we 
always produce the negatively-charged state at $e^- \, p$ colliders when produced singly. The decay of $H^{--}$ into two 
same-sign dileptons leads to the following signals, which we explore in detail:
\begin{itemize}
\item
{\bf Signal\,-\,I} : $3\,l$ + $\, \geq 1\,j$\, (with at least two negatively-charged leptons of same flavor),
\item
{\bf Signal\,-\,II} : $\geq 2\,l\, +\, \geq 1\,j$\, (where at least  two negatively-charged leptons are of same flavor).
\end{itemize}

Note that Signal-I with a trilepton final state comes from the subprocesses listed as 
{\it Process-I} in Fig.~\ref{fig:feyn12} while Signal-II, which has at least two charged leptons 
can get contributions from all the subprocesses in Fig. \ref{fig:feyn12}, {\it i.e.} {\it Process-I}
and {\it Process-II}. Signal-II also has substantial MET due 
to the presence of neutrinos in the final state as shown in  {\it Process-II}. Although a trigger 
upon the MET could reduce the contributions from subprocesses in {\it Process-I} which will 
have very little MET, it could prove useful in SM background suppression.

The production cross-sections for $e^{-} p \rightarrow e^{+}\,j\,H^{--}$ and $e^{-} p \rightarrow j\,\nu\,H^{--}$ depend on the Yukawa coupling 
strength $Y^{\Delta}_{ee}$ quadratically. So we must choose large diagonal entries for $Y^\Delta$, as mentioned before. The off-diagonal Yukawa couplings are then required to be quite small, so as to satisfy the LFV constraints from $\mu \to 3e$, $\mu \to e \gamma$, etc. 
The current bounds on the branching ratio of $\mu \to e \gamma$ and $\mu \to eee$ are $4.2 \times 10^{-13}$~\cite{TheMEG:2016wtm}
 and $10^{-12}$~\cite{Bellgardt:1987du}, respectively. For our model, if the doubly-charged 
Higgs mass is around 1 TeV, then the corresponding bounds on the elements of 
$Y^{\Delta}_{\alpha}$ matrix are given by~\cite{Akeroyd:2009nu, Dev:2017ouk, Dev:2018kpa}
\begin{align}
 Y^{\Delta}_{\mu e} Y^{\Delta}_{ee} \ < \ 2.38 \times 10^{-5}   &&{\rm and}&&
 Y^{\Delta}_{ee} Y^{\Delta}_{e\mu} +
Y^{\Delta}_{e\mu} Y^{\Delta}_{\mu\mu} +
Y^{\Delta}_{e\tau} Y^{\Delta}_{\mu\tau} \ < \ 2.42 \times 10^{-4}\,.  
\label{eq:mu23e-and-m2eg}
 \end{align}
Note that the diagonal elements $Y^\Delta_{ii}$ cannot be arbitrarily large either. Specifically, a large $Y^\Delta_{ee}$ would significantly contribute to the Bhabha scattering $e^+e^-\to e^+e^-$ via the $t$-channel $H^{\pm\pm}$, altering both the total cross section and the differential distributions~\cite{Abdallah:2002qj, Abbiendi:2003pr, Achard:2003mv}. Using the LEP data on the $e^+e^-\to e^+e^-$ cross-section measurements~\cite{Abbiendi:2003pr, Abdallah:2005ph}, we obtain a 90\% CL upper limit~\cite{Dev:2018upe}
\begin{align}
\frac{|Y^\Delta_{ee}|^2}{M^2_{H^{\pm\pm}}} \ < \ 1.2\times 10^{-7} \, . 
\label{eq:LEP}
\end{align}
Thus, for a TeV-scale $H^{\pm\pm}$, we must have $|Y^\Delta_{ee}|<0.35$.\footnote{Note that the neutrinoless double beta decay limits are not relevant here, because the amplitude of the $H^{--}$-mediated process depends on the $H^{--}W^+W^+$ coupling, which is suppressed by the small triplet VEV $v_\Delta$.}  Similarly, both $Y^\Delta_{ee}$ and $Y^\Delta_{\mu\mu}$ contribute to the muonium-antimuonium oscillation, i.e. the LFV conversion of the bound states $(\mu^+e^-)\leftrightarrow (\mu^-e^+)$, whose oscillation probability has been constrained by the MACS experiment~\cite{Willmann:1998gd}, and leads to the 90\% CL upper limit~\cite{Dev:2018upe}
 \begin{align}
\frac{|(Y^\Delta)^\dagger_{ee}Y^\Delta_{\mu\mu}|}{M^2_{H^{\pm\pm}}} \ < \ 1.2\times 10^{-7} \, . 
\label{eq:muonium}
\end{align}
There exist other constraints involving the diagonal elements from $\mu\to e\gamma$ and electron/muon anomalous magnetic moment~\cite{Dev:2018upe}, which are however weaker than the two constraints mentioned above.

\begin{figure}[t!]
	\centering
	\includegraphics[width=0.45\textwidth]{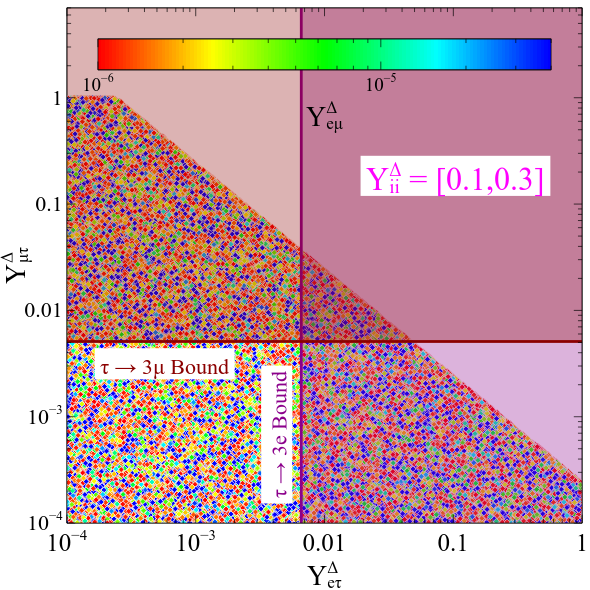}
	\caption{Allowed off-diagonal Yukawa couplings $Y^{\Delta}_{\mu \tau}$ and $Y^{\Delta}_{e \tau }$ as a function of $Y^{\Delta}_{e\mu}$ (shown by the color code). The diagonal couplings $Y^{\Delta}_{ee,\mu\mu}$ are randomly varied between the range of 0.1 and 0.3, for a fixed  $M_{H^{\pm\pm}}=1$ TeV. The vertical and horizontal shaded regions are excluded from the experimental constraints on the tau-LFV decay modes $\tau\to 3e$ and $\tau\to 3\mu$, respectively, for fixed $Y^{\Delta}_{ee,\mu\mu}=0.3$.}
	\label{fig:mu23e}
\end{figure}

By choosing the diagonal elements $Y^{\Delta}_{ee,\mu\mu}\in [0.1,0.3]$ to be consistent with the bounds in ~\eqref{eq:LEP} and \eqref{eq:muonium}, and assuming $Y^\Delta_{\tau\tau}$ to be small (since it does not play any role in our subsequent collider analysis), 
we show the allowed values of the off-diagonal Yukawa coupling parameters $Y^\Delta_{e\mu,e\tau,\mu\tau}$ as a scatter plot in 
Fig.\,\ref{fig:mu23e} which satisfy the LFV constraints shown in \eqref{eq:mu23e-and-m2eg}. Here we have chosen $M_{H^{\pm\pm}}=1$ TeV as a benchmark value. The vertical and horizontal shaded regions are excluded from the experimental constraints on the tau-LFV decay modes $\tau\to 3e$ and $\tau\to 3\mu$, respectively~\cite{Amhis:2016xyh}, for fixed $Y^{\Delta}_{ee,\mu\mu}=0.3$.  It is clear that the off-diagonal Yukawa couplings are constrained to be small for our choice of large diagonal Yukawa couplings.  

To simulate the signal events, we implement the model described in Section~\ref{model} in {\tt Feynrules}~\cite{Alloul:2013bka} to create the UFO model files.  
The package {\tt Madgraph (v2.4.3)} \cite{Alwall:2014hca} is used to 
generate parton-level events using {\tt CTEQ6L1}~\cite{Pumplin:2002vw} 
parton distribution functions for the colliding proton beam. The showering and 
subsequent hadronization has been done with {\tt Pythia (v6.4)}~\cite{Sjostrand:2006za}. 
Finally, we pass the generated events through 
{\tt Delphes (v3.4.1)} \cite{deFavereau:2013fsa} 
for detector simulation. To reconstruct 
jets in the final state, we have used the anti-$k_t$ jet clustering algorithm~\cite{Cacciari:2008gp} with a radius parameter $R=0.4$ in {\tt Fastjet}~\cite{Cacciari:2011ma} for jet formation. 

\begin{figure}[h!]
	\centering
	\includegraphics[width=0.5\textwidth]{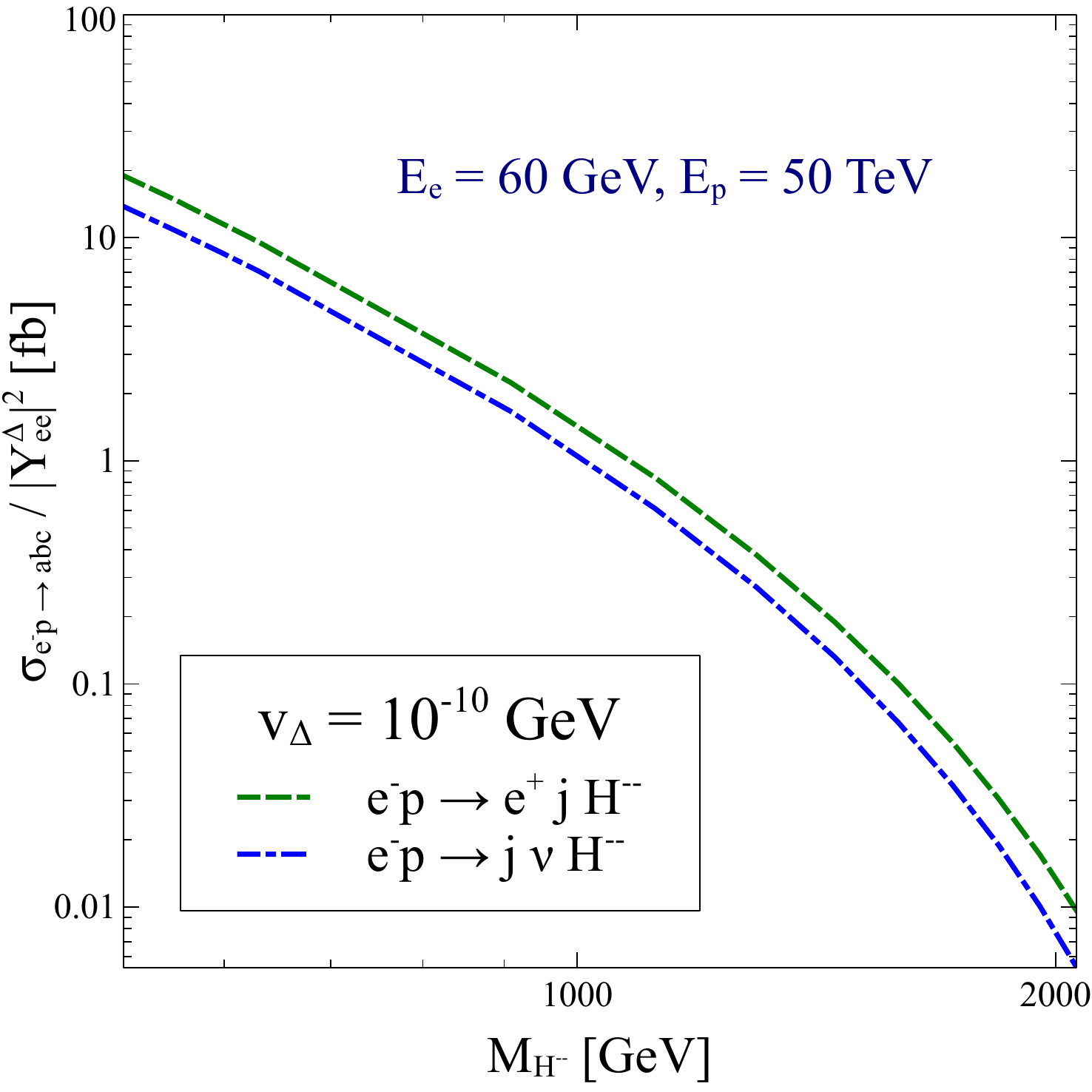}
	\caption{Parton-level production cross-sections (normalized to $|Y^\Delta_{ee}|^2=1$) of the doubly-charged Higgs for {\it Process-I} (green, dashed) and {\it Process-II} (blue, dot dashed) shown in Fig.~\ref{fig:feyn12} at $e^-p$
	collider with beam energies of 60 GeV (for electron) and 50 TeV (for proton). We have chosen the triplet VEV as $v_{\Delta} = 0.1$ eV. }
	\label{fig:prodcs}
\end{figure}
 We  show the variation of the single production cross-section
 of the doubly-charged Higgs as a function of its mass in Fig.\,\ref{fig:prodcs} for both 
{\it Process-I} and {\it Process-II}. Note that {\it Process-I} with a larger multiplicity of 
subprocesses as well as the enhanced coupling of the photon to the two units of charge 
of the doubly-charged scalar is significantly larger than the {\it Process-II} with 
$W$ boson exchange, where the second (VBF) subprocess hardly contributes to the 
production cross-section due to the very small VEV. The mass range of interest for 
the doubly-charged Higgs for our analysis is dictated by the limits from $\sqrt s=13$ TeV LHC which is about 800 GeV~\cite{CMS-PAS-HIG-16-036, Aaboud:2017qph}. 
As shown in Fig.~\ref{fig:prodcs}, for $M_{H^{\pm \pm}}$ in between 500 GeV and 1 TeV, the cross-section varies between 20 fb and 2 fb.  For larger mass values the cross-section falls rapidly, as the mass
approaches the center-of-mass energy of $\sqrt s=3.46$ TeV for the FCC-eh. 

In Table~\ref{tab:benchmarks} we give the choice of our benchmark points which give us three different mass values for the doubly-charged Higgs boson. A closer look at Eq.~\eqref{eq:mass_dch} shows that 
for a given choice of triplet VEV and $\lambda_4$, the value of $\mu$ would dictate the mass values. For simplicity, we just vary $\mu$ to generate different values for the doubly-charged 
scalar mass, while making sure that the quartic coupling values are consistent with all the theoretical constraints~\cite{Arhrib:2011uy, Chun:2012jw, Dev:2013ff, Kobakhidze:2013pya, Chabab:2015nel, Haba:2016zbu, Das:2016bir, Xu:2016klg, Dev:2017ouk, Ghosh:2017pxl}. The first benchmark point we choose has a doubly-charged scalar mass close to the current LHC bound, which is kind of the most optimistic scenario for its discovery at FCC-eh. The $Y^\Delta_{ee}$ value chosen for each benchmark is the maximum value allowed by the LEP limit [cf.~Eq.~\eqref{eq:LEP}] for the corresponding doubly-charged scalar mass. We also give the total decay width of the doubly-charged scalar in each case, which is simply given by $\Gamma_{H^{--}}\simeq \frac{|Y^\Delta_{ee}|^2}{8\pi}M_{H^{--}}$, assuming that it dominantly decays to the $ee$ final state only.

\begin{table}[t!]
\begin{tabular}{||c|c|c|c|c|c|c||}
    \hline
    \hline
    Benchmark Points & $v_{\Delta}$ [GeV] & $Y^\Delta_{ee}$ & $\mu$ [GeV] &$M_{H^{--}}$ [GeV] & $M_{H^{-}}$ [GeV] & $\Gamma_{H^{--}}$ [GeV]\\ 
    \hline
    BP1 & 10$^{-10}$ & 0.31 & $2 \times 10^{-9} $ & 908.6 & 916.9 & 3.5\\ 
\hline
    BP2 & 10$^{-10}$ & 0.39 & $3 \times 10^{-9} $ & 1119.6 & 1126.2 & 6.8\\
    \hline
    BP3 & 10$^{-10}$ & 0.45 & $4 \times 10^{-9} $ & 1296.7 & 1302.5 & 10.4 \\
\hline
\hline   
    \end{tabular}
\caption{Representative benchmark points used in our analysis for studying the 
doubly-charged scalar production at the FCC-eh. }
\label{tab:benchmarks}
\end{table}

In what follows, we perform a detailed cut-based analysis for the signal and background for both Signal-I and Signal-II identified above.\footnote{Since we consider the on-shell production of the doubly-charged scalar, and its width/mass is less than 1\% (see Table~\ref{tab:benchmarks}), the interference between signal and background is negligible.} 

\subsection{Signal\,-\,I ($3\,l$ + $\, \geq 1\,j$)\,:}
The trilepton signal containing at least a pair of same-sign, same-flavor charged leptons
can come from {\it Process-I} only. The SM background for the signal can have several 
sources, depending on their rate of production. We list the dominant and relevant 
SM processes below: 
\begin{enumerate}
\item The most dominant SM background for the above final state, at an $e^- \, p$ collider is the 
irreducible background given by $e^{-} p \rightarrow j\, e^-\, l^+\, l^-$.

\item A subdominant contribution can also come from the reducible,  on-shell $Z$ boson 
production which then decays to a pair of charged leptons via the process 
$e^{-} p \rightarrow e^{-} j Z$. Note that the requirement of a same-sign, same-flavor pair of 
charged leptons in the final state ensures that only the $Z$ decay to $e^- \, e^+$ contributes.

\item A relatively weaker background contribution could also come from an $\alpha_{\rm em}$-suppressed process $e^{-} p \rightarrow e^{-} j \gamma Z$ where the $Z$ boson again decays leptonically and specifically to $e^- \, e^+$ only, while the photon is misidentified as a lepton. However, the photon misidentification rate is less than $10^{-4}$ for electron and $10^{-3}$ for muon~\cite{Aad:2010fh}\footnote{These numbers from Ref.~\cite{Aad:2010fh} include both photons and jets misidentified as leptons. So the photon-only contribution will be even smaller.}, so this background will be negligible, as we will see below (in Table~\ref{tab:smbkg_3l}). 

\item One should also account for the possibility of contributions to the SM background where a jet may fake a charged lepton. Notwithstanding the fact that the fake-rate for such events would be significantly smaller, one begs the question whether a relatively sizable cross-section for jet-enriched final state leaves an imprint on the signal under consideration. For the estimation of fake-rate 
induced background, we  have considered the process   $e^{-} p \rightarrow e^{-} j j$ and assumed the rate for a jet faking a charged lepton as  0.1$\%$~\cite{ATLAS:2016iqc}. 
\end{enumerate}

To begin with, all the above channels for the background contribute substantially with basic acceptance cuts for the subprocess under consideration. However, the contributions to the actual 
signal topology starts showing the relative importance of the backgrounds that would eventually 
contribute to the final analysis. We begin the event generation by demanding:
\begin{enumerate}
\item[{\bf A0.}] 
The final state leptons and jets satisfy a minimum requirement 
for the transverse momenta and their pseudorapidity, given by  
$p_{T}^j > 40 $ GeV, $p_{T}^{l} > 10$ GeV,
$|\eta^{j,l}| \leq 4.5$.
\end{enumerate}

In Table \ref{tab:smbkg_3l} and Table \ref{tab:signal_3l} we present the cut-flow estimates for both the 
SM background and the signal cross-section at the proposed FCC-eh machine. The {\bf A0} cuts 
are simply aimed at understanding the signal and background characteristics in a few kinematic 
variables which could be then utilized  to device suitable cuts that help in improving the 
significance of the signal. Note that for the signal, the decay of the doubly-charged Higgs boson 
is an important variable. Since the off-diagonal elements of $Y^\Delta$ are assumed to be small, only the lepton-flavor-conserving modes $H^{\pm\pm}\to l^\pm l^\pm$ occur. As mentioned before, we shall
neglect the $\tau$ decay mode and focus only on the $e^-$ and $\mu^-$ modes. We first consider the simple case where $H^{\pm\pm}$ decays to $e^\pm e^\pm$ with 100\% branching ratio, {\it i.e.} only the $ee$ element of $Y^\Delta$ is large and all other elements small. Later in Section~\ref{sec:mumu}, we will also consider a case where $H^{\pm\pm}$ decays to both $e^\pm e^\pm$ and $\mu^\pm \mu^\pm$, each with 50\% branching ratio, {\it i.e.} both $Y^\Delta_{ee}$ and $Y^\Delta_{\mu\mu}$ are of same size. 
This would help in generalizing the
result when exploring the sensitivity for heavier mass values of the doubly-charged Higgs.  To ensure that we only select  electron or muon-flavored  leptons, we 
impose a further requirement: 
\begin{enumerate}
\item[{\bf A1.}] The final state events must contain at least three charged leptons, among which we 
have $N(e^{+}) = 1$, $N(\mu^{+}$) = 0, and either a pair of $e^-$ or $\mu^-$   {\it i.e.}, 
$N$($e^{-}$ or $\mu^{-}$) = 2 while the number of jets in the final state must be greater than 
or equal to one,  {\it i.e.} $N(j) \geq 1$.
\end{enumerate}
\begin{figure}[t!]
	\centering
	\includegraphics[angle=0,height=6cm,width=8.5cm]{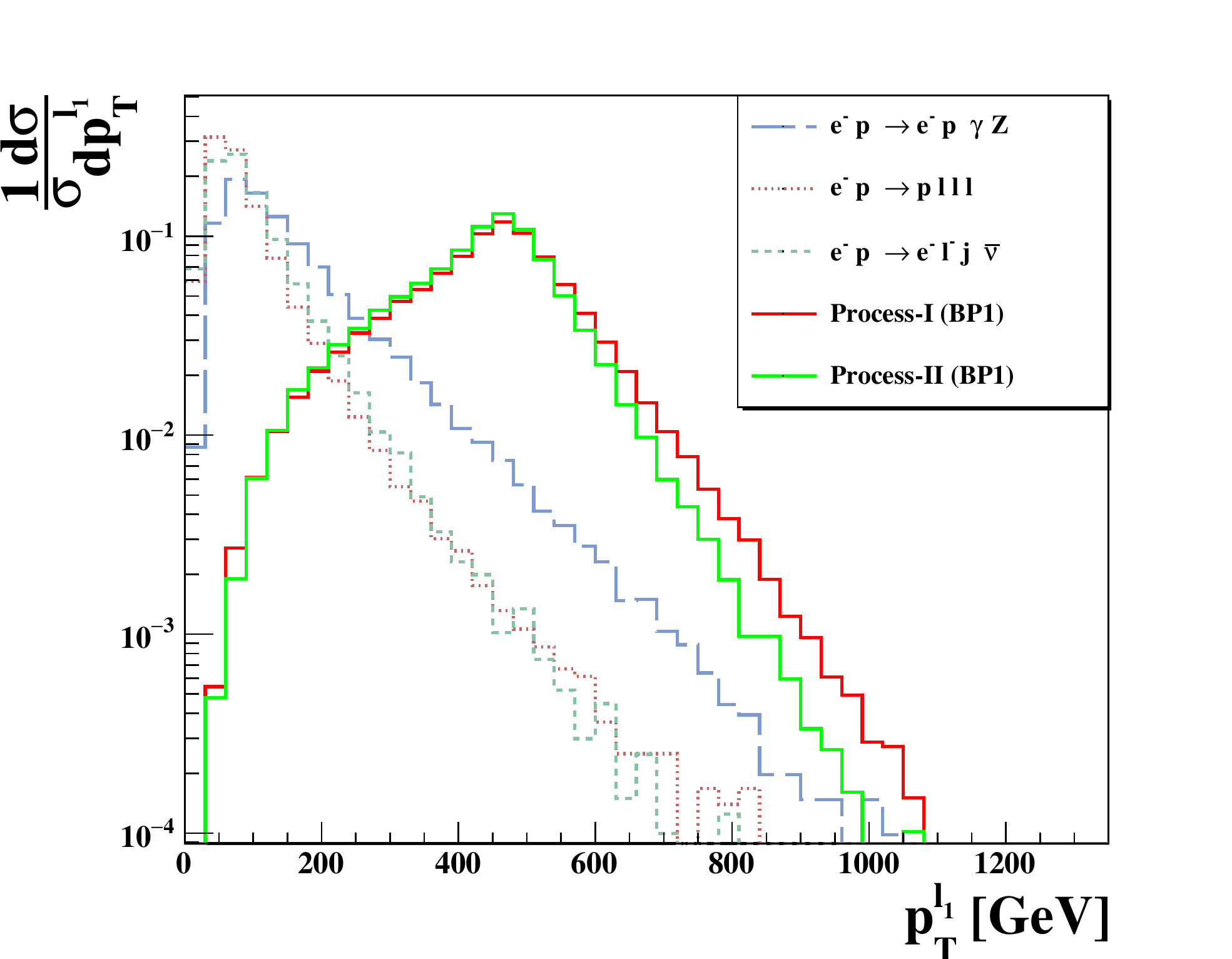}
    \includegraphics[angle=0,height=6cm,width=8.5cm]{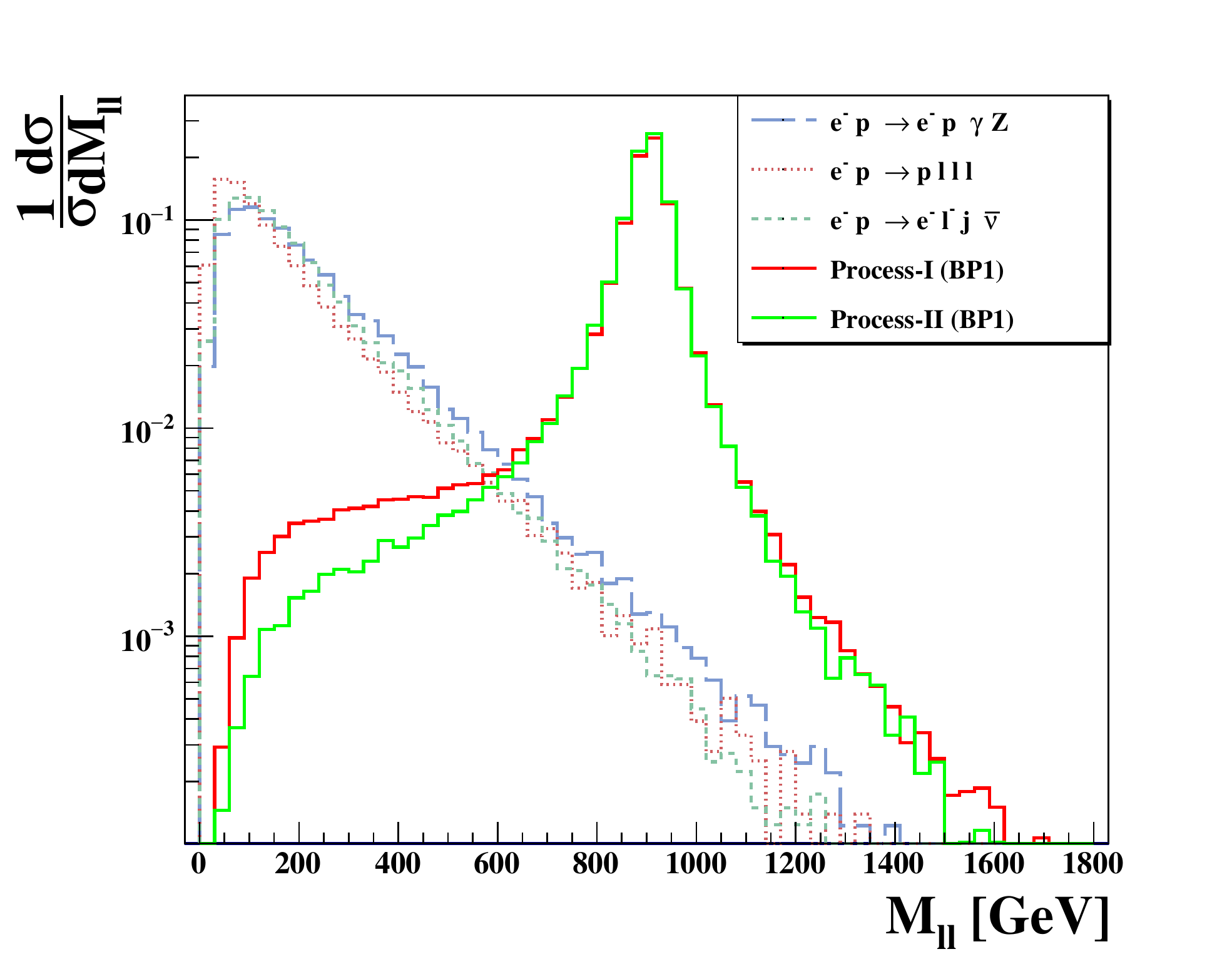}	
	\caption{The normalized distribution of $p_T$ of leading charged lepton (left) and the invariant mass 
	($M_{ll}$) of the same-sign lepton pair (right) for both the signal and SM background.} 
	\label{fig:kin_dist}
\end{figure}
This requirement is good enough to suppress most of the SM background, especially the jet-faking-lepton channel as
seen from Table \ref{tab:smbkg_3l}. However, note that the doubly-charged Higgs signal strength is totally dependent on
the production cross-section which depends on its mass. As LHC already puts strong limits, the mass range of our interest would entail small production cross-section even at the FCC-eh machine, evident in Fig. \ref{fig:prodcs}. Thus we need to 
find methods of suppressing the background further to ensure sensitivity to heavier mass values of the doubly-charged Higgs. The two very significant characteristics one expects is that the charged leptons coming from the decay of the heavy scalar would carry a significantly large transverse momentum and the invariant mass of the same-sign, same-flavor lepton 
pair would be concentrated in regions of very high mass, specific to the mass of the doubly-charged Higgs. In comparison, 
the SM background events should have leptons with relatively softer $p_T$, as well as a more continuum invariant mass distribution peaked at smaller values of $M_{ll}$.

In Fig.\,\ref{fig:kin_dist}, we show the transverse momentum ($p_T$) distribution of the leading charged lepton, both for  signal and background. Additionally, we also show the distribution of the invariant mass of the same-sign same-flavor leptons which shows expected differences between the signal and background events. We therefore impose the 
following cuts ({\bf A2} and {\bf A3}), which turn out to be very much useful in reducing the backgrounds events. 
\begin{enumerate}
\item[{\bf A2.}] We impose cuts on the transverse momentum of the leading and sub-leading charged lepton 
and also on the leading jet as:
 $p_{T}^{l_i} > 50$ GeV ($i = 1, 2$) and $p^{j_1}_{T} > 50$ GeV. 
\item[{\bf A3.}] We also demand that  the invariant mass $M_{ll}$ of same-sign lepton pair is large, 
with  $M_{ll} >  800\,\, {\rm GeV}$. 
 This cut reduces the background significantly. 
\end{enumerate}
%
\def\I{i}
\begin{center}
\begin{table}[t!]
\begin{tabular}{||c|c||} \hline \hline
\begin{tabular}{c}
    \multicolumn{1}{c}{SM Backgrounds for FCC-eh}\\    \hline
    Channels  \\     \hline
    $e^{-}$ $p$ $\rightarrow$ $e^{-}$ $j$ $\gamma$ $Z$\,($\rightarrow l^+\,l^-$)  \\         \hline    
    $e^{-}$ $p$ $\rightarrow$ $j\, e^-\, l^+\, l^-$ \\ \hline
    $e^{-}$ $p$ $\rightarrow$ $e^{-}$ $j\,\, j$  \\      \hline
    Total Background \\ 
\end{tabular}
&
\begin{tabular}{c|c|c|c}
    \multicolumn{3}{c}{~~~~~~~~~~Effective Cross-section after applying cuts (fb)}\\       \hline 
   ~~~~~~{\bf A0}~~~~~~ & \,\,~~~~~~~~{\bf A1}~~~~~~~\,\, &\,\, ~~~~~~{\bf A2}~~~~~~\,\, & ~~~~~~{\bf A3}~~~~~~\\    \hline
    0.12 & 0.03 & 0.02 & -   \\     \hline
53.13 & 12.57 & 3.63 & 0.08  \\ \hline
    $104.08 \times 10^3$  & 0.2 & - & -   \\    \hline
    ~ &  ~ & ~ & 0.08 \\
\end{tabular}\\ \hline \hline
\end{tabular}
\caption{Cut-flow table of the cross-section for the
relevant SM background channels for the cuts {\bf A0 - A3} as mentioned in the text at the FCC-eh collider with 3.46 TeV center-of-mass energy. We assume $l = e$ or $\mu$.}
\label{tab:smbkg_3l}
\end{table}
\end{center}
\def\I{i}
\begin{center}
\begin{table}[h!]
\begin{tabular}{||c|c||} \hline \hline
\begin{tabular}{c|c|c}
    \multicolumn{3}{c}{Signal for FCC-eh}\\     \hline
    Process & Signal & BP \\     \hline
    I & 3\,$l$ + $\geq 1j$  & BP1  \\     \hline
    I & 3\,$l$ + $\geq 1j$  & BP2  \\        \hline
    I & 3\,$l$ + $\geq 1j$  & BP3  \\
    \end{tabular}
&
\begin{tabular}{c|c|c|c}
    \multicolumn{3}{c}{~~~~~~~~~Effective Cross-section after applying cuts (fb)}\\       \hline 
    ~~~~~~~~{\bf A0}~~~~~~~~ & \,\,~~~~~~~~{\bf A1}~~~~~~~~\,\, & ~~~~~{\bf A2}~~~~~ & ~~~~~{\bf A3}~~~~~\\
    \hline
    0.21 & 0.12 & 0.11 & 0.09   \\    \hline
    0.13 & 0.07 & 0.07 & 0.06   \\ \hline
    0.08 & 0.05 & 0.04 & 0.04  \\
\end{tabular}\\ \hline \hline
\end{tabular}
\caption{Cut-flow table of cross-section for signal
for the cuts {\bf A0 - A3} as mentioned in the text 
at the FCC-eh collider with 3.46 TeV center-of-mass energy. Here we  assume  ${\rm BR}(H^{\pm\pm}\to e^{\pm}e^{\pm})=100\%$.}
\label{tab:signal_3l}
\end{table}
\end{center}
In fact, the only relevant SM background events that remain after cut {\bf A3} are from the process 
$e^- \, p \to e^- \, l^+l^- j$.  The suppression in the signal cross-section for all the three benchmark points is 
about 50\% for the event selection cuts, as can be seen in Table\,\ref{tab:signal_3l}.

\subsection{Signal\,-\,II ($\geq 2\,l\, +\, \geq 1\,j$)\,:}

We now consider the multi-lepton channel where we have at least two charged leptons in the final state and both need to 
be of the same sign and flavor. This would correspond to the inclusive search of the signal for a singly-produced doubly-charged Higgs coming from both {\it Process-I} and {\it Process-II}. The combination of both the processes for the inclusive search 
would mean that one avoids kinematic selection of events that could potentially reduce contribution from any of the particular processes.  This on the other hand would mean that there would be additional sources of background for the 
final state under consideration from the ones discussed in the previous subsection which are listed below.
\begin{enumerate}
\item The irreducible background coming from $e^{-} p \rightarrow e^{-} l^{-} j \bar{\nu_l}$.
Again, for the same-sign, same-flavor requirement of the charged-lepton pair, the signal topology ensures 
that $l=e$. 


\item To add to the fake-rate contribution, we must consider the process $e^{-} p \rightarrow e^{-} j \gamma$
where $\gamma$ can  fake the charged lepton $e^-$.

\item In addition to fake rates, we note that charge mismeasurement can also lead to potential backgrounds which mimic 
the signal.  We evaluate $e^{-} p \rightarrow e^{-} e^{+} j\, \nu$ process, where we consider a conservative $0.1\%$ \cite{Khachatryan:2016olu} as the charge mismeasurement or charge-flip rate/probability in the detector.

\item We also estimate the background that  arises due to the jet faking a lepton 
which as before is assumed to be 0.1$\%$ .
\end{enumerate}

As before, {\bf B0} and {\bf B1} discussed below define the selection of events for the signal in consideration.
\begin{enumerate}
\item[{\bf B0.}] As in {\bf Signal-I}, the final state leptons and jets must satisfy a minimum requirement 
for the transverse momenta and their pseudorapidity, given by  
$p_{T}^j > 40 $ GeV, $p_{T}^{l} > 10$ GeV,
$|\eta^{j,l}| \leq 4.5$.
\item[{\bf B1.}] We then demand that we select only those events which have at least 2 charged leptons and a minimum 
of 1 jet, which defines the signal as $\geq 2l + \geq 1j$.
\end{enumerate}

Similar to the previous analysis, we again implement a few similar cuts on the events to suppress the background and 
improve signal sensitivity.

\begin{enumerate}
\item[{\bf B2.}] The events satisfying {\bf B0} and {\bf B1} are then required to contain two 
leptons of same sign and same flavor. The kinematic properties of these charged leptons is expected to 
be similar to what was observed in Fig.\,\ref{fig:kin_dist} and therefore we impose a cut on the $p_T$ of the
leading and sub-leading charged lepton as $p_{T}^{l_i} > 50$ GeV ($i = 1, 2$) while the leading 
jet must satisfy $p^{j_1}_{T} > 50$ GeV. 
\item[{\bf B3.}] To utilize the hardness of the invariant mass distribution of the same sign-same-flavor dilepton for the 
signal in suppressing the background, we further demand  that $M_{ll} >  800\,\, {\rm GeV}$\,.
\end{enumerate}

\def\I{i}
\begin{center}
\begin{table}[h!]
\begin{tabular}{||c|c||}
\hline
\hline
\begin{tabular}{c}
    \multicolumn{1}{c}{SM Backgrounds for FCC-eh }\\ 
    \hline
    Channels  \\ 
    \hline
    e$^{-}$ p $\rightarrow$ $e^{-} l^{-}\, j\, \bar{\nu}$  \\ 
 \hline
    $e^{-}$ $p$ $\rightarrow$ $e^{-}$ $j$ $\gamma$   \\     
    \hline
    
   e$^{-}$ p $\rightarrow$ $j\,l\,l\,l$  \\
    \hline
$e^{-}$ $p$ $\rightarrow$ $e^{+}\,e^{-}\,j\,\nu$ \\
    \hline
$e^{-}$ $p$ $\rightarrow$ $e^{-}$ $j$\,\, $j$   \\     
    \hline
    Total Background \\ 
\end{tabular}
&
\begin{tabular}{c|c|c|c}
    \multicolumn{3}{c}{Effective Cross-section after applying cuts (fb)}\\   
    \hline 
    ~~~~~~~~{\bf B0}~~~~~~~~ & \,\,~~~~~~~~{\bf B1}~~~~~~~~\,\, & ~~~~~{\bf B2}~~~~~ \,\, & ~~~~~{\bf B3}~~~~~\\
    \hline
    659.2 & 447.5 & 56.7 & 1.24   \\    \hline
    $2.24 \times 10^{3}$ & 0.06 & - & -   \\    \hline
    53.13 & 41.2 & 40.4 & 0.1   \\    \hline
346.0 & 0.22 & 0.04 & $6.8 \times 10^{-4}$  \\   \hline
    $104.08 \times 10^{3}$  & 327.18 & 37.0 & 0.37   \\    \hline
    ~ &  ~ & ~& 1.71 \\
\end{tabular} \\ \hline \hline
\end{tabular}
\caption{Cut-flow table of the cross-section for the
relevant SM background channels for the cuts {\bf B0 - B3} as mentioned in the text at the FCC-eh collider with 3.46 TeV center-of-mass energy. We assume
$l = e$ or $\mu$.}
\label{tab:smbkg_2l}
\end{table}
\end{center}
%
\def\I{i}
\begin{center}
\begin{table}[h!]
\begin{tabular}{||c|c||}
\hline
\hline
\begin{tabular}{c|c|c}
    \multicolumn{3}{c}{Signal for FCC-eh}\\ 
    \hline
    Process & Signal & BP \\ 
    \hline
    I + II & $\geq$ 2\,$l$ + $\geq$1\,$j$ & BP1  \\ 
    \hline
    I + II & $\geq$2\,$l$ + $\geq$1\,$j$ & BP2  \\
        \hline
    I + II& $\geq$2\,$l$ + $\geq$1\,$j$ & BP3  \\
    \end{tabular}
&
\begin{tabular}{c|c|c|c}
    \multicolumn{3}{c}{Effective Cross-section after applying cuts (fb)}\\   
    \hline 
    ~~~~~~~~{\bf B0}~~~~~~~~ & \,\,~~~~~~~~{\bf B1}~~~~~~~~\,\, & ~~~~~{\bf B2}~~~~~ \,\, & ~~~~~{\bf B3}~~~~~\\
    \hline
    0.36 & 0.29 & 0.24 & 0.21   \\    \hline
    0.22 & 0.18 & 0.14 & 0.13   \\ \hline
    0.13 & 0.11 & 0.09 & 0.08   \\
\end{tabular}\\ \hline \hline
\end{tabular}
\caption{Cut-flow table of cross-section for signal
for the cuts {\bf B0 - B3} as mentioned in the text 
at the FCC-eh collider with 3.46 TeV center-of-mass energy. Here we  consider  ${\rm BR}(H^{\pm\pm}\to e^{\pm} e^{\pm})=100\%$.}
\label{tab:signal_2l}
\end{table}
\end{center}

In Table\,\ref{tab:smbkg_2l} and Table\,\ref{tab:signal_2l} we show the cut-flow of the  SM background and signal for the 
aforementioned cuts. As before, the most selective of all the cuts which improves the signal over the SM background is 
{\bf B3}. However, although the ($\geq 2l + \geq 1j$) inclusive channel gets contributions from both the processes shown in 
Fig. \ref{fig:feyn12}, the signal-to-background ratio is still inferior to the trilepton channel. The reason for this is because of the size of the 
continuum background $e^{-} l^{-}\, j\, \bar{\nu}$. 

\subsection{$\mu\mu$ Channel} \label{sec:mumu}
An important point which we should highlight here is the fact that so far we have only considered the electron flavor for the same-sign leptons in the decay of the doubly-charged Higgs, while in principle, it could decay to any of the lepton flavors depending on the Yukawa structure $Y^\Delta_{ij}$. However, at an $e^- \, p$ machine, the SM background with two same-sign negatively-charged leptons would be dominated by the electron pair. Thus, for a suitable choice of $Y^\Delta_{ee}$ and $Y^\Delta_{\mu\mu}$, we could essentially have the signal containing two same-sign muons in the final state which would be relatively background-free after the cuts we put. To put this fact in perspective, we list in Table~\ref{tab:list} the signal events for both the final states in consideration, written explicitly as  $e^+  + 2 \mu^- + \geq 1j$ and $2\mu^- + \geq 1j + \geq 0l$, after all the cuts mentioned before. Here, we assume that both the Yukawa couplings  $Y^\Delta_{ee}$  and $Y^\Delta_{\mu\mu}$  are non-zero and of equal strength with the values given by Table~\ref{tab:benchmarks} for each benchmark point, while satisfying the constraint given in Eq.~\eqref{eq:muonium}, such that the $H^{--}$ decay is now equally shared between the electron and muon modes, which means a 50\% branching ratio for the $\mu\mu$ channel. The presence of a large  $Y^{\Delta}_{ee}$, on the other hand, ensures a significant production cross-section section of the doubly-charged Higgs boson.  

Note that the leading contribution for the SM background in the $\mu\mu$ channel would arise from the 
subprocesses: $e^- \, p \to \nu_l \, Z \, W^- \, j$ and $e^- \, p \to e^- \, Z \, W^- \, j$, with both the $W$ and $Z$ bosons decaying into the muon channels. The parton-level cross section for these processes is itself at the level of $\sim 10^{-2}$ fb and assuming that similar cut efficiencies for the background would apply as in the case of the $ee$ mode, the background numbers would get further suppressed 
by a factor of $\sim \mathcal{O}(10^{-3})$, rendering the background almost negligible for the $\mu\mu$ channel.
\begin{table}[t!]
\begin{center}
\begin{tabular}{||c|c|c|c||}\hline\hline
Signal & \multicolumn{3}{c||}{$\sigma$ (fb)}\\ \cline{2-4}
& BP1 & BP2 & BP3 \\ \hline\hline
$e^+  + 2 \mu^- + \geq 1j$ & 0.06 & 0.04 & 0.02 \\
$2\mu^- + \geq 1j + \geq 0l$ &  0.12 & 0.08 & 0.05 \\ \hline\hline
\end{tabular}
\end{center}
\caption{Signal cross-section after all the cuts for $e^+  + 2 \mu^- + \geq 1j$ ({\bf A0-A3}) and 
$2\mu^- + \geq 1j + \geq 0l$ ({\bf B0-B3}) mentioned in the text. Here we  consider  
${\rm BR}(H^{\pm\pm}\to e^{\pm} e^{\pm})={\rm BR}(H^{\pm\pm}\to \mu^{\pm} \mu^{\pm})=50\%$.} \label{tab:list}
\end{table}

Therefore, the cross-sections for the signal events given in Table~\ref{tab:list} imply a nearly background-free signal of $\sim 5-10$ events with an integrated luminosity of about 100 ${\rm fb}^{-1}$. Thus our predictions are more conservative when including only electron flavor in our analysis, which is still significantly robust in the trilepton channel for mass values of the doubly-charged scalar we have chosen. We show this by presenting the signal significance in the following subsection. 

\def\I{i}
\begin{center}
\begin{table}[t!]
\begin{tabular}{||c|c||} \hline \hline
\begin{tabular}{c}
    \multicolumn{1}{c}{Benchmark}\\ 
    Points  \\        \hline
    BP1  \\     \hline
    BP2  \\     \hline
    BP3  \\          
\end{tabular}
&
\begin{tabular}{c|c}
    \multicolumn{2}{c}{$\int\mathcal{L}dt$ (${\rm fb}^{-1}$)}\\  \hline
    $\mathcal{S} = 3\,\sigma$ & $\mathcal{S} = 5\,\sigma$\\    \hline
    118 (354) & 328 (982)   \\ \hline
    245 (941) & 681 (2615)  \\ \hline
    520 (2034) & 1444   (5650)  \\ \end{tabular}\\ \hline \hline
\end{tabular}
\caption{Required integrated luminosity for achieving $3\,\sigma$ and $5\,\sigma$ excess of the
$3\,l$ ($\geq 2l$) signal, where the signal contains at least two electrons ($e^-e^-$). Here we  consider 
${\rm BR}(H^{\pm\pm}\to e^{\pm} e^{\pm})=100\%$.}
\label{tab:significance}
\end{table}
\end{center}
\subsection{Signal Significance}
In determining the statistical significance of the signal we have used the following general 
expression~\cite{Cowan:2010js}: 
\begin{eqnarray}
\mathcal{S} \ = \ \sqrt{2 \left[ \left( s + b\right)\,{\rm ln}\left( 1 + \frac{s}{b}\right) -s\right]}\,,
\label{eq:sb}
\end{eqnarray}
where $s$ and $b$ denote the number of signal and background events ({\it i.e.}~cross section times integrated luminosity) respectively, after all selection cuts are applied. Note that the formula~\eqref{eq:sb} reduces to the more well-known form $\mathcal{S}=s/\sqrt{b}$ in the limit $s\ll b$.
In Table \ref{tab:significance} we show the required integrated luminosity for either $3\,l$ or $\geq 2l$ channel to be observed with $3\sigma$ and $5\sigma$ statistical significance, where we consider $l=e$ only.  Owing to the better 
$s/b$ ratio for the trilepton channel, one is clearly sensitive to the new physics signal in the
trilepton channel with a relatively smaller data set when compared to the inclusive channel.
However, both the channels provide a robust discovery prospect of the doubly-charged Higgs boson at FCC-eh, 
provided we have a sizable Yukawa coupling to the electron flavor, $Y^{\Delta}_{ee}\sim {\cal O}(1)$.  Now, if the signal is driven by the $\mu\mu$ channel too, with the decay probability of the doubly-charged 
Higgs boson being 50\% in the muon mode, then the required integrated luminosity for discovery reduces drastically. In 
fact, assuming a conservative background of around $10^{-5}$ fb in the $\mu\mu$ channel as discussed above and using the trilepton 
signal data from Table~\ref{tab:list}, we find that for BP1 discovery at $5\sigma$  statistical significance, the required 
integrated luminosity is just 27 ${\rm fb}^{-1}$, while it is  43 ${\rm fb}^{-1}$ and 95 ${\rm fb}^{-1}$ for BP2 and BP3, respectively.
\begin{figure}[t!]
	\centering
	\includegraphics[width=0.45\textwidth]{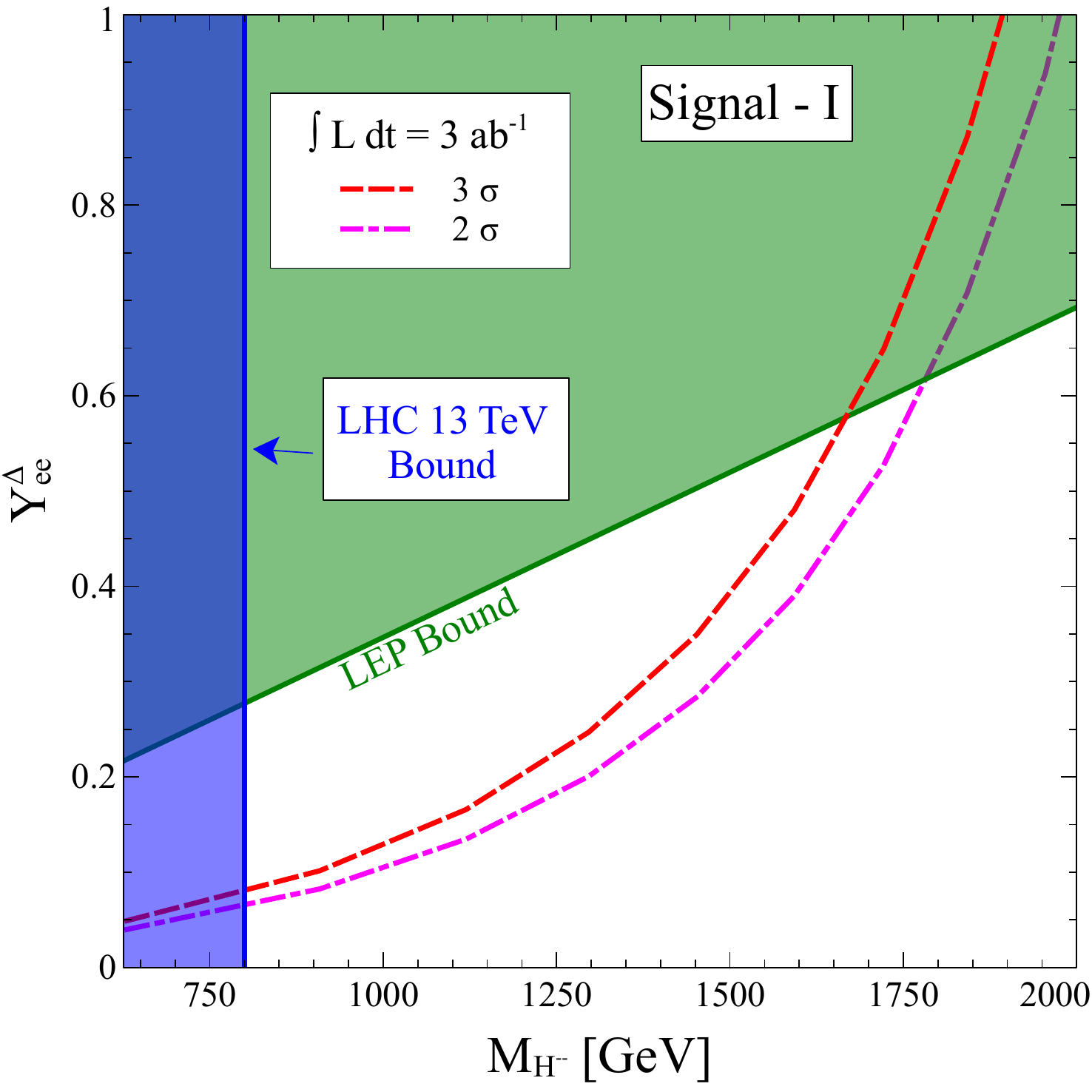}
	\includegraphics[width=0.45\textwidth]{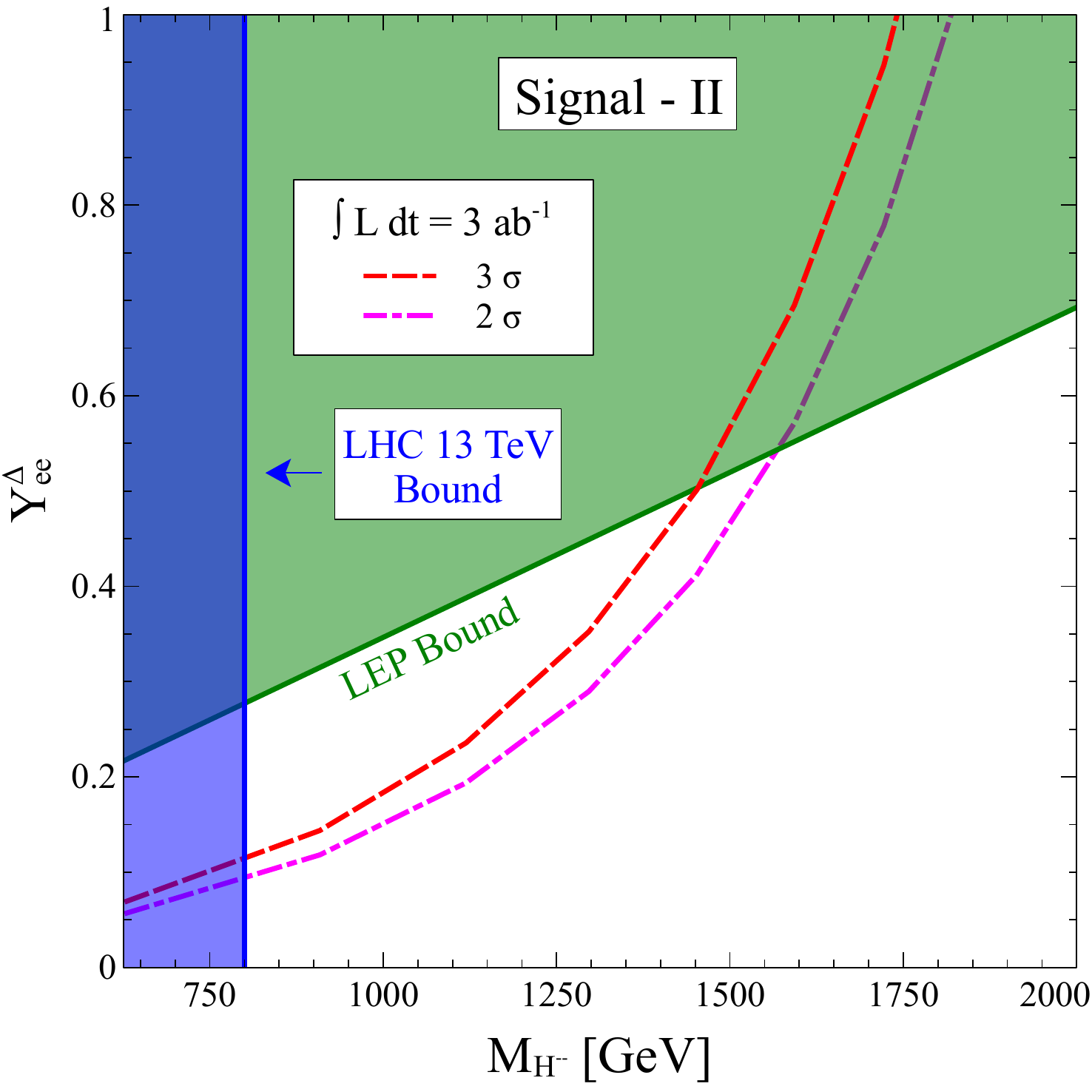}
	\caption{The $2\sigma$ and $3\sigma$   sensitivity reach of $Y^{\Delta}_{ee}$ at FCC-eh as a function of the doubly-charged Higgs mass $M_{H^{--}}$. The left panel is for Signal-I and the right panel is for Signal-II discussed in the text. The vertical (blue) shaded region is excluded from the current lower bound on $M_{H^{\pm\pm}}$ from the $\sqrt s=13$ TeV LHC data. The green-shaded region is excluded by the LEP constraints on the Bhabha scattering process. Here we have considered ${\rm BR}(H^{\pm\pm}\to e^{\pm} e^{\pm})=100\%$.}
	\label{fig:yukawa}
\end{figure}

To highlight the sensitivity of the FCC-eh to a doubly-charged Higgs mass at a given integrated luminosity, we should 
know what values of the Yukawa coupling $Y^{\Delta}_{ee}$ yield the required $H^{--}$ production cross-section. This is shown in Fig.\,\ref{fig:yukawa} where we plot the $2\sigma$ and $3\sigma$  reach for the 
doubly-charged Higgs boson in the trilepton (Signal-I) and the inclusive dilepton (Signal-II) 
channels, and show the range of the Yukawa coupling values which would achieve the 
signal significance for a fixed integrated luminosity of 3 ${\rm ab}^{-1}$. This is for the scenario where we demand that $H^{--}$ decays to $e^{-}e^{-}$ with 100 $\%$ branching ratio. The sensitivity curves are obtained using Eq.~\eqref{eq:sb} with the signal and background values obtained from the cut-based analysis described above. Fig.\,\ref{fig:yukawa} in turn also gives us information on the maximum allowed value of the triplet VEV (corresponding to the minimum value of $Y^\Delta_{ee}$ accessible), so as to observe a signal for the doubly-charged Higgs boson at the FCC-eh. The vertical line in Fig.\,\ref{fig:yukawa} shows the current lower limit on the doubly-charged scalar mass from the $\sqrt s=13$ TeV LHC data, assuming 100\% branching ratio of the $H^{--}$ decay to same-sign electron-pair~\cite{CMS-PAS-HIG-16-036, Aaboud:2017qph}. Strictly speaking, this bound is applicable only when the $H^{\pm\pm}\to e^\pm e^\pm$ decay is prompt, which requires $Y^\Delta_{ee}\gtrsim 10^{-7}$ (but practically indistinguishable from zero in the linear $Y^\Delta_{ee}$-scale of Fig.\,\ref{fig:yukawa}). The green-shaded region is excluded by LEP data on the Bhabha scattering process $e^+e^-\to e^+e^-$~\cite{Abbiendi:2003pr, Abdallah:2005ph} as given by Eq.~\eqref{eq:LEP}. The direct search limits from HERA~\cite{Aktas:2006nu}, LEP~\cite{Abdallah:2002qj, Abbiendi:2003pr, Achard:2003mv} and Tevatron~\cite{Acosta:2004uj, Abazov:2008ab} are only relevant for smaller $M_{H^{\pm\pm}}$ well below the mass range shown here and are anyway superseded by the LHC limits. It is worth noting that even the high-luminosity phase of the $\sqrt s=14$ TeV LHC cannot improve the mass reach beyond 1 TeV~\cite{Mitra:2016wpr}, and therefore, the FCC-eh provides a unique opportunity to directly probe the Yukawa coupling of heavier doubly-charged scalars to electrons. Only the proton-proton mode of FCC (FCC-hh) will be able to achieve a higher sensitivity in $M_{H^{\pm\pm}}$ up to about 6 TeV~\cite{Du:2018eaw}, independent of $Y^\Delta_{ee}$, as long as $H^{\pm\pm}$ dominantly decays to same-sign lepton pairs. On the other hand,  the future $e^+e^-$ colliders like ILC (or CLIC) might be able to cover a similar (or larger) parameter space in the direct (or indirect) search channel, as compared to the FCC-eh reach shown in Fig.\,\ref{fig:yukawa}, depending on the center-of-mass energy~\cite{Dev:2018upe}.  

\begin{figure}[t!]
	\centering
	\includegraphics[width=0.45\textwidth]{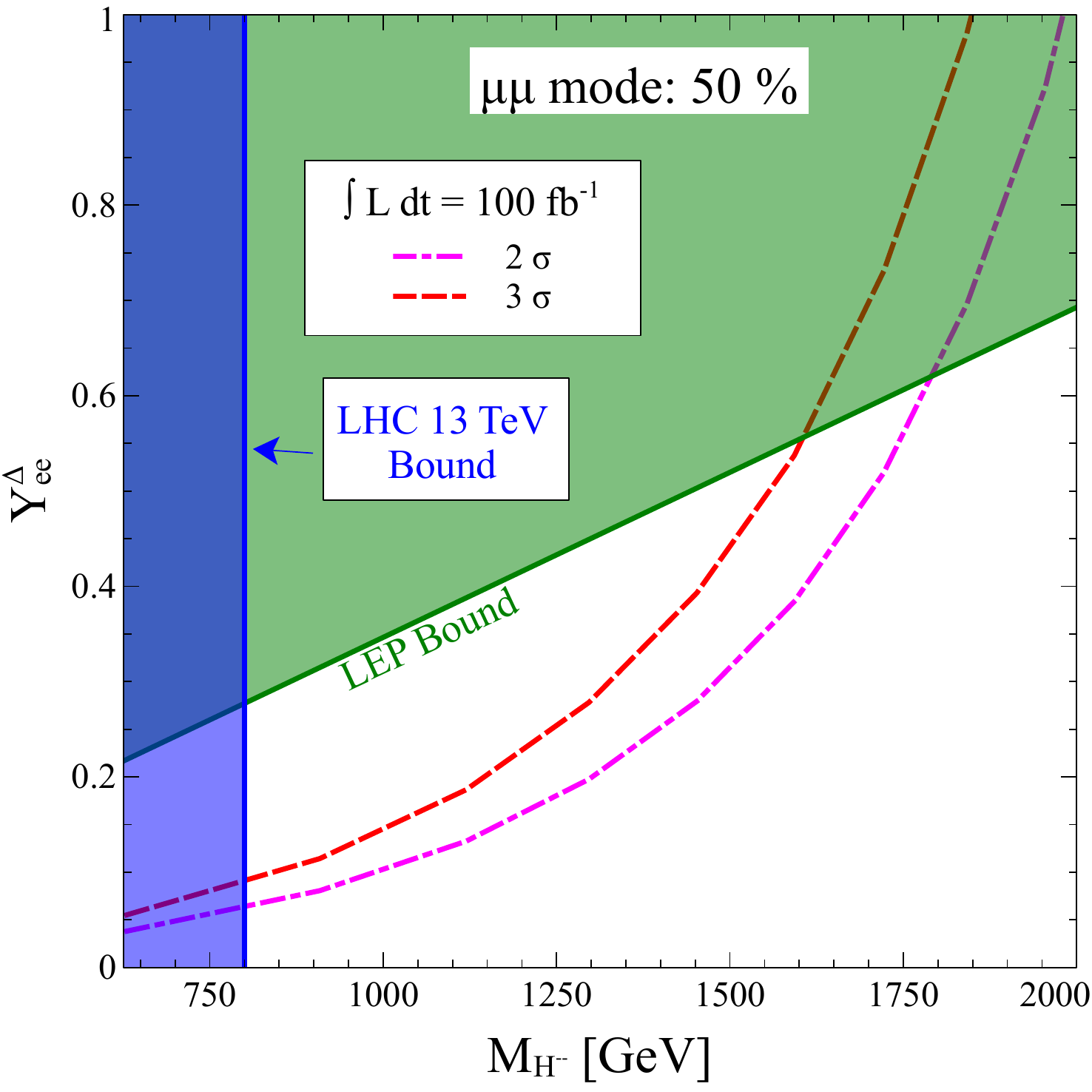}
	\caption{The  sensitivity reach of $Y^{\Delta}_{ee}$ at FCC-eh as a function of the doubly-charged Higgs mass $M_{H^{\pm \pm}}$ for Signal-I with BR$(H^{--} \to \mu^{-} \mu^{-})=50\%$ . The vertical (blue) shaded region is excluded from the current lower bound on $M_{H^{\pm\pm}}$ from the $\sqrt s=13$ TeV LHC data. The green-shaded region is excluded by the LEP constraints on the Bhabha scattering process.}
	\label{fig:yukawamumu}
\end{figure}

The more robust and striking channel of discovery would however be the $\mu^-\mu^-$ channel, as illustrated in Fig.~\ref{fig:yukawamumu}, which is relatively 
background free, barring very low rates from associated $W^- \, Z$ productions. As a large production rate for the 
doubly-charged Higgs is a necessity, we ensure a 50\% decay probability of the $H^{\pm\pm}$ in the $\mu\mu$
mode while respecting the bound of Eq. (\ref{eq:muonium}). We show the much improved sensitivity reach using the 
signal rates for $e^{+} +2 \mu^{-} + \geq 1j$ final state, where a similar $2\sigma$ or $3\sigma$ range as in Fig.~\ref{fig:yukawa} can now be attained with a much smaller integrated luminosity of 100 fb$^{-1}$. A similar sensitivity is also expected in the inclusive $2\mu^-+\geq 1j+\geq 0l$ channel. This reach can be further improved if the signal significance of 
both the $ee$ and $\mu\mu$ channels (observed independently) are combined together.

   
\section{Conclusion \label{conclu}}
We have analyzed the discovery prospect of a doubly-charged Higgs at the newly proposed FCC-eh  electron-proton collider, operating with beam energies $E_e=60$ GeV for electron beam and $E_p=50$ TeV for proton beam. As the LHC would have limited sensitivity to a doubly-charged Higgs mass, even with the high-luminosity phase, searching for heavier doubly-charged scalars would be a 
challenge. A type-II seesaw-based scenario for neutrino mass generation with very tiny triplet VEV would imply a substantially large Yukawa coupling of the scalar to the charged leptons. Neither the LHC nor an $e^+e^-$ machine could gain from this, unless the collisions are of $e^-e^-$ 
type~\cite{Mukhopadhyaya:2005vf}, where the production of doubly charged Higgs  has direct dependency on  Yukawa. However, the proposed FCC machine, where the 
collisions are between an electron and a proton beam (FCC-eh) would open up the 
possibility of looking for these heavier doubly-charged states through their single on-shell production in association with lepton and a jet. 

We have studied two types of final states, namely, $3l+\ge1j$ and inclusive $\ge2l+\ge1j$, to probe TeV-scale doubly-charged scalars at FCC-eh. We consider a few 
representative benchmark points for a hybrid type-I+type-II seesaw with a small triplet VEV $v_\Delta\sim {\cal O}(0.1)$ eV and large diagonal Yukawa couplings $Y^\Delta_{ee}\sim {\cal O}(1)$.  The model is consistent with  neutrino oscillation data, as well as, in agreement with the 
bounds from lepton flavor violating processes, such as, $\mu \to e\gamma$ and $\mu \to eee$.  
We find that a doubly-charged Higgs with mass around a TeV can be probed at $3\sigma$ significance  in  
FCC-eh with  $\mathcal{L} \sim 245\, \rm{fb}^{-1}$ in the $3l+\ge1j$ ($l=e$) mode. Moreover, higher mass values, up to about 2 TeV, could also be probed within a few years'  
of running of the FCC-eh. A more robust and striking channel would be observed with only 
$\mathcal{L}=100 \, \rm{fb}^{-1}$ of data  if  the doubly-charged Higgs has a decay probability in the $\mu\mu$ mode, 
which is comparable, if not more than the $ee$ mode. This reach can be further improved if the signal significance of 
both the $ee$ and $\mu\mu$ channels (observed independently) are combined. 






\section*{Acknowledgments} 
BD, MM and SKR thank the organizers of WHEPP-XIV at IIT, Kanpur where this work was initiated, and the organizers of SUSY 2017 at TIFR, Mumbai where part of this work was  done, for the local hospitality. SK would like to thank Subhadeep Mondal for help with numerical aspects 
at the initial stage of the work. The work of BD is supported by the US Department of Energy under Grant No. DE-SC0017987. MM acknowledges the support of DST-INSPIRE research grant IFA14-PH-99. The work of SKR was partially supported by funding available from the Department of Atomic Energy, Government of India, for the Regional Centre for Accelerator-based Particle Physics (RECAPP), Harish-Chandra Research Institute. SK would like to acknowledge support from the Department of Atomic Energy (DAE) Neutrino Project under  
the  plan  project  of  Harish-Chandra  Research  Institute and the European Union’s Horizon 2020 research and 
innovation programme Invisibles Plus RISE under the Marie Sklodowska-Curie grant agreement No 690575.
\bibliography{reference.bib}

\begin{thebibliography}{130}%
\makeatletter
\providecommand \@ifxundefined [1]{%
 \@ifx{#1\undefined}
}%
\providecommand \@ifnum [1]{%
 \ifnum #1\expandafter \@firstoftwo
 \else \expandafter \@secondoftwo
 \fi
}%
\providecommand \@ifx [1]{%
 \ifx #1\expandafter \@firstoftwo
 \else \expandafter \@secondoftwo
 \fi
}%
\providecommand \natexlab [1]{#1}%
\providecommand \enquote  [1]{``#1''}%
\providecommand \bibnamefont  [1]{#1}%
\providecommand \bibfnamefont [1]{#1}%
\providecommand \citenamefont [1]{#1}%
\providecommand \href@noop [0]{\@secondoftwo}%
\providecommand \href [0]{\begingroup \@sanitize@url \@href}%
\providecommand \@href[1]{\@@startlink{#1}\@@href}%
\providecommand \@@href[1]{\endgroup#1\@@endlink}%
\providecommand \@sanitize@url [0]{\catcode `\\12\catcode `\$12\catcode
  `\&12\catcode `\#12\catcode `\^12\catcode `\_12\catcode `\%12\relax}%
\providecommand \@@startlink[1]{}%
\providecommand \@@endlink[0]{}%
\providecommand \url  [0]{\begingroup\@sanitize@url \@url }%
\providecommand \@url [1]{\endgroup\@href {#1}{\urlprefix }}%
\providecommand \urlprefix  [0]{URL }%
\providecommand \Eprint [0]{\href }%
\providecommand \doibase [0]{http://dx.doi.org/}%
\providecommand \selectlanguage [0]{\@gobble}%
\providecommand \bibinfo  [0]{\@secondoftwo}%
\providecommand \bibfield  [0]{\@secondoftwo}%
\providecommand \translation [1]{[#1]}%
\providecommand \BibitemOpen [0]{}%
\providecommand \bibitemStop [0]{}%
\providecommand \bibitemNoStop [0]{.\EOS\space}%
\providecommand \EOS [0]{\spacefactor3000\relax}%
\providecommand \BibitemShut  [1]{\csname bibitem#1\endcsname}%
\let\auto@bib@innerbib\@empty
\bibitem [{\citenamefont {De~Salas}\ \emph {et~al.}(2018)\citenamefont
  {De~Salas}, \citenamefont {Gariazzo}, \citenamefont {Mena}, \citenamefont
  {Ternes},\ and\ \citenamefont {Tortola}}]{deSalas:2018bym}%
  \BibitemOpen
  \bibfield  {author} {\bibinfo {author} {\bibfnamefont {P.~F.}\ \bibnamefont
  {De~Salas}}, \bibinfo {author} {\bibfnamefont {S.}~\bibnamefont {Gariazzo}},
  \bibinfo {author} {\bibfnamefont {O.}~\bibnamefont {Mena}}, \bibinfo {author}
  {\bibfnamefont {C.~A.}\ \bibnamefont {Ternes}}, \ and\ \bibinfo {author}
  {\bibfnamefont {M.}~\bibnamefont {Tortola}},\ }\href@noop {} {\  (\bibinfo
  {year} {2018})},\ \Eprint {http://arxiv.org/abs/1806.11051} {arXiv:1806.11051
  [hep-ph]} \BibitemShut {NoStop}%
\bibitem [{\citenamefont {Mohapatra}\ and\ \citenamefont
  {Smirnov}(2006)}]{Mohapatra:2006gs}%
  \BibitemOpen
  \bibfield  {author} {\bibinfo {author} {\bibfnamefont {R.~N.}\ \bibnamefont
  {Mohapatra}}\ and\ \bibinfo {author} {\bibfnamefont {A.~Y.}\ \bibnamefont
  {Smirnov}},\ }\href {\doibase 10.1146/annurev.nucl.56.080805.140534}
  {\bibfield  {journal} {\bibinfo  {journal} {Ann. Rev. Nucl. Part. Sci.}\
  }\textbf {\bibinfo {volume} {56}},\ \bibinfo {pages} {569} (\bibinfo {year}
  {2006})},\ \Eprint {http://arxiv.org/abs/hep-ph/0603118}
  {arXiv:hep-ph/0603118 [hep-ph]} \BibitemShut {NoStop}%
\bibitem [{\citenamefont {Weinberg}(1979)}]{Weinberg:1979sa}%
  \BibitemOpen
  \bibfield  {author} {\bibinfo {author} {\bibfnamefont {S.}~\bibnamefont
  {Weinberg}},\ }\href {\doibase 10.1103/PhysRevLett.43.1566} {\bibfield
  {journal} {\bibinfo  {journal} {Phys. Rev. Lett.}\ }\textbf {\bibinfo
  {volume} {43}},\ \bibinfo {pages} {1566} (\bibinfo {year}
  {1979})}\BibitemShut {NoStop}%
\bibitem [{\citenamefont {Wilczek}\ and\ \citenamefont
  {Zee}(1979)}]{Wilczek:1979hc}%
  \BibitemOpen
  \bibfield  {author} {\bibinfo {author} {\bibfnamefont {F.}~\bibnamefont
  {Wilczek}}\ and\ \bibinfo {author} {\bibfnamefont {A.}~\bibnamefont {Zee}},\
  }\href {\doibase 10.1103/PhysRevLett.43.1571} {\bibfield  {journal} {\bibinfo
   {journal} {Phys. Rev. Lett.}\ }\textbf {\bibinfo {volume} {43}},\ \bibinfo
  {pages} {1571} (\bibinfo {year} {1979})}\BibitemShut {NoStop}%
\bibitem [{\citenamefont {Ma}(1998)}]{Ma:1998dn}%
  \BibitemOpen
  \bibfield  {author} {\bibinfo {author} {\bibfnamefont {E.}~\bibnamefont
  {Ma}},\ }\href {\doibase 10.1103/PhysRevLett.81.1171} {\bibfield  {journal}
  {\bibinfo  {journal} {Phys. Rev. Lett.}\ }\textbf {\bibinfo {volume} {81}},\
  \bibinfo {pages} {1171} (\bibinfo {year} {1998})},\ \Eprint
  {http://arxiv.org/abs/hep-ph/9805219} {arXiv:hep-ph/9805219 [hep-ph]}
  \BibitemShut {NoStop}%
\bibitem [{\citenamefont {Minkowski}(1977)}]{Minkowski:1977sc}%
  \BibitemOpen
  \bibfield  {author} {\bibinfo {author} {\bibfnamefont {P.}~\bibnamefont
  {Minkowski}},\ }\href {\doibase 10.1016/0370-2693(77)90435-X} {\bibfield
  {journal} {\bibinfo  {journal} {Phys. Lett.}\ }\textbf {\bibinfo {volume}
  {B67}},\ \bibinfo {pages} {421} (\bibinfo {year} {1977})}\BibitemShut
  {NoStop}%
\bibitem [{\citenamefont {Mohapatra}\ and\ \citenamefont
  {Senjanovi\'{c}}(1980)}]{Mohapatra:1979ia}%
  \BibitemOpen
  \bibfield  {author} {\bibinfo {author} {\bibfnamefont {R.~N.}\ \bibnamefont
  {Mohapatra}}\ and\ \bibinfo {author} {\bibfnamefont {G.}~\bibnamefont
  {Senjanovi\'{c}}},\ }\href {\doibase 10.1103/PhysRevLett.44.912} {\bibfield
  {journal} {\bibinfo  {journal} {Phys. Rev. Lett.}\ }\textbf {\bibinfo
  {volume} {44}},\ \bibinfo {pages} {912} (\bibinfo {year} {1980})}\BibitemShut
  {NoStop}%
\bibitem [{\citenamefont {Yanagida}(1979)}]{Yanagida:1979as}%
  \BibitemOpen
  \bibfield  {author} {\bibinfo {author} {\bibfnamefont {T.}~\bibnamefont
  {Yanagida}},\ }\href@noop {} {\bibfield  {journal} {\bibinfo  {journal}
  {Conf. Proc.}\ }\textbf {\bibinfo {volume} {C7902131}},\ \bibinfo {pages}
  {95} (\bibinfo {year} {1979})}\BibitemShut {NoStop}%
\bibitem [{\citenamefont {Gell-Mann}\ \emph {et~al.}(1979)\citenamefont
  {Gell-Mann}, \citenamefont {Ramond},\ and\ \citenamefont
  {Slansky}}]{GellMann:1980vs}%
  \BibitemOpen
  \bibfield  {author} {\bibinfo {author} {\bibfnamefont {M.}~\bibnamefont
  {Gell-Mann}}, \bibinfo {author} {\bibfnamefont {P.}~\bibnamefont {Ramond}}, \
  and\ \bibinfo {author} {\bibfnamefont {R.}~\bibnamefont {Slansky}},\
  }\href@noop {} {\bibfield  {journal} {\bibinfo  {journal} {Conf. Proc.}\
  }\textbf {\bibinfo {volume} {C790927}},\ \bibinfo {pages} {315} (\bibinfo
  {year} {1979})},\ \Eprint {http://arxiv.org/abs/1306.4669} {arXiv:1306.4669
  [hep-th]} \BibitemShut {NoStop}%
\bibitem [{\citenamefont {Glashow}(1980)}]{Glashow:1979nm}%
  \BibitemOpen
  \bibfield  {author} {\bibinfo {author} {\bibfnamefont {S.~L.}\ \bibnamefont
  {Glashow}},\ }\href {\doibase 10.1007/978-1-4684-7197-7_15} {\bibfield
  {journal} {\bibinfo  {journal} {NATO Sci. Ser. B}\ }\textbf {\bibinfo
  {volume} {61}},\ \bibinfo {pages} {687} (\bibinfo {year} {1980})}\BibitemShut
  {NoStop}%
\bibitem [{\citenamefont {Konetschny}\ and\ \citenamefont
  {Kummer}(1977)}]{Konetschny:1977bn}%
  \BibitemOpen
  \bibfield  {author} {\bibinfo {author} {\bibfnamefont {W.}~\bibnamefont
  {Konetschny}}\ and\ \bibinfo {author} {\bibfnamefont {W.}~\bibnamefont
  {Kummer}},\ }\href {\doibase 10.1016/0370-2693(77)90407-5} {\bibfield
  {journal} {\bibinfo  {journal} {Phys. Lett.}\ }\textbf {\bibinfo {volume}
  {70B}},\ \bibinfo {pages} {433} (\bibinfo {year} {1977})}\BibitemShut
  {NoStop}%
\bibitem [{\citenamefont {Schechter}\ and\ \citenamefont
  {Valle}(1980)}]{Schechter:1980gr}%
  \BibitemOpen
  \bibfield  {author} {\bibinfo {author} {\bibfnamefont {J.}~\bibnamefont
  {Schechter}}\ and\ \bibinfo {author} {\bibfnamefont {J.~W.~F.}\ \bibnamefont
  {Valle}},\ }\href {\doibase 10.1103/PhysRevD.22.2227} {\bibfield  {journal}
  {\bibinfo  {journal} {Phys. Rev.}\ }\textbf {\bibinfo {volume} {D22}},\
  \bibinfo {pages} {2227} (\bibinfo {year} {1980})}\BibitemShut {NoStop}%
\bibitem [{\citenamefont {Magg}\ and\ \citenamefont
  {Wetterich}(1980)}]{Magg:1980ut}%
  \BibitemOpen
  \bibfield  {author} {\bibinfo {author} {\bibfnamefont {M.}~\bibnamefont
  {Magg}}\ and\ \bibinfo {author} {\bibfnamefont {C.}~\bibnamefont
  {Wetterich}},\ }\href {\doibase 10.1016/0370-2693(80)90825-4} {\bibfield
  {journal} {\bibinfo  {journal} {Phys. Lett.}\ }\textbf {\bibinfo {volume}
  {B94}},\ \bibinfo {pages} {61} (\bibinfo {year} {1980})}\BibitemShut
  {NoStop}%
\bibitem [{\citenamefont {Cheng}\ and\ \citenamefont
  {Li}(1980)}]{Cheng:1980qt}%
  \BibitemOpen
  \bibfield  {author} {\bibinfo {author} {\bibfnamefont {T.~P.}\ \bibnamefont
  {Cheng}}\ and\ \bibinfo {author} {\bibfnamefont {L.-F.}\ \bibnamefont {Li}},\
  }\href {\doibase 10.1103/PhysRevD.22.2860} {\bibfield  {journal} {\bibinfo
  {journal} {Phys. Rev.}\ }\textbf {\bibinfo {volume} {D22}},\ \bibinfo {pages}
  {2860} (\bibinfo {year} {1980})}\BibitemShut {NoStop}%
\bibitem [{\citenamefont {Lazarides}\ \emph {et~al.}(1981)\citenamefont
  {Lazarides}, \citenamefont {Shafi},\ and\ \citenamefont
  {Wetterich}}]{Lazarides:1980nt}%
  \BibitemOpen
  \bibfield  {author} {\bibinfo {author} {\bibfnamefont {G.}~\bibnamefont
  {Lazarides}}, \bibinfo {author} {\bibfnamefont {Q.}~\bibnamefont {Shafi}}, \
  and\ \bibinfo {author} {\bibfnamefont {C.}~\bibnamefont {Wetterich}},\ }\href
  {\doibase 10.1016/0550-3213(81)90354-0} {\bibfield  {journal} {\bibinfo
  {journal} {Nucl. Phys.}\ }\textbf {\bibinfo {volume} {B181}},\ \bibinfo
  {pages} {287} (\bibinfo {year} {1981})}\BibitemShut {NoStop}%
\bibitem [{\citenamefont {Mohapatra}\ and\ \citenamefont
  {Senjanovi\'{c}}(1981)}]{Mohapatra:1980yp}%
  \BibitemOpen
  \bibfield  {author} {\bibinfo {author} {\bibfnamefont {R.~N.}\ \bibnamefont
  {Mohapatra}}\ and\ \bibinfo {author} {\bibfnamefont {G.}~\bibnamefont
  {Senjanovi\'{c}}},\ }\href {\doibase 10.1103/PhysRevD.23.165} {\bibfield
  {journal} {\bibinfo  {journal} {Phys. Rev.}\ }\textbf {\bibinfo {volume}
  {D23}},\ \bibinfo {pages} {165} (\bibinfo {year} {1981})}\BibitemShut
  {NoStop}%
\bibitem [{\citenamefont {Foot}\ \emph {et~al.}(1989)\citenamefont {Foot},
  \citenamefont {Lew}, \citenamefont {He},\ and\ \citenamefont
  {Joshi}}]{Foot:1988aq}%
  \BibitemOpen
  \bibfield  {author} {\bibinfo {author} {\bibfnamefont {R.}~\bibnamefont
  {Foot}}, \bibinfo {author} {\bibfnamefont {H.}~\bibnamefont {Lew}}, \bibinfo
  {author} {\bibfnamefont {X.~G.}\ \bibnamefont {He}}, \ and\ \bibinfo {author}
  {\bibfnamefont {G.~C.}\ \bibnamefont {Joshi}},\ }\href {\doibase
  10.1007/BF01415558} {\bibfield  {journal} {\bibinfo  {journal} {Z. Phys.}\
  }\textbf {\bibinfo {volume} {C44}},\ \bibinfo {pages} {441} (\bibinfo {year}
  {1989})}\BibitemShut {NoStop}%
\bibitem [{\citenamefont {Mohapatra}\ and\ \citenamefont
  {Pati}(1975{\natexlab{a}})}]{Mohapatra:1974hk}%
  \BibitemOpen
  \bibfield  {author} {\bibinfo {author} {\bibfnamefont {R.~N.}\ \bibnamefont
  {Mohapatra}}\ and\ \bibinfo {author} {\bibfnamefont {J.~C.}\ \bibnamefont
  {Pati}},\ }\href {\doibase 10.1103/PhysRevD.11.566} {\bibfield  {journal}
  {\bibinfo  {journal} {Phys. Rev.}\ }\textbf {\bibinfo {volume} {D11}},\
  \bibinfo {pages} {566} (\bibinfo {year} {1975}{\natexlab{a}})}\BibitemShut
  {NoStop}%
\bibitem [{\citenamefont {Mohapatra}\ and\ \citenamefont
  {Pati}(1975{\natexlab{b}})}]{Mohapatra:1974gc}%
  \BibitemOpen
  \bibfield  {author} {\bibinfo {author} {\bibfnamefont {R.~N.}\ \bibnamefont
  {Mohapatra}}\ and\ \bibinfo {author} {\bibfnamefont {J.~C.}\ \bibnamefont
  {Pati}},\ }\href {\doibase 10.1103/PhysRevD.11.2558} {\bibfield  {journal}
  {\bibinfo  {journal} {Phys. Rev.}\ }\textbf {\bibinfo {volume} {D11}},\
  \bibinfo {pages} {2558} (\bibinfo {year} {1975}{\natexlab{b}})}\BibitemShut
  {NoStop}%
\bibitem [{\citenamefont {Senjanovi\'{c}}\ and\ \citenamefont
  {Mohapatra}(1975)}]{Senjanovic:1975rk}%
  \BibitemOpen
  \bibfield  {author} {\bibinfo {author} {\bibfnamefont {G.}~\bibnamefont
  {Senjanovi\'{c}}}\ and\ \bibinfo {author} {\bibfnamefont {R.~N.}\
  \bibnamefont {Mohapatra}},\ }\href {\doibase 10.1103/PhysRevD.12.1502}
  {\bibfield  {journal} {\bibinfo  {journal} {Phys. Rev.}\ }\textbf {\bibinfo
  {volume} {D12}},\ \bibinfo {pages} {1502} (\bibinfo {year}
  {1975})}\BibitemShut {NoStop}%
\bibitem [{\citenamefont {Mohapatra}(1986)}]{Mohapatra:1986aw}%
  \BibitemOpen
  \bibfield  {author} {\bibinfo {author} {\bibfnamefont {R.~N.}\ \bibnamefont
  {Mohapatra}},\ }\href {\doibase 10.1103/PhysRevLett.56.561} {\bibfield
  {journal} {\bibinfo  {journal} {Phys. Rev. Lett.}\ }\textbf {\bibinfo
  {volume} {56}},\ \bibinfo {pages} {561} (\bibinfo {year} {1986})}\BibitemShut
  {NoStop}%
\bibitem [{\citenamefont {Nandi}\ and\ \citenamefont
  {Sarkar}(1986)}]{Nandi:1985uh}%
  \BibitemOpen
  \bibfield  {author} {\bibinfo {author} {\bibfnamefont {S.}~\bibnamefont
  {Nandi}}\ and\ \bibinfo {author} {\bibfnamefont {U.}~\bibnamefont {Sarkar}},\
  }\href {\doibase 10.1103/PhysRevLett.56.564} {\bibfield  {journal} {\bibinfo
  {journal} {Phys. Rev. Lett.}\ }\textbf {\bibinfo {volume} {56}},\ \bibinfo
  {pages} {564} (\bibinfo {year} {1986})}\BibitemShut {NoStop}%
\bibitem [{\citenamefont {Mohapatra}\ and\ \citenamefont
  {Valle}(1986)}]{Mohapatra:1986bd}%
  \BibitemOpen
  \bibfield  {author} {\bibinfo {author} {\bibfnamefont {R.~N.}\ \bibnamefont
  {Mohapatra}}\ and\ \bibinfo {author} {\bibfnamefont {J.~W.~F.}\ \bibnamefont
  {Valle}},\ }\href {\doibase 10.1103/PhysRevD.34.1642} {\bibfield  {journal}
  {\bibinfo  {journal} {Phys. Rev.}\ }\textbf {\bibinfo {volume} {D34}},\
  \bibinfo {pages} {1642} (\bibinfo {year} {1986})}\BibitemShut {NoStop}%
\bibitem [{\citenamefont {Malinsky}\ \emph {et~al.}(2005)\citenamefont
  {Malinsky}, \citenamefont {Romao},\ and\ \citenamefont
  {Valle}}]{Malinsky:2005bi}%
  \BibitemOpen
  \bibfield  {author} {\bibinfo {author} {\bibfnamefont {M.}~\bibnamefont
  {Malinsky}}, \bibinfo {author} {\bibfnamefont {J.~C.}\ \bibnamefont {Romao}},
  \ and\ \bibinfo {author} {\bibfnamefont {J.~W.~F.}\ \bibnamefont {Valle}},\
  }\href {\doibase 10.1103/PhysRevLett.95.161801} {\bibfield  {journal}
  {\bibinfo  {journal} {Phys. Rev. Lett.}\ }\textbf {\bibinfo {volume} {95}},\
  \bibinfo {pages} {161801} (\bibinfo {year} {2005})},\ \Eprint
  {http://arxiv.org/abs/hep-ph/0506296} {arXiv:hep-ph/0506296 [hep-ph]}
  \BibitemShut {NoStop}%
\bibitem [{\citenamefont {Kang}\ and\ \citenamefont {Kim}(2007)}]{Kang:2006sn}%
  \BibitemOpen
  \bibfield  {author} {\bibinfo {author} {\bibfnamefont {S.~K.}\ \bibnamefont
  {Kang}}\ and\ \bibinfo {author} {\bibfnamefont {C.~S.}\ \bibnamefont {Kim}},\
  }\href {\doibase 10.1016/j.physletb.2006.12.071} {\bibfield  {journal}
  {\bibinfo  {journal} {Phys. Lett.}\ }\textbf {\bibinfo {volume} {B646}},\
  \bibinfo {pages} {248} (\bibinfo {year} {2007})},\ \Eprint
  {http://arxiv.org/abs/hep-ph/0607072} {arXiv:hep-ph/0607072 [hep-ph]}
  \BibitemShut {NoStop}%
\bibitem [{\citenamefont {Gavela}\ \emph {et~al.}(2009)\citenamefont {Gavela},
  \citenamefont {Hambye}, \citenamefont {Hernandez},\ and\ \citenamefont
  {Hernandez}}]{Gavela:2009cd}%
  \BibitemOpen
  \bibfield  {author} {\bibinfo {author} {\bibfnamefont {M.~B.}\ \bibnamefont
  {Gavela}}, \bibinfo {author} {\bibfnamefont {T.}~\bibnamefont {Hambye}},
  \bibinfo {author} {\bibfnamefont {D.}~\bibnamefont {Hernandez}}, \ and\
  \bibinfo {author} {\bibfnamefont {P.}~\bibnamefont {Hernandez}},\ }\href
  {\doibase 10.1088/1126-6708/2009/09/038} {\bibfield  {journal} {\bibinfo
  {journal} {JHEP}\ }\textbf {\bibinfo {volume} {09}},\ \bibinfo {pages} {038}
  (\bibinfo {year} {2009})},\ \Eprint {http://arxiv.org/abs/0906.1461}
  {arXiv:0906.1461 [hep-ph]} \BibitemShut {NoStop}%
\bibitem [{\citenamefont {Dev}\ and\ \citenamefont
  {Pilaftsis}(2012)}]{Dev:2012sg}%
  \BibitemOpen
  \bibfield  {author} {\bibinfo {author} {\bibfnamefont {P.~S.~B.}\
  \bibnamefont {Dev}}\ and\ \bibinfo {author} {\bibfnamefont {A.}~\bibnamefont
  {Pilaftsis}},\ }\href {\doibase 10.1103/PhysRevD.86.113001} {\bibfield
  {journal} {\bibinfo  {journal} {Phys. Rev.}\ }\textbf {\bibinfo {volume}
  {D86}},\ \bibinfo {pages} {113001} (\bibinfo {year} {2012})},\ \Eprint
  {http://arxiv.org/abs/1209.4051} {arXiv:1209.4051 [hep-ph]} \BibitemShut
  {NoStop}%
\bibitem [{\citenamefont {Law}\ and\ \citenamefont
  {McDonald}(2013)}]{Law:2013gma}%
  \BibitemOpen
  \bibfield  {author} {\bibinfo {author} {\bibfnamefont {S.~S.~C.}\
  \bibnamefont {Law}}\ and\ \bibinfo {author} {\bibfnamefont {K.~L.}\
  \bibnamefont {McDonald}},\ }\href {\doibase 10.1103/PhysRevD.87.113003}
  {\bibfield  {journal} {\bibinfo  {journal} {Phys. Rev.}\ }\textbf {\bibinfo
  {volume} {D87}},\ \bibinfo {pages} {113003} (\bibinfo {year} {2013})},\
  \Eprint {http://arxiv.org/abs/1303.4887} {arXiv:1303.4887 [hep-ph]}
  \BibitemShut {NoStop}%
\bibitem [{\citenamefont {Cai}\ \emph {et~al.}(2017)\citenamefont {Cai},
  \citenamefont {Herrero-Garcia}, \citenamefont {Schmidt}, \citenamefont
  {Vicente},\ and\ \citenamefont {Volkas}}]{Cai:2017jrq}%
  \BibitemOpen
  \bibfield  {author} {\bibinfo {author} {\bibfnamefont {Y.}~\bibnamefont
  {Cai}}, \bibinfo {author} {\bibfnamefont {J.}~\bibnamefont {Herrero-Garcia}},
  \bibinfo {author} {\bibfnamefont {M.~A.}\ \bibnamefont {Schmidt}}, \bibinfo
  {author} {\bibfnamefont {A.}~\bibnamefont {Vicente}}, \ and\ \bibinfo
  {author} {\bibfnamefont {R.~R.}\ \bibnamefont {Volkas}},\ }\href {\doibase
  10.3389/fphy.2017.00063} {\bibfield  {journal} {\bibinfo  {journal} {Front.in
  Phys.}\ }\textbf {\bibinfo {volume} {5}},\ \bibinfo {pages} {63} (\bibinfo
  {year} {2017})},\ \Eprint {http://arxiv.org/abs/1706.08524} {arXiv:1706.08524
  [hep-ph]} \BibitemShut {NoStop}%
\bibitem [{\citenamefont {Atre}\ \emph {et~al.}(2009)\citenamefont {Atre},
  \citenamefont {Han}, \citenamefont {Pascoli},\ and\ \citenamefont
  {Zhang}}]{Atre:2009rg}%
  \BibitemOpen
  \bibfield  {author} {\bibinfo {author} {\bibfnamefont {A.}~\bibnamefont
  {Atre}}, \bibinfo {author} {\bibfnamefont {T.}~\bibnamefont {Han}}, \bibinfo
  {author} {\bibfnamefont {S.}~\bibnamefont {Pascoli}}, \ and\ \bibinfo
  {author} {\bibfnamefont {B.}~\bibnamefont {Zhang}},\ }\href {\doibase
  10.1088/1126-6708/2009/05/030} {\bibfield  {journal} {\bibinfo  {journal}
  {JHEP}\ }\textbf {\bibinfo {volume} {05}},\ \bibinfo {pages} {030} (\bibinfo
  {year} {2009})},\ \Eprint {http://arxiv.org/abs/0901.3589} {arXiv:0901.3589
  [hep-ph]} \BibitemShut {NoStop}%
\bibitem [{\citenamefont {Deppisch}\ \emph {et~al.}(2015)\citenamefont
  {Deppisch}, \citenamefont {Bhupal~Dev},\ and\ \citenamefont
  {Pilaftsis}}]{Deppisch:2015qwa}%
  \BibitemOpen
  \bibfield  {author} {\bibinfo {author} {\bibfnamefont {F.~F.}\ \bibnamefont
  {Deppisch}}, \bibinfo {author} {\bibfnamefont {P.~S.}\ \bibnamefont
  {Bhupal~Dev}}, \ and\ \bibinfo {author} {\bibfnamefont {A.}~\bibnamefont
  {Pilaftsis}},\ }\href {\doibase 10.1088/1367-2630/17/7/075019} {\bibfield
  {journal} {\bibinfo  {journal} {New J. Phys.}\ }\textbf {\bibinfo {volume}
  {17}},\ \bibinfo {pages} {075019} (\bibinfo {year} {2015})},\ \Eprint
  {http://arxiv.org/abs/1502.06541} {arXiv:1502.06541 [hep-ph]} \BibitemShut
  {NoStop}%
\bibitem [{\citenamefont {Golling}\ \emph {et~al.}(2017)\citenamefont {Golling}
  \emph {et~al.}}]{Golling:2016gvc}%
  \BibitemOpen
  \bibfield  {author} {\bibinfo {author} {\bibfnamefont {T.}~\bibnamefont
  {Golling}} \emph {et~al.},\ }\href {\doibase 10.23731/CYRM-2017-003.441}
  {\bibfield  {journal} {\bibinfo  {journal} {CERN Yellow Report}\ ,\ \bibinfo
  {pages} {441}} (\bibinfo {year} {2017})},\ \Eprint
  {http://arxiv.org/abs/1606.00947} {arXiv:1606.00947 [hep-ph]} \BibitemShut
  {NoStop}%
\bibitem [{\citenamefont {Cai}\ \emph {et~al.}(2018)\citenamefont {Cai},
  \citenamefont {Han}, \citenamefont {Li},\ and\ \citenamefont
  {Ruiz}}]{Cai:2017mow}%
  \BibitemOpen
  \bibfield  {author} {\bibinfo {author} {\bibfnamefont {Y.}~\bibnamefont
  {Cai}}, \bibinfo {author} {\bibfnamefont {T.}~\bibnamefont {Han}}, \bibinfo
  {author} {\bibfnamefont {T.}~\bibnamefont {Li}}, \ and\ \bibinfo {author}
  {\bibfnamefont {R.}~\bibnamefont {Ruiz}},\ }\href {\doibase
  10.3389/fphy.2018.00040} {\bibfield  {journal} {\bibinfo  {journal} {Front.in
  Phys.}\ }\textbf {\bibinfo {volume} {6}},\ \bibinfo {pages} {40} (\bibinfo
  {year} {2018})},\ \Eprint {http://arxiv.org/abs/1711.02180} {arXiv:1711.02180
  [hep-ph]} \BibitemShut {NoStop}%
\bibitem [{\citenamefont {de~Blas}\ \emph {et~al.}(2018)\citenamefont {de~Blas}
  \emph {et~al.}}]{deBlas:2018mhx}%
  \BibitemOpen
  \bibfield  {author} {\bibinfo {author} {\bibfnamefont {J.}~\bibnamefont
  {de~Blas}} \emph {et~al.},\ }\href@noop {} {\  (\bibinfo {year} {2018})},\
  \Eprint {http://arxiv.org/abs/1812.02093} {arXiv:1812.02093 [hep-ph]}
  \BibitemShut {NoStop}%
\bibitem [{\citenamefont {Abelleira~Fernandez}\ \emph
  {et~al.}(2012)\citenamefont {Abelleira~Fernandez} \emph
  {et~al.}}]{AbelleiraFernandez:2012cc}%
  \BibitemOpen
  \bibfield  {author} {\bibinfo {author} {\bibfnamefont {J.~L.}\ \bibnamefont
  {Abelleira~Fernandez}} \emph {et~al.} (\bibinfo {collaboration} {LHeC Study
  Group}),\ }\href {\doibase 10.1088/0954-3899/39/7/075001} {\bibfield
  {journal} {\bibinfo  {journal} {J. Phys.}\ }\textbf {\bibinfo {volume}
  {G39}},\ \bibinfo {pages} {075001} (\bibinfo {year} {2012})},\ \Eprint
  {http://arxiv.org/abs/1206.2913} {arXiv:1206.2913 [physics.acc-ph]}
  \BibitemShut {NoStop}%
\bibitem [{\citenamefont {Acar}\ \emph {et~al.}(2017)\citenamefont {Acar},
  \citenamefont {Akay}, \citenamefont {Beser}, \citenamefont {Canbay},
  \citenamefont {Karadeniz}, \citenamefont {Kaya}, \citenamefont {Oner},\ and\
  \citenamefont {Sultansoy}}]{Acar:2016rde}%
  \BibitemOpen
  \bibfield  {author} {\bibinfo {author} {\bibfnamefont {Y.~C.}\ \bibnamefont
  {Acar}}, \bibinfo {author} {\bibfnamefont {A.~N.}\ \bibnamefont {Akay}},
  \bibinfo {author} {\bibfnamefont {S.}~\bibnamefont {Beser}}, \bibinfo
  {author} {\bibfnamefont {A.~C.}\ \bibnamefont {Canbay}}, \bibinfo {author}
  {\bibfnamefont {H.}~\bibnamefont {Karadeniz}}, \bibinfo {author}
  {\bibfnamefont {U.}~\bibnamefont {Kaya}}, \bibinfo {author} {\bibfnamefont
  {B.~B.}\ \bibnamefont {Oner}}, \ and\ \bibinfo {author} {\bibfnamefont
  {S.}~\bibnamefont {Sultansoy}},\ }\href {\doibase 10.1016/j.nima.2017.07.041}
  {\bibfield  {journal} {\bibinfo  {journal} {Nucl. Instrum. Meth.}\ }\textbf
  {\bibinfo {volume} {A871}},\ \bibinfo {pages} {47} (\bibinfo {year}
  {2017})},\ \Eprint {http://arxiv.org/abs/1608.02190} {arXiv:1608.02190
  [physics.acc-ph]} \BibitemShut {NoStop}%
\bibitem [{\citenamefont {Azuelos}\ \emph {et~al.}(2018)\citenamefont
  {Azuelos}, \citenamefont {D'Onofrio}, \citenamefont {Fischer},\ and\
  \citenamefont {Zurita}}]{Azuelos:2018syu}%
  \BibitemOpen
  \bibfield  {author} {\bibinfo {author} {\bibfnamefont {G.}~\bibnamefont
  {Azuelos}}, \bibinfo {author} {\bibfnamefont {M.}~\bibnamefont {D'Onofrio}},
  \bibinfo {author} {\bibfnamefont {O.}~\bibnamefont {Fischer}}, \ and\
  \bibinfo {author} {\bibfnamefont {J.}~\bibnamefont {Zurita}},\ }\href
  {\doibase 10.22323/1.316.0190} {\bibfield  {journal} {\bibinfo  {journal}
  {PoS}\ }\textbf {\bibinfo {volume} {DIS2018}},\ \bibinfo {pages} {190}
  (\bibinfo {year} {2018})},\ \Eprint {http://arxiv.org/abs/1807.01618}
  {arXiv:1807.01618 [hep-ph]} \BibitemShut {NoStop}%
\bibitem [{\citenamefont {Liang}\ \emph {et~al.}(2010)\citenamefont {Liang},
  \citenamefont {He}, \citenamefont {Ma}, \citenamefont {Wang},\ and\
  \citenamefont {Zhang}}]{Liang:2010gm}%
  \BibitemOpen
  \bibfield  {author} {\bibinfo {author} {\bibfnamefont {H.}~\bibnamefont
  {Liang}}, \bibinfo {author} {\bibfnamefont {X.-G.}\ \bibnamefont {He}},
  \bibinfo {author} {\bibfnamefont {W.-G.}\ \bibnamefont {Ma}}, \bibinfo
  {author} {\bibfnamefont {S.-M.}\ \bibnamefont {Wang}}, \ and\ \bibinfo
  {author} {\bibfnamefont {R.-Y.}\ \bibnamefont {Zhang}},\ }\href {\doibase
  10.1007/JHEP09(2010)023} {\bibfield  {journal} {\bibinfo  {journal} {JHEP}\
  }\textbf {\bibinfo {volume} {09}},\ \bibinfo {pages} {023} (\bibinfo {year}
  {2010})},\ \Eprint {http://arxiv.org/abs/1006.5534} {arXiv:1006.5534
  [hep-ph]} \BibitemShut {NoStop}%
\bibitem [{\citenamefont {Blaksley}\ \emph {et~al.}(2011)\citenamefont
  {Blaksley}, \citenamefont {Blennow}, \citenamefont {Bonnet}, \citenamefont
  {Coloma},\ and\ \citenamefont {Fernandez-Martinez}}]{Blaksley:2011ey}%
  \BibitemOpen
  \bibfield  {author} {\bibinfo {author} {\bibfnamefont {C.}~\bibnamefont
  {Blaksley}}, \bibinfo {author} {\bibfnamefont {M.}~\bibnamefont {Blennow}},
  \bibinfo {author} {\bibfnamefont {F.}~\bibnamefont {Bonnet}}, \bibinfo
  {author} {\bibfnamefont {P.}~\bibnamefont {Coloma}}, \ and\ \bibinfo {author}
  {\bibfnamefont {E.}~\bibnamefont {Fernandez-Martinez}},\ }\href {\doibase
  10.1016/j.nuclphysb.2011.06.021} {\bibfield  {journal} {\bibinfo  {journal}
  {Nucl. Phys.}\ }\textbf {\bibinfo {volume} {B852}},\ \bibinfo {pages} {353}
  (\bibinfo {year} {2011})},\ \Eprint {http://arxiv.org/abs/1105.0308}
  {arXiv:1105.0308 [hep-ph]} \BibitemShut {NoStop}%
\bibitem [{\citenamefont {Duarte}\ \emph {et~al.}(2015)\citenamefont {Duarte},
  \citenamefont {González-Sprinberg},\ and\ \citenamefont
  {Sampayo}}]{Duarte:2014zea}%
  \BibitemOpen
  \bibfield  {author} {\bibinfo {author} {\bibfnamefont {L.}~\bibnamefont
  {Duarte}}, \bibinfo {author} {\bibfnamefont {G.~A.}\ \bibnamefont
  {González-Sprinberg}}, \ and\ \bibinfo {author} {\bibfnamefont {O.~A.}\
  \bibnamefont {Sampayo}},\ }\href {\doibase 10.1103/PhysRevD.91.053007}
  {\bibfield  {journal} {\bibinfo  {journal} {Phys. Rev.}\ }\textbf {\bibinfo
  {volume} {D91}},\ \bibinfo {pages} {053007} (\bibinfo {year} {2015})},\
  \Eprint {http://arxiv.org/abs/1412.1433} {arXiv:1412.1433 [hep-ph]}
  \BibitemShut {NoStop}%
\bibitem [{\citenamefont {Mondal}\ and\ \citenamefont
  {Rai}(2016{\natexlab{a}})}]{Mondal:2015zba}%
  \BibitemOpen
  \bibfield  {author} {\bibinfo {author} {\bibfnamefont {S.}~\bibnamefont
  {Mondal}}\ and\ \bibinfo {author} {\bibfnamefont {S.~K.}\ \bibnamefont
  {Rai}},\ }\href {\doibase 10.1103/PhysRevD.93.011702} {\bibfield  {journal}
  {\bibinfo  {journal} {Phys. Rev.}\ }\textbf {\bibinfo {volume} {D93}},\
  \bibinfo {pages} {011702} (\bibinfo {year} {2016}{\natexlab{a}})},\ \Eprint
  {http://arxiv.org/abs/1510.08632} {arXiv:1510.08632 [hep-ph]} \BibitemShut
  {NoStop}%
\bibitem [{\citenamefont {Lindner}\ \emph {et~al.}(2016)\citenamefont
  {Lindner}, \citenamefont {Queiroz}, \citenamefont {Rodejohann},\ and\
  \citenamefont {Yaguna}}]{Lindner:2016lxq}%
  \BibitemOpen
  \bibfield  {author} {\bibinfo {author} {\bibfnamefont {M.}~\bibnamefont
  {Lindner}}, \bibinfo {author} {\bibfnamefont {F.~S.}\ \bibnamefont
  {Queiroz}}, \bibinfo {author} {\bibfnamefont {W.}~\bibnamefont {Rodejohann}},
  \ and\ \bibinfo {author} {\bibfnamefont {C.~E.}\ \bibnamefont {Yaguna}},\
  }\href {\doibase 10.1007/JHEP06(2016)140} {\bibfield  {journal} {\bibinfo
  {journal} {JHEP}\ }\textbf {\bibinfo {volume} {06}},\ \bibinfo {pages} {140}
  (\bibinfo {year} {2016})},\ \Eprint {http://arxiv.org/abs/1604.08596}
  {arXiv:1604.08596 [hep-ph]} \BibitemShut {NoStop}%
\bibitem [{\citenamefont {Mondal}\ and\ \citenamefont
  {Rai}(2016{\natexlab{b}})}]{Mondal:2016kof}%
  \BibitemOpen
  \bibfield  {author} {\bibinfo {author} {\bibfnamefont {S.}~\bibnamefont
  {Mondal}}\ and\ \bibinfo {author} {\bibfnamefont {S.~K.}\ \bibnamefont
  {Rai}},\ }\href {\doibase 10.1103/PhysRevD.94.033008} {\bibfield  {journal}
  {\bibinfo  {journal} {Phys. Rev.}\ }\textbf {\bibinfo {volume} {D94}},\
  \bibinfo {pages} {033008} (\bibinfo {year} {2016}{\natexlab{b}})},\ \Eprint
  {http://arxiv.org/abs/1605.04508} {arXiv:1605.04508 [hep-ph]} \BibitemShut
  {NoStop}%
\bibitem [{\citenamefont {Antusch}\ \emph {et~al.}(2017)\citenamefont
  {Antusch}, \citenamefont {Cazzato},\ and\ \citenamefont
  {Fischer}}]{Antusch:2016ejd}%
  \BibitemOpen
  \bibfield  {author} {\bibinfo {author} {\bibfnamefont {S.}~\bibnamefont
  {Antusch}}, \bibinfo {author} {\bibfnamefont {E.}~\bibnamefont {Cazzato}}, \
  and\ \bibinfo {author} {\bibfnamefont {O.}~\bibnamefont {Fischer}},\ }\href
  {\doibase 10.1142/S0217751X17500786} {\bibfield  {journal} {\bibinfo
  {journal} {Int. J. Mod. Phys.}\ }\textbf {\bibinfo {volume} {A32}},\ \bibinfo
  {pages} {1750078} (\bibinfo {year} {2017})},\ \Eprint
  {http://arxiv.org/abs/1612.02728} {arXiv:1612.02728 [hep-ph]} \BibitemShut
  {NoStop}%
\bibitem [{\citenamefont {Curtin}\ \emph {et~al.}(2018)\citenamefont {Curtin},
  \citenamefont {Deshpande}, \citenamefont {Fischer},\ and\ \citenamefont
  {Zurita}}]{Curtin:2017bxr}%
  \BibitemOpen
  \bibfield  {author} {\bibinfo {author} {\bibfnamefont {D.}~\bibnamefont
  {Curtin}}, \bibinfo {author} {\bibfnamefont {K.}~\bibnamefont {Deshpande}},
  \bibinfo {author} {\bibfnamefont {O.}~\bibnamefont {Fischer}}, \ and\
  \bibinfo {author} {\bibfnamefont {J.}~\bibnamefont {Zurita}},\ }\href
  {\doibase 10.1007/JHEP07(2018)024} {\bibfield  {journal} {\bibinfo  {journal}
  {JHEP}\ }\textbf {\bibinfo {volume} {07}},\ \bibinfo {pages} {024} (\bibinfo
  {year} {2018})},\ \Eprint {http://arxiv.org/abs/1712.07135} {arXiv:1712.07135
  [hep-ph]} \BibitemShut {NoStop}%
\bibitem [{\citenamefont {Mandal}\ \emph {et~al.}(2018)\citenamefont {Mandal},
  \citenamefont {Mitra},\ and\ \citenamefont {Sinha}}]{Mandal:2018qpg}%
  \BibitemOpen
  \bibfield  {author} {\bibinfo {author} {\bibfnamefont {S.}~\bibnamefont
  {Mandal}}, \bibinfo {author} {\bibfnamefont {M.}~\bibnamefont {Mitra}}, \
  and\ \bibinfo {author} {\bibfnamefont {N.}~\bibnamefont {Sinha}},\ }\href
  {\doibase 10.1103/PhysRevD.98.095004} {\bibfield  {journal} {\bibinfo
  {journal} {Phys. Rev.}\ }\textbf {\bibinfo {volume} {D98}},\ \bibinfo {pages}
  {095004} (\bibinfo {year} {2018})},\ \Eprint
  {http://arxiv.org/abs/1807.06455} {arXiv:1807.06455 [hep-ph]} \BibitemShut
  {NoStop}%
\bibitem [{\citenamefont {Das}\ \emph {et~al.}(2018)\citenamefont {Das},
  \citenamefont {Jana}, \citenamefont {Mandal},\ and\ \citenamefont
  {Nandi}}]{Das:2018usr}%
  \BibitemOpen
  \bibfield  {author} {\bibinfo {author} {\bibfnamefont {A.}~\bibnamefont
  {Das}}, \bibinfo {author} {\bibfnamefont {S.}~\bibnamefont {Jana}}, \bibinfo
  {author} {\bibfnamefont {S.}~\bibnamefont {Mandal}}, \ and\ \bibinfo {author}
  {\bibfnamefont {S.}~\bibnamefont {Nandi}},\ }\href@noop {} {\  (\bibinfo
  {year} {2018})},\ \Eprint {http://arxiv.org/abs/1811.04291} {arXiv:1811.04291
  [hep-ph]} \BibitemShut {NoStop}%
\bibitem [{\citenamefont {Li}\ \emph {et~al.}(2018)\citenamefont {Li},
  \citenamefont {Si},\ and\ \citenamefont {Yang}}]{Li:2018wut}%
  \BibitemOpen
  \bibfield  {author} {\bibinfo {author} {\bibfnamefont {S.-Y.}\ \bibnamefont
  {Li}}, \bibinfo {author} {\bibfnamefont {Z.-G.}\ \bibnamefont {Si}}, \ and\
  \bibinfo {author} {\bibfnamefont {X.-H.}\ \bibnamefont {Yang}},\ }\href@noop
  {} {\  (\bibinfo {year} {2018})},\ \Eprint {http://arxiv.org/abs/1811.10313}
  {arXiv:1811.10313 [hep-ph]} \BibitemShut {NoStop}%
\bibitem [{\citenamefont {Accomando}\ and\ \citenamefont
  {Petrarca}(1994)}]{Accomando:1993ar}%
  \BibitemOpen
  \bibfield  {author} {\bibinfo {author} {\bibfnamefont {E.}~\bibnamefont
  {Accomando}}\ and\ \bibinfo {author} {\bibfnamefont {S.}~\bibnamefont
  {Petrarca}},\ }\href {\doibase 10.1016/0370-2693(94)90293-3} {\bibfield
  {journal} {\bibinfo  {journal} {Phys. Lett.}\ }\textbf {\bibinfo {volume}
  {B323}},\ \bibinfo {pages} {212} (\bibinfo {year} {1994})},\ \Eprint
  {http://arxiv.org/abs/hep-ph/9401242} {arXiv:hep-ph/9401242 [hep-ph]}
  \BibitemShut {NoStop}%
\bibitem [{\citenamefont {Aktas}\ \emph {et~al.}(2006)\citenamefont {Aktas}
  \emph {et~al.}}]{Aktas:2006nu}%
  \BibitemOpen
  \bibfield  {author} {\bibinfo {author} {\bibfnamefont {A.}~\bibnamefont
  {Aktas}} \emph {et~al.} (\bibinfo {collaboration} {H1}),\ }\href {\doibase
  10.1016/j.physletb.2006.05.061} {\bibfield  {journal} {\bibinfo  {journal}
  {Phys. Lett.}\ }\textbf {\bibinfo {volume} {B638}},\ \bibinfo {pages} {432}
  (\bibinfo {year} {2006})},\ \Eprint {http://arxiv.org/abs/hep-ex/0604027}
  {arXiv:hep-ex/0604027 [hep-ex]} \BibitemShut {NoStop}%
\bibitem [{\citenamefont {Yue}\ \emph {et~al.}(2007)\citenamefont {Yue},
  \citenamefont {Zhao},\ and\ \citenamefont {Ma}}]{Yue:2007ym}%
  \BibitemOpen
  \bibfield  {author} {\bibinfo {author} {\bibfnamefont {C.-X.}\ \bibnamefont
  {Yue}}, \bibinfo {author} {\bibfnamefont {S.}~\bibnamefont {Zhao}}, \ and\
  \bibinfo {author} {\bibfnamefont {W.}~\bibnamefont {Ma}},\ }\href {\doibase
  10.1016/j.nuclphysb.2007.06.003} {\bibfield  {journal} {\bibinfo  {journal}
  {Nucl. Phys.}\ }\textbf {\bibinfo {volume} {B784}},\ \bibinfo {pages} {36}
  (\bibinfo {year} {2007})},\ \Eprint {http://arxiv.org/abs/0706.0232}
  {arXiv:0706.0232 [hep-ph]} \BibitemShut {NoStop}%
\bibitem [{\citenamefont {Ade}\ \emph {et~al.}(2016)\citenamefont {Ade} \emph
  {et~al.}}]{Ade:2015xua}%
  \BibitemOpen
  \bibfield  {author} {\bibinfo {author} {\bibfnamefont {P.~A.~R.}\
  \bibnamefont {Ade}} \emph {et~al.} (\bibinfo {collaboration} {Planck}),\
  }\href {\doibase 10.1051/0004-6361/201525830} {\bibfield  {journal} {\bibinfo
   {journal} {Astron. Astrophys.}\ }\textbf {\bibinfo {volume} {594}},\
  \bibinfo {pages} {A13} (\bibinfo {year} {2016})},\ \Eprint
  {http://arxiv.org/abs/1502.01589} {arXiv:1502.01589 [astro-ph.CO]}
  \BibitemShut {NoStop}%
\bibitem [{\citenamefont {Grifols}\ \emph {et~al.}(1989)\citenamefont
  {Grifols}, \citenamefont {Mendez},\ and\ \citenamefont
  {Schuler}}]{Grifols:1989xe}%
  \BibitemOpen
  \bibfield  {author} {\bibinfo {author} {\bibfnamefont {J.~A.}\ \bibnamefont
  {Grifols}}, \bibinfo {author} {\bibfnamefont {A.}~\bibnamefont {Mendez}}, \
  and\ \bibinfo {author} {\bibfnamefont {G.~A.}\ \bibnamefont {Schuler}},\
  }\href {\doibase 10.1142/S0217732389001702} {\bibfield  {journal} {\bibinfo
  {journal} {Mod. Phys. Lett.}\ }\textbf {\bibinfo {volume} {A4}},\ \bibinfo
  {pages} {1485} (\bibinfo {year} {1989})}\BibitemShut {NoStop}%
\bibitem [{\citenamefont {Gunion}\ \emph {et~al.}(1989)\citenamefont {Gunion},
  \citenamefont {Grifols}, \citenamefont {Mendez}, \citenamefont {Kayser},\
  and\ \citenamefont {Olness}}]{Gunion:1989in}%
  \BibitemOpen
  \bibfield  {author} {\bibinfo {author} {\bibfnamefont {J.~F.}\ \bibnamefont
  {Gunion}}, \bibinfo {author} {\bibfnamefont {J.}~\bibnamefont {Grifols}},
  \bibinfo {author} {\bibfnamefont {A.}~\bibnamefont {Mendez}}, \bibinfo
  {author} {\bibfnamefont {B.}~\bibnamefont {Kayser}}, \ and\ \bibinfo {author}
  {\bibfnamefont {F.~I.}\ \bibnamefont {Olness}},\ }\href {\doibase
  10.1103/PhysRevD.40.1546} {\bibfield  {journal} {\bibinfo  {journal} {Phys.
  Rev.}\ }\textbf {\bibinfo {volume} {D40}},\ \bibinfo {pages} {1546} (\bibinfo
  {year} {1989})}\BibitemShut {NoStop}%
\bibitem [{\citenamefont {Muhlleitner}\ and\ \citenamefont
  {Spira}(2003)}]{Muhlleitner:2003me}%
  \BibitemOpen
  \bibfield  {author} {\bibinfo {author} {\bibfnamefont {M.}~\bibnamefont
  {Muhlleitner}}\ and\ \bibinfo {author} {\bibfnamefont {M.}~\bibnamefont
  {Spira}},\ }\href {\doibase 10.1103/PhysRevD.68.117701} {\bibfield  {journal}
  {\bibinfo  {journal} {Phys. Rev.}\ }\textbf {\bibinfo {volume} {D68}},\
  \bibinfo {pages} {117701} (\bibinfo {year} {2003})},\ \Eprint
  {http://arxiv.org/abs/hep-ph/0305288} {arXiv:hep-ph/0305288 [hep-ph]}
  \BibitemShut {NoStop}%
\bibitem [{\citenamefont {Akeroyd}\ and\ \citenamefont
  {Aoki}(2005)}]{Akeroyd:2005gt}%
  \BibitemOpen
  \bibfield  {author} {\bibinfo {author} {\bibfnamefont {A.~G.}\ \bibnamefont
  {Akeroyd}}\ and\ \bibinfo {author} {\bibfnamefont {M.}~\bibnamefont {Aoki}},\
  }\href {\doibase 10.1103/PhysRevD.72.035011} {\bibfield  {journal} {\bibinfo
  {journal} {Phys. Rev.}\ }\textbf {\bibinfo {volume} {D72}},\ \bibinfo {pages}
  {035011} (\bibinfo {year} {2005})},\ \Eprint
  {http://arxiv.org/abs/hep-ph/0506176} {arXiv:hep-ph/0506176 [hep-ph]}
  \BibitemShut {NoStop}%
\bibitem [{\citenamefont {Han}\ \emph {et~al.}(2007)\citenamefont {Han},
  \citenamefont {Mukhopadhyaya}, \citenamefont {Si},\ and\ \citenamefont
  {Wang}}]{Han:2007bk}%
  \BibitemOpen
  \bibfield  {author} {\bibinfo {author} {\bibfnamefont {T.}~\bibnamefont
  {Han}}, \bibinfo {author} {\bibfnamefont {B.}~\bibnamefont {Mukhopadhyaya}},
  \bibinfo {author} {\bibfnamefont {Z.}~\bibnamefont {Si}}, \ and\ \bibinfo
  {author} {\bibfnamefont {K.}~\bibnamefont {Wang}},\ }\href {\doibase
  10.1103/PhysRevD.76.075013} {\bibfield  {journal} {\bibinfo  {journal} {Phys.
  Rev.}\ }\textbf {\bibinfo {volume} {D76}},\ \bibinfo {pages} {075013}
  (\bibinfo {year} {2007})},\ \Eprint {http://arxiv.org/abs/0706.0441}
  {arXiv:0706.0441 [hep-ph]} \BibitemShut {NoStop}%
\bibitem [{\citenamefont {Chao}\ \emph {et~al.}(2008)\citenamefont {Chao},
  \citenamefont {Si}, \citenamefont {Xing},\ and\ \citenamefont
  {Zhou}}]{Chao:2008mq}%
  \BibitemOpen
  \bibfield  {author} {\bibinfo {author} {\bibfnamefont {W.}~\bibnamefont
  {Chao}}, \bibinfo {author} {\bibfnamefont {Z.-G.}\ \bibnamefont {Si}},
  \bibinfo {author} {\bibfnamefont {Z.-z.}\ \bibnamefont {Xing}}, \ and\
  \bibinfo {author} {\bibfnamefont {S.}~\bibnamefont {Zhou}},\ }\href {\doibase
  10.1016/j.physletb.2008.08.003} {\bibfield  {journal} {\bibinfo  {journal}
  {Phys. Lett.}\ }\textbf {\bibinfo {volume} {B666}},\ \bibinfo {pages} {451}
  (\bibinfo {year} {2008})},\ \Eprint {http://arxiv.org/abs/0804.1265}
  {arXiv:0804.1265 [hep-ph]} \BibitemShut {NoStop}%
\bibitem [{\citenamefont {Fileviez~Perez}\ \emph {et~al.}(2008)\citenamefont
  {Fileviez~Perez}, \citenamefont {Han}, \citenamefont {Huang}, \citenamefont
  {Li},\ and\ \citenamefont {Wang}}]{Perez:2008ha}%
  \BibitemOpen
  \bibfield  {author} {\bibinfo {author} {\bibfnamefont {P.}~\bibnamefont
  {Fileviez~Perez}}, \bibinfo {author} {\bibfnamefont {T.}~\bibnamefont {Han}},
  \bibinfo {author} {\bibfnamefont {G.-y.}\ \bibnamefont {Huang}}, \bibinfo
  {author} {\bibfnamefont {T.}~\bibnamefont {Li}}, \ and\ \bibinfo {author}
  {\bibfnamefont {K.}~\bibnamefont {Wang}},\ }\href {\doibase
  10.1103/PhysRevD.78.015018} {\bibfield  {journal} {\bibinfo  {journal} {Phys.
  Rev.}\ }\textbf {\bibinfo {volume} {D78}},\ \bibinfo {pages} {015018}
  (\bibinfo {year} {2008})},\ \Eprint {http://arxiv.org/abs/0805.3536}
  {arXiv:0805.3536 [hep-ph]} \BibitemShut {NoStop}%
\bibitem [{\citenamefont {del Aguila}\ and\ \citenamefont
  {Aguilar-Saavedra}(2009)}]{delAguila:2008cj}%
  \BibitemOpen
  \bibfield  {author} {\bibinfo {author} {\bibfnamefont {F.}~\bibnamefont {del
  Aguila}}\ and\ \bibinfo {author} {\bibfnamefont {J.~A.}\ \bibnamefont
  {Aguilar-Saavedra}},\ }\href {\doibase 10.1016/j.nuclphysb.2008.12.029}
  {\bibfield  {journal} {\bibinfo  {journal} {Nucl. Phys.}\ }\textbf {\bibinfo
  {volume} {B813}},\ \bibinfo {pages} {22} (\bibinfo {year} {2009})},\ \Eprint
  {http://arxiv.org/abs/0808.2468} {arXiv:0808.2468 [hep-ph]} \BibitemShut
  {NoStop}%
\bibitem [{\citenamefont {Akeroyd}\ and\ \citenamefont
  {Chiang}(2009)}]{Akeroyd:2009hb}%
  \BibitemOpen
  \bibfield  {author} {\bibinfo {author} {\bibfnamefont {A.~G.}\ \bibnamefont
  {Akeroyd}}\ and\ \bibinfo {author} {\bibfnamefont {C.-W.}\ \bibnamefont
  {Chiang}},\ }\href {\doibase 10.1103/PhysRevD.80.113010} {\bibfield
  {journal} {\bibinfo  {journal} {Phys. Rev.}\ }\textbf {\bibinfo {volume}
  {D80}},\ \bibinfo {pages} {113010} (\bibinfo {year} {2009})},\ \Eprint
  {http://arxiv.org/abs/0909.4419} {arXiv:0909.4419 [hep-ph]} \BibitemShut
  {NoStop}%
\bibitem [{\citenamefont {Akeroyd}\ \emph {et~al.}(2010)\citenamefont
  {Akeroyd}, \citenamefont {Chiang},\ and\ \citenamefont
  {Gaur}}]{Akeroyd:2010ip}%
  \BibitemOpen
  \bibfield  {author} {\bibinfo {author} {\bibfnamefont {A.~G.}\ \bibnamefont
  {Akeroyd}}, \bibinfo {author} {\bibfnamefont {C.-W.}\ \bibnamefont {Chiang}},
  \ and\ \bibinfo {author} {\bibfnamefont {N.}~\bibnamefont {Gaur}},\ }\href
  {\doibase 10.1007/JHEP11(2010)005} {\bibfield  {journal} {\bibinfo  {journal}
  {JHEP}\ }\textbf {\bibinfo {volume} {11}},\ \bibinfo {pages} {005} (\bibinfo
  {year} {2010})},\ \Eprint {http://arxiv.org/abs/1009.2780} {arXiv:1009.2780
  [hep-ph]} \BibitemShut {NoStop}%
\bibitem [{\citenamefont {Melfo}\ \emph {et~al.}(2012)\citenamefont {Melfo},
  \citenamefont {Nemevsek}, \citenamefont {Nesti}, \citenamefont
  {Senjanovi\'{c}},\ and\ \citenamefont {Zhang}}]{Melfo:2011nx}%
  \BibitemOpen
  \bibfield  {author} {\bibinfo {author} {\bibfnamefont {A.}~\bibnamefont
  {Melfo}}, \bibinfo {author} {\bibfnamefont {M.}~\bibnamefont {Nemevsek}},
  \bibinfo {author} {\bibfnamefont {F.}~\bibnamefont {Nesti}}, \bibinfo
  {author} {\bibfnamefont {G.}~\bibnamefont {Senjanovi\'{c}}}, \ and\ \bibinfo
  {author} {\bibfnamefont {Y.}~\bibnamefont {Zhang}},\ }\href {\doibase
  10.1103/PhysRevD.85.055018} {\bibfield  {journal} {\bibinfo  {journal} {Phys.
  Rev.}\ }\textbf {\bibinfo {volume} {D85}},\ \bibinfo {pages} {055018}
  (\bibinfo {year} {2012})},\ \Eprint {http://arxiv.org/abs/1108.4416}
  {arXiv:1108.4416 [hep-ph]} \BibitemShut {NoStop}%
\bibitem [{\citenamefont {Alloul}\ \emph {et~al.}(2013)\citenamefont {Alloul},
  \citenamefont {Frank}, \citenamefont {Fuks},\ and\ \citenamefont {Rausch~de
  Traubenberg}}]{Alloul:2013raa}%
  \BibitemOpen
  \bibfield  {author} {\bibinfo {author} {\bibfnamefont {A.}~\bibnamefont
  {Alloul}}, \bibinfo {author} {\bibfnamefont {M.}~\bibnamefont {Frank}},
  \bibinfo {author} {\bibfnamefont {B.}~\bibnamefont {Fuks}}, \ and\ \bibinfo
  {author} {\bibfnamefont {M.}~\bibnamefont {Rausch~de Traubenberg}},\ }\href
  {\doibase 10.1103/PhysRevD.88.075004} {\bibfield  {journal} {\bibinfo
  {journal} {Phys. Rev.}\ }\textbf {\bibinfo {volume} {D88}},\ \bibinfo {pages}
  {075004} (\bibinfo {year} {2013})},\ \Eprint {http://arxiv.org/abs/1307.1711}
  {arXiv:1307.1711 [hep-ph]} \BibitemShut {NoStop}%
\bibitem [{\citenamefont {Chun}\ and\ \citenamefont
  {Sharma}(2014)}]{Chun:2013vma}%
  \BibitemOpen
  \bibfield  {author} {\bibinfo {author} {\bibfnamefont {E.~J.}\ \bibnamefont
  {Chun}}\ and\ \bibinfo {author} {\bibfnamefont {P.}~\bibnamefont {Sharma}},\
  }\href {\doibase 10.1016/j.physletb.2013.11.056} {\bibfield  {journal}
  {\bibinfo  {journal} {Phys. Lett.}\ }\textbf {\bibinfo {volume} {B728}},\
  \bibinfo {pages} {256} (\bibinfo {year} {2014})},\ \Eprint
  {http://arxiv.org/abs/1309.6888} {arXiv:1309.6888 [hep-ph]} \BibitemShut
  {NoStop}%
\bibitem [{\citenamefont {del Águila}\ and\ \citenamefont
  {Chala}(2014)}]{delAguila:2013mia}%
  \BibitemOpen
  \bibfield  {author} {\bibinfo {author} {\bibfnamefont {F.}~\bibnamefont {del
  Águila}}\ and\ \bibinfo {author} {\bibfnamefont {M.}~\bibnamefont {Chala}},\
  }\href {\doibase 10.1007/JHEP03(2014)027} {\bibfield  {journal} {\bibinfo
  {journal} {JHEP}\ }\textbf {\bibinfo {volume} {03}},\ \bibinfo {pages} {027}
  (\bibinfo {year} {2014})},\ \Eprint {http://arxiv.org/abs/1311.1510}
  {arXiv:1311.1510 [hep-ph]} \BibitemShut {NoStop}%
\bibitem [{\citenamefont {Bambhaniya}\ \emph {et~al.}(2014)\citenamefont
  {Bambhaniya}, \citenamefont {Chakrabortty}, \citenamefont {Gluza},
  \citenamefont {Kordiaczyńska},\ and\ \citenamefont
  {Szafron}}]{Bambhaniya:2013wza}%
  \BibitemOpen
  \bibfield  {author} {\bibinfo {author} {\bibfnamefont {G.}~\bibnamefont
  {Bambhaniya}}, \bibinfo {author} {\bibfnamefont {J.}~\bibnamefont
  {Chakrabortty}}, \bibinfo {author} {\bibfnamefont {J.}~\bibnamefont {Gluza}},
  \bibinfo {author} {\bibfnamefont {M.}~\bibnamefont {Kordiaczyńska}}, \ and\
  \bibinfo {author} {\bibfnamefont {R.}~\bibnamefont {Szafron}},\ }\href
  {\doibase 10.1007/JHEP05(2014)033} {\bibfield  {journal} {\bibinfo  {journal}
  {JHEP}\ }\textbf {\bibinfo {volume} {05}},\ \bibinfo {pages} {033} (\bibinfo
  {year} {2014})},\ \Eprint {http://arxiv.org/abs/1311.4144} {arXiv:1311.4144
  [hep-ph]} \BibitemShut {NoStop}%
\bibitem [{\citenamefont {Dev}\ \emph {et~al.}(2016)\citenamefont {Dev},
  \citenamefont {Mohapatra},\ and\ \citenamefont {Zhang}}]{Dev:2016dja}%
  \BibitemOpen
  \bibfield  {author} {\bibinfo {author} {\bibfnamefont {P.~S.~B.}\
  \bibnamefont {Dev}}, \bibinfo {author} {\bibfnamefont {R.~N.}\ \bibnamefont
  {Mohapatra}}, \ and\ \bibinfo {author} {\bibfnamefont {Y.}~\bibnamefont
  {Zhang}},\ }\href {\doibase 10.1007/JHEP05(2016)174} {\bibfield  {journal}
  {\bibinfo  {journal} {JHEP}\ }\textbf {\bibinfo {volume} {05}},\ \bibinfo
  {pages} {174} (\bibinfo {year} {2016})},\ \Eprint
  {http://arxiv.org/abs/1602.05947} {arXiv:1602.05947 [hep-ph]} \BibitemShut
  {NoStop}%
\bibitem [{\citenamefont {Mitra}\ \emph {et~al.}(2017)\citenamefont {Mitra},
  \citenamefont {Niyogi},\ and\ \citenamefont {Spannowsky}}]{Mitra:2016wpr}%
  \BibitemOpen
  \bibfield  {author} {\bibinfo {author} {\bibfnamefont {M.}~\bibnamefont
  {Mitra}}, \bibinfo {author} {\bibfnamefont {S.}~\bibnamefont {Niyogi}}, \
  and\ \bibinfo {author} {\bibfnamefont {M.}~\bibnamefont {Spannowsky}},\
  }\href {\doibase 10.1103/PhysRevD.95.035042} {\bibfield  {journal} {\bibinfo
  {journal} {Phys. Rev.}\ }\textbf {\bibinfo {volume} {D95}},\ \bibinfo {pages}
  {035042} (\bibinfo {year} {2017})},\ \Eprint
  {http://arxiv.org/abs/1611.09594} {arXiv:1611.09594 [hep-ph]} \BibitemShut
  {NoStop}%
\bibitem [{\citenamefont {Ghosh}\ \emph {et~al.}(2018)\citenamefont {Ghosh},
  \citenamefont {Ghosh}, \citenamefont {Saha},\ and\ \citenamefont
  {Shaw}}]{Ghosh:2017pxl}%
  \BibitemOpen
  \bibfield  {author} {\bibinfo {author} {\bibfnamefont {D.~K.}\ \bibnamefont
  {Ghosh}}, \bibinfo {author} {\bibfnamefont {N.}~\bibnamefont {Ghosh}},
  \bibinfo {author} {\bibfnamefont {I.}~\bibnamefont {Saha}}, \ and\ \bibinfo
  {author} {\bibfnamefont {A.}~\bibnamefont {Shaw}},\ }\href {\doibase
  10.1103/PhysRevD.97.115022} {\bibfield  {journal} {\bibinfo  {journal} {Phys.
  Rev.}\ }\textbf {\bibinfo {volume} {D97}},\ \bibinfo {pages} {115022}
  (\bibinfo {year} {2018})},\ \Eprint {http://arxiv.org/abs/1711.06062}
  {arXiv:1711.06062 [hep-ph]} \BibitemShut {NoStop}%
\bibitem [{\citenamefont {Li}(2018)}]{Li:2018jns}%
  \BibitemOpen
  \bibfield  {author} {\bibinfo {author} {\bibfnamefont {T.}~\bibnamefont
  {Li}},\ }\href {\doibase 10.1007/JHEP09(2018)079} {\bibfield  {journal}
  {\bibinfo  {journal} {JHEP}\ }\textbf {\bibinfo {volume} {09}},\ \bibinfo
  {pages} {079} (\bibinfo {year} {2018})},\ \Eprint
  {http://arxiv.org/abs/1802.00945} {arXiv:1802.00945 [hep-ph]} \BibitemShut
  {NoStop}%
\bibitem [{\citenamefont {Borah}\ \emph {et~al.}(2018)\citenamefont {Borah},
  \citenamefont {Fuks}, \citenamefont {Goswami},\ and\ \citenamefont
  {Poulose}}]{Borah:2018yxd}%
  \BibitemOpen
  \bibfield  {author} {\bibinfo {author} {\bibfnamefont {D.}~\bibnamefont
  {Borah}}, \bibinfo {author} {\bibfnamefont {B.}~\bibnamefont {Fuks}},
  \bibinfo {author} {\bibfnamefont {D.}~\bibnamefont {Goswami}}, \ and\
  \bibinfo {author} {\bibfnamefont {P.}~\bibnamefont {Poulose}},\ }\href
  {\doibase 10.1103/PhysRevD.98.035008} {\bibfield  {journal} {\bibinfo
  {journal} {Phys. Rev.}\ }\textbf {\bibinfo {volume} {D98}},\ \bibinfo {pages}
  {035008} (\bibinfo {year} {2018})},\ \Eprint
  {http://arxiv.org/abs/1805.06910} {arXiv:1805.06910 [hep-ph]} \BibitemShut
  {NoStop}%
\bibitem [{\citenamefont {Antusch}\ \emph
  {et~al.}(2018{\natexlab{a}})\citenamefont {Antusch}, \citenamefont {Cazzato},
  \citenamefont {Fischer}, \citenamefont {Hammad},\ and\ \citenamefont
  {Wang}}]{Antusch:2018bgr}%
  \BibitemOpen
  \bibfield  {author} {\bibinfo {author} {\bibfnamefont {S.}~\bibnamefont
  {Antusch}}, \bibinfo {author} {\bibfnamefont {E.}~\bibnamefont {Cazzato}},
  \bibinfo {author} {\bibfnamefont {O.}~\bibnamefont {Fischer}}, \bibinfo
  {author} {\bibfnamefont {A.}~\bibnamefont {Hammad}}, \ and\ \bibinfo {author}
  {\bibfnamefont {K.}~\bibnamefont {Wang}},\ }\href {\doibase
  10.1007/JHEP10(2018)067} {\bibfield  {journal} {\bibinfo  {journal} {JHEP}\
  }\textbf {\bibinfo {volume} {10}},\ \bibinfo {pages} {067} (\bibinfo {year}
  {2018}{\natexlab{a}})},\ \Eprint {http://arxiv.org/abs/1805.11400}
  {arXiv:1805.11400 [hep-ph]} \BibitemShut {NoStop}%
\bibitem [{\citenamefont {Crivellin}\ \emph {et~al.}(2019)\citenamefont
  {Crivellin}, \citenamefont {Ghezzi}, \citenamefont {Panizzi}, \citenamefont
  {Pruna},\ and\ \citenamefont {Signer}}]{Crivellin:2018ahj}%
  \BibitemOpen
  \bibfield  {author} {\bibinfo {author} {\bibfnamefont {A.}~\bibnamefont
  {Crivellin}}, \bibinfo {author} {\bibfnamefont {M.}~\bibnamefont {Ghezzi}},
  \bibinfo {author} {\bibfnamefont {L.}~\bibnamefont {Panizzi}}, \bibinfo
  {author} {\bibfnamefont {G.~M.}\ \bibnamefont {Pruna}}, \ and\ \bibinfo
  {author} {\bibfnamefont {A.}~\bibnamefont {Signer}},\ }\href {\doibase
  10.1103/PhysRevD.99.035004} {\bibfield  {journal} {\bibinfo  {journal} {Phys.
  Rev.}\ }\textbf {\bibinfo {volume} {D99}},\ \bibinfo {pages} {035004}
  (\bibinfo {year} {2019})},\ \Eprint {http://arxiv.org/abs/1807.10224}
  {arXiv:1807.10224 [hep-ph]} \BibitemShut {NoStop}%
\bibitem [{\citenamefont {Bhupal~Dev}\ and\ \citenamefont
  {Zhang}(2018)}]{Dev:2018kpa}%
  \BibitemOpen
  \bibfield  {author} {\bibinfo {author} {\bibfnamefont {P.~S.}\ \bibnamefont
  {Bhupal~Dev}}\ and\ \bibinfo {author} {\bibfnamefont {Y.}~\bibnamefont
  {Zhang}},\ }\href {\doibase 10.1007/JHEP10(2018)199} {\bibfield  {journal}
  {\bibinfo  {journal} {JHEP}\ }\textbf {\bibinfo {volume} {10}},\ \bibinfo
  {pages} {199} (\bibinfo {year} {2018})},\ \Eprint
  {http://arxiv.org/abs/1808.00943} {arXiv:1808.00943 [hep-ph]} \BibitemShut
  {NoStop}%
\bibitem [{\citenamefont {Du}\ \emph {et~al.}(2018)\citenamefont {Du},
  \citenamefont {Dunbrack}, \citenamefont {Ramsey-Musolf},\ and\ \citenamefont
  {Yu}}]{Du:2018eaw}%
  \BibitemOpen
  \bibfield  {author} {\bibinfo {author} {\bibfnamefont {Y.}~\bibnamefont
  {Du}}, \bibinfo {author} {\bibfnamefont {A.}~\bibnamefont {Dunbrack}},
  \bibinfo {author} {\bibfnamefont {M.~J.}\ \bibnamefont {Ramsey-Musolf}}, \
  and\ \bibinfo {author} {\bibfnamefont {J.-H.}\ \bibnamefont {Yu}},\
  }\href@noop {} {\  (\bibinfo {year} {2018})},\ \Eprint
  {http://arxiv.org/abs/1810.09450} {arXiv:1810.09450 [hep-ph]} \BibitemShut
  {NoStop}%
\bibitem [{\citenamefont {Babu}\ and\ \citenamefont
  {Jana}(2017)}]{Babu:2016rcr}%
  \BibitemOpen
  \bibfield  {author} {\bibinfo {author} {\bibfnamefont {K.~S.}\ \bibnamefont
  {Babu}}\ and\ \bibinfo {author} {\bibfnamefont {S.}~\bibnamefont {Jana}},\
  }\href {\doibase 10.1103/PhysRevD.95.055020} {\bibfield  {journal} {\bibinfo
  {journal} {Phys. Rev.}\ }\textbf {\bibinfo {volume} {D95}},\ \bibinfo {pages}
  {055020} (\bibinfo {year} {2017})},\ \Eprint
  {http://arxiv.org/abs/1612.09224} {arXiv:1612.09224 [hep-ph]} \BibitemShut
  {NoStop}%
\bibitem [{\citenamefont {Huitu}\ \emph {et~al.}(1997)\citenamefont {Huitu},
  \citenamefont {Maalampi}, \citenamefont {Pietila},\ and\ \citenamefont
  {Raidal}}]{Huitu:1996su}%
  \BibitemOpen
  \bibfield  {author} {\bibinfo {author} {\bibfnamefont {K.}~\bibnamefont
  {Huitu}}, \bibinfo {author} {\bibfnamefont {J.}~\bibnamefont {Maalampi}},
  \bibinfo {author} {\bibfnamefont {A.}~\bibnamefont {Pietila}}, \ and\
  \bibinfo {author} {\bibfnamefont {M.}~\bibnamefont {Raidal}},\ }\href
  {\doibase 10.1016/S0550-3213(97)87466-4} {\bibfield  {journal} {\bibinfo
  {journal} {Nucl. Phys.}\ }\textbf {\bibinfo {volume} {B487}},\ \bibinfo
  {pages} {27} (\bibinfo {year} {1997})},\ \Eprint
  {http://arxiv.org/abs/hep-ph/9606311} {arXiv:hep-ph/9606311 [hep-ph]}
  \BibitemShut {NoStop}%
\bibitem [{\citenamefont {Maalampi}\ and\ \citenamefont
  {Romanenko}(2002)}]{Maalampi:2002vx}%
  \BibitemOpen
  \bibfield  {author} {\bibinfo {author} {\bibfnamefont {J.}~\bibnamefont
  {Maalampi}}\ and\ \bibinfo {author} {\bibfnamefont {N.}~\bibnamefont
  {Romanenko}},\ }\href {\doibase 10.1016/S0370-2693(02)01549-6} {\bibfield
  {journal} {\bibinfo  {journal} {Phys. Lett.}\ }\textbf {\bibinfo {volume}
  {B532}},\ \bibinfo {pages} {202} (\bibinfo {year} {2002})},\ \Eprint
  {http://arxiv.org/abs/hep-ph/0201196} {arXiv:hep-ph/0201196 [hep-ph]}
  \BibitemShut {NoStop}%
\bibitem [{\citenamefont {Kanemura}\ \emph {et~al.}(2013)\citenamefont
  {Kanemura}, \citenamefont {Yagyu},\ and\ \citenamefont
  {Yokoya}}]{Kanemura:2013vxa}%
  \BibitemOpen
  \bibfield  {author} {\bibinfo {author} {\bibfnamefont {S.}~\bibnamefont
  {Kanemura}}, \bibinfo {author} {\bibfnamefont {K.}~\bibnamefont {Yagyu}}, \
  and\ \bibinfo {author} {\bibfnamefont {H.}~\bibnamefont {Yokoya}},\ }\href
  {\doibase 10.1016/j.physletb.2013.08.054} {\bibfield  {journal} {\bibinfo
  {journal} {Phys. Lett.}\ }\textbf {\bibinfo {volume} {B726}},\ \bibinfo
  {pages} {316} (\bibinfo {year} {2013})},\ \Eprint
  {http://arxiv.org/abs/1305.2383} {arXiv:1305.2383 [hep-ph]} \BibitemShut
  {NoStop}%
\bibitem [{\citenamefont {Englert}\ \emph {et~al.}(2013)\citenamefont
  {Englert}, \citenamefont {Re},\ and\ \citenamefont
  {Spannowsky}}]{Englert:2013wga}%
  \BibitemOpen
  \bibfield  {author} {\bibinfo {author} {\bibfnamefont {C.}~\bibnamefont
  {Englert}}, \bibinfo {author} {\bibfnamefont {E.}~\bibnamefont {Re}}, \ and\
  \bibinfo {author} {\bibfnamefont {M.}~\bibnamefont {Spannowsky}},\ }\href
  {\doibase 10.1103/PhysRevD.88.035024} {\bibfield  {journal} {\bibinfo
  {journal} {Phys. Rev.}\ }\textbf {\bibinfo {volume} {D88}},\ \bibinfo {pages}
  {035024} (\bibinfo {year} {2013})},\ \Eprint {http://arxiv.org/abs/1306.6228}
  {arXiv:1306.6228 [hep-ph]} \BibitemShut {NoStop}%
\bibitem [{\citenamefont {Dutta}\ \emph {et~al.}(2014)\citenamefont {Dutta},
  \citenamefont {Eusebi}, \citenamefont {Gao}, \citenamefont {Ghosh},\ and\
  \citenamefont {Kamon}}]{Dutta:2014dba}%
  \BibitemOpen
  \bibfield  {author} {\bibinfo {author} {\bibfnamefont {B.}~\bibnamefont
  {Dutta}}, \bibinfo {author} {\bibfnamefont {R.}~\bibnamefont {Eusebi}},
  \bibinfo {author} {\bibfnamefont {Y.}~\bibnamefont {Gao}}, \bibinfo {author}
  {\bibfnamefont {T.}~\bibnamefont {Ghosh}}, \ and\ \bibinfo {author}
  {\bibfnamefont {T.}~\bibnamefont {Kamon}},\ }\href {\doibase
  10.1103/PhysRevD.90.055015} {\bibfield  {journal} {\bibinfo  {journal} {Phys.
  Rev.}\ }\textbf {\bibinfo {volume} {D90}},\ \bibinfo {pages} {055015}
  (\bibinfo {year} {2014})},\ \Eprint {http://arxiv.org/abs/1404.0685}
  {arXiv:1404.0685 [hep-ph]} \BibitemShut {NoStop}%
\bibitem [{\citenamefont {Kanemura}\ \emph {et~al.}(2014)\citenamefont
  {Kanemura}, \citenamefont {Kikuchi}, \citenamefont {Yagyu},\ and\
  \citenamefont {Yokoya}}]{Kanemura:2014goa}%
  \BibitemOpen
  \bibfield  {author} {\bibinfo {author} {\bibfnamefont {S.}~\bibnamefont
  {Kanemura}}, \bibinfo {author} {\bibfnamefont {M.}~\bibnamefont {Kikuchi}},
  \bibinfo {author} {\bibfnamefont {K.}~\bibnamefont {Yagyu}}, \ and\ \bibinfo
  {author} {\bibfnamefont {H.}~\bibnamefont {Yokoya}},\ }\href {\doibase
  10.1103/PhysRevD.90.115018} {\bibfield  {journal} {\bibinfo  {journal} {Phys.
  Rev.}\ }\textbf {\bibinfo {volume} {D90}},\ \bibinfo {pages} {115018}
  (\bibinfo {year} {2014})},\ \Eprint {http://arxiv.org/abs/1407.6547}
  {arXiv:1407.6547 [hep-ph]} \BibitemShut {NoStop}%
\bibitem [{\citenamefont {Bambhaniya}\ \emph {et~al.}(2015)\citenamefont
  {Bambhaniya}, \citenamefont {Chakrabortty}, \citenamefont {Gluza},
  \citenamefont {Jelinski},\ and\ \citenamefont
  {Szafron}}]{Bambhaniya:2015wna}%
  \BibitemOpen
  \bibfield  {author} {\bibinfo {author} {\bibfnamefont {G.}~\bibnamefont
  {Bambhaniya}}, \bibinfo {author} {\bibfnamefont {J.}~\bibnamefont
  {Chakrabortty}}, \bibinfo {author} {\bibfnamefont {J.}~\bibnamefont {Gluza}},
  \bibinfo {author} {\bibfnamefont {T.}~\bibnamefont {Jelinski}}, \ and\
  \bibinfo {author} {\bibfnamefont {R.}~\bibnamefont {Szafron}},\ }\href
  {\doibase 10.1103/PhysRevD.92.015016} {\bibfield  {journal} {\bibinfo
  {journal} {Phys. Rev.}\ }\textbf {\bibinfo {volume} {D92}},\ \bibinfo {pages}
  {015016} (\bibinfo {year} {2015})},\ \Eprint
  {http://arxiv.org/abs/1504.03999} {arXiv:1504.03999 [hep-ph]} \BibitemShut
  {NoStop}%
\bibitem [{\citenamefont {Sirunyan}\ \emph {et~al.}(2018)\citenamefont
  {Sirunyan} \emph {et~al.}}]{Sirunyan:2017ret}%
  \BibitemOpen
  \bibfield  {author} {\bibinfo {author} {\bibfnamefont {A.~M.}\ \bibnamefont
  {Sirunyan}} \emph {et~al.} (\bibinfo {collaboration} {CMS}),\ }\href
  {\doibase 10.1103/PhysRevLett.120.081801} {\bibfield  {journal} {\bibinfo
  {journal} {Phys. Rev. Lett.}\ }\textbf {\bibinfo {volume} {120}},\ \bibinfo
  {pages} {081801} (\bibinfo {year} {2018})},\ \Eprint
  {http://arxiv.org/abs/1709.05822} {arXiv:1709.05822 [hep-ex]} \BibitemShut
  {NoStop}%
\bibitem [{\citenamefont {Mukhopadhyaya}\ and\ \citenamefont
  {Rai}(2006)}]{Mukhopadhyaya:2005vf}%
  \BibitemOpen
  \bibfield  {author} {\bibinfo {author} {\bibfnamefont {B.}~\bibnamefont
  {Mukhopadhyaya}}\ and\ \bibinfo {author} {\bibfnamefont {S.~K.}\ \bibnamefont
  {Rai}},\ }\href {\doibase 10.1016/j.physletb.2005.12.008} {\bibfield
  {journal} {\bibinfo  {journal} {Phys. Lett.}\ }\textbf {\bibinfo {volume}
  {B633}},\ \bibinfo {pages} {519} (\bibinfo {year} {2006})},\ \Eprint
  {http://arxiv.org/abs/hep-ph/0508290} {arXiv:hep-ph/0508290 [hep-ph]}
  \BibitemShut {NoStop}%
\bibitem [{\citenamefont {Chen}\ \emph {et~al.}(2009)\citenamefont {Chen},
  \citenamefont {Geng},\ and\ \citenamefont {Zhuridov}}]{Chen:2008jh}%
  \BibitemOpen
  \bibfield  {author} {\bibinfo {author} {\bibfnamefont {C.-S.}\ \bibnamefont
  {Chen}}, \bibinfo {author} {\bibfnamefont {C.-Q.}\ \bibnamefont {Geng}}, \
  and\ \bibinfo {author} {\bibfnamefont {D.~V.}\ \bibnamefont {Zhuridov}},\
  }\href {\doibase 10.1140/epjc/s10052-008-0856-3} {\bibfield  {journal}
  {\bibinfo  {journal} {Eur. Phys. J.}\ }\textbf {\bibinfo {volume} {C60}},\
  \bibinfo {pages} {119} (\bibinfo {year} {2009})},\ \Eprint
  {http://arxiv.org/abs/0803.1556} {arXiv:0803.1556 [hep-ph]} \BibitemShut
  {NoStop}%
\bibitem [{\citenamefont {Rodejohann}(2010)}]{Rodejohann:2010jh}%
  \BibitemOpen
  \bibfield  {author} {\bibinfo {author} {\bibfnamefont {W.}~\bibnamefont
  {Rodejohann}},\ }\href {\doibase 10.1103/PhysRevD.81.114001} {\bibfield
  {journal} {\bibinfo  {journal} {Phys. Rev.}\ }\textbf {\bibinfo {volume}
  {D81}},\ \bibinfo {pages} {114001} (\bibinfo {year} {2010})},\ \Eprint
  {http://arxiv.org/abs/1005.2854} {arXiv:1005.2854 [hep-ph]} \BibitemShut
  {NoStop}%
\bibitem [{\citenamefont {Yue}\ \emph {et~al.}(2011)\citenamefont {Yue},
  \citenamefont {Su}, \citenamefont {Zhang},\ and\ \citenamefont
  {Wang}}]{Yue:2010zu}%
  \BibitemOpen
  \bibfield  {author} {\bibinfo {author} {\bibfnamefont {C.-X.}\ \bibnamefont
  {Yue}}, \bibinfo {author} {\bibfnamefont {X.-S.}\ \bibnamefont {Su}},
  \bibinfo {author} {\bibfnamefont {J.}~\bibnamefont {Zhang}}, \ and\ \bibinfo
  {author} {\bibfnamefont {J.}~\bibnamefont {Wang}},\ }\href {\doibase
  10.1088/0253-6102/56/4/20} {\bibfield  {journal} {\bibinfo  {journal}
  {Commun. Theor. Phys.}\ }\textbf {\bibinfo {volume} {56}},\ \bibinfo {pages}
  {709} (\bibinfo {year} {2011})},\ \Eprint {http://arxiv.org/abs/1010.4633}
  {arXiv:1010.4633 [hep-ph]} \BibitemShut {NoStop}%
\bibitem [{\citenamefont {Rodejohann}\ and\ \citenamefont
  {Zhang}(2011)}]{Rodejohann:2010bv}%
  \BibitemOpen
  \bibfield  {author} {\bibinfo {author} {\bibfnamefont {W.}~\bibnamefont
  {Rodejohann}}\ and\ \bibinfo {author} {\bibfnamefont {H.}~\bibnamefont
  {Zhang}},\ }\href {\doibase 10.1103/PhysRevD.83.073005} {\bibfield  {journal}
  {\bibinfo  {journal} {Phys. Rev.}\ }\textbf {\bibinfo {volume} {D83}},\
  \bibinfo {pages} {073005} (\bibinfo {year} {2011})},\ \Eprint
  {http://arxiv.org/abs/1011.3606} {arXiv:1011.3606 [hep-ph]} \BibitemShut
  {NoStop}%
\bibitem [{\citenamefont {Nomura}\ \emph {et~al.}(2018)\citenamefont {Nomura},
  \citenamefont {Okada},\ and\ \citenamefont {Yokoya}}]{Nomura:2017abh}%
  \BibitemOpen
  \bibfield  {author} {\bibinfo {author} {\bibfnamefont {T.}~\bibnamefont
  {Nomura}}, \bibinfo {author} {\bibfnamefont {H.}~\bibnamefont {Okada}}, \
  and\ \bibinfo {author} {\bibfnamefont {H.}~\bibnamefont {Yokoya}},\ }\href
  {\doibase 10.1016/j.nuclphysb.2018.02.011} {\bibfield  {journal} {\bibinfo
  {journal} {Nucl. Phys.}\ }\textbf {\bibinfo {volume} {B929}},\ \bibinfo
  {pages} {193} (\bibinfo {year} {2018})},\ \Eprint
  {http://arxiv.org/abs/1702.03396} {arXiv:1702.03396 [hep-ph]} \BibitemShut
  {NoStop}%
\bibitem [{\citenamefont {Agrawal}\ \emph {et~al.}(2018)\citenamefont
  {Agrawal}, \citenamefont {Mitra}, \citenamefont {Niyogi}, \citenamefont
  {Shil},\ and\ \citenamefont {Spannowsky}}]{Agrawal:2018pci}%
  \BibitemOpen
  \bibfield  {author} {\bibinfo {author} {\bibfnamefont {P.}~\bibnamefont
  {Agrawal}}, \bibinfo {author} {\bibfnamefont {M.}~\bibnamefont {Mitra}},
  \bibinfo {author} {\bibfnamefont {S.}~\bibnamefont {Niyogi}}, \bibinfo
  {author} {\bibfnamefont {S.}~\bibnamefont {Shil}}, \ and\ \bibinfo {author}
  {\bibfnamefont {M.}~\bibnamefont {Spannowsky}},\ }\href {\doibase
  10.1103/PhysRevD.98.015024} {\bibfield  {journal} {\bibinfo  {journal} {Phys.
  Rev.}\ }\textbf {\bibinfo {volume} {D98}},\ \bibinfo {pages} {015024}
  (\bibinfo {year} {2018})},\ \Eprint {http://arxiv.org/abs/1803.00677}
  {arXiv:1803.00677 [hep-ph]} \BibitemShut {NoStop}%
\bibitem [{\citenamefont {Bhupal~Dev}\ \emph {et~al.}(2018)\citenamefont
  {Bhupal~Dev}, \citenamefont {Mohapatra},\ and\ \citenamefont
  {Zhang}}]{Dev:2018upe}%
  \BibitemOpen
  \bibfield  {author} {\bibinfo {author} {\bibfnamefont {P.~S.}\ \bibnamefont
  {Bhupal~Dev}}, \bibinfo {author} {\bibfnamefont {R.~N.}\ \bibnamefont
  {Mohapatra}}, \ and\ \bibinfo {author} {\bibfnamefont {Y.}~\bibnamefont
  {Zhang}},\ }\href {\doibase 10.1103/PhysRevD.98.075028} {\bibfield  {journal}
  {\bibinfo  {journal} {Phys. Rev.}\ }\textbf {\bibinfo {volume} {D98}},\
  \bibinfo {pages} {075028} (\bibinfo {year} {2018})},\ \Eprint
  {http://arxiv.org/abs/1803.11167} {arXiv:1803.11167 [hep-ph]} \BibitemShut
  {NoStop}%
\bibitem [{\citenamefont {Dev}\ \emph {et~al.}(2018)\citenamefont {Dev},
  \citenamefont {Ramsey-Musolf},\ and\ \citenamefont {Zhang}}]{Dev:2018sel}%
  \BibitemOpen
  \bibfield  {author} {\bibinfo {author} {\bibfnamefont {P.~S.~B.}\
  \bibnamefont {Dev}}, \bibinfo {author} {\bibfnamefont {M.~J.}\ \bibnamefont
  {Ramsey-Musolf}}, \ and\ \bibinfo {author} {\bibfnamefont {Y.}~\bibnamefont
  {Zhang}},\ }\href {\doibase 10.1103/PhysRevD.98.055013} {\bibfield  {journal}
  {\bibinfo  {journal} {Phys. Rev.}\ }\textbf {\bibinfo {volume} {D98}},\
  \bibinfo {pages} {055013} (\bibinfo {year} {2018})},\ \Eprint
  {http://arxiv.org/abs/1806.08499} {arXiv:1806.08499 [hep-ph]} \BibitemShut
  {NoStop}%
\bibitem [{\citenamefont {Akeroyd}\ \emph {et~al.}(2009)\citenamefont
  {Akeroyd}, \citenamefont {Aoki},\ and\ \citenamefont
  {Sugiyama}}]{Akeroyd:2009nu}%
  \BibitemOpen
  \bibfield  {author} {\bibinfo {author} {\bibfnamefont {A.~G.}\ \bibnamefont
  {Akeroyd}}, \bibinfo {author} {\bibfnamefont {M.}~\bibnamefont {Aoki}}, \
  and\ \bibinfo {author} {\bibfnamefont {H.}~\bibnamefont {Sugiyama}},\ }\href
  {\doibase 10.1103/PhysRevD.79.113010} {\bibfield  {journal} {\bibinfo
  {journal} {Phys. Rev.}\ }\textbf {\bibinfo {volume} {D79}},\ \bibinfo {pages}
  {113010} (\bibinfo {year} {2009})},\ \Eprint {http://arxiv.org/abs/0904.3640}
  {arXiv:0904.3640 [hep-ph]} \BibitemShut {NoStop}%
\bibitem [{\citenamefont {Dev}\ \emph {et~al.}(2017)\citenamefont {Dev},
  \citenamefont {Vila},\ and\ \citenamefont {Rodejohann}}]{Dev:2017ouk}%
  \BibitemOpen
  \bibfield  {author} {\bibinfo {author} {\bibfnamefont {P.~S.~B.}\
  \bibnamefont {Dev}}, \bibinfo {author} {\bibfnamefont {C.~M.}\ \bibnamefont
  {Vila}}, \ and\ \bibinfo {author} {\bibfnamefont {W.}~\bibnamefont
  {Rodejohann}},\ }\href {\doibase 10.1016/j.nuclphysb.2017.06.007} {\bibfield
  {journal} {\bibinfo  {journal} {Nucl. Phys.}\ }\textbf {\bibinfo {volume}
  {B921}},\ \bibinfo {pages} {436} (\bibinfo {year} {2017})},\ \Eprint
  {http://arxiv.org/abs/1703.00828} {arXiv:1703.00828 [hep-ph]} \BibitemShut
  {NoStop}%
\bibitem [{CMS(2017)}]{CMS-PAS-HIG-16-036}%
  \BibitemOpen
  \href@noop {} {\emph {\bibinfo {title} {{A search for doubly-charged Higgs
  boson production in three and four lepton final states at
  $\sqrt{s}=13~\mathrm{TeV}$}}}},\ \bibinfo {type} {Tech. Rep.}\ \bibinfo
  {number} {CMS-PAS-HIG-16-036}\ (\bibinfo  {institution} {CERN},\ \bibinfo
  {address} {Geneva},\ \bibinfo {year} {2017})\BibitemShut {NoStop}%
\bibitem [{\citenamefont {Aaboud}\ \emph {et~al.}(2018)\citenamefont {Aaboud}
  \emph {et~al.}}]{Aaboud:2017qph}%
  \BibitemOpen
  \bibfield  {author} {\bibinfo {author} {\bibfnamefont {M.}~\bibnamefont
  {Aaboud}} \emph {et~al.} (\bibinfo {collaboration} {ATLAS}),\ }\href
  {\doibase 10.1140/EPJC/S10052-018-5661-Z, 10.1140/epjc/s10052-018-5661-z}
  {\bibfield  {journal} {\bibinfo  {journal} {Eur. Phys. J.}\ }\textbf
  {\bibinfo {volume} {C78}},\ \bibinfo {pages} {199} (\bibinfo {year}
  {2018})},\ \Eprint {http://arxiv.org/abs/1710.09748} {arXiv:1710.09748
  [hep-ex]} \BibitemShut {NoStop}%
\bibitem [{\citenamefont {Barry}\ and\ \citenamefont
  {Rodejohann}(2013)}]{Barry:2013xxa}%
  \BibitemOpen
  \bibfield  {author} {\bibinfo {author} {\bibfnamefont {J.}~\bibnamefont
  {Barry}}\ and\ \bibinfo {author} {\bibfnamefont {W.}~\bibnamefont
  {Rodejohann}},\ }\href {\doibase 10.1007/JHEP09(2013)153} {\bibfield
  {journal} {\bibinfo  {journal} {JHEP}\ }\textbf {\bibinfo {volume} {09}},\
  \bibinfo {pages} {153} (\bibinfo {year} {2013})},\ \Eprint
  {http://arxiv.org/abs/1303.6324} {arXiv:1303.6324 [hep-ph]} \BibitemShut
  {NoStop}%
\bibitem [{\citenamefont {Bambhaniya}\ \emph {et~al.}(2016)\citenamefont
  {Bambhaniya}, \citenamefont {Dev}, \citenamefont {Goswami},\ and\
  \citenamefont {Mitra}}]{Bambhaniya:2015ipg}%
  \BibitemOpen
  \bibfield  {author} {\bibinfo {author} {\bibfnamefont {G.}~\bibnamefont
  {Bambhaniya}}, \bibinfo {author} {\bibfnamefont {P.~S.~B.}\ \bibnamefont
  {Dev}}, \bibinfo {author} {\bibfnamefont {S.}~\bibnamefont {Goswami}}, \ and\
  \bibinfo {author} {\bibfnamefont {M.}~\bibnamefont {Mitra}},\ }\href
  {\doibase 10.1007/JHEP04(2016)046} {\bibfield  {journal} {\bibinfo  {journal}
  {JHEP}\ }\textbf {\bibinfo {volume} {04}},\ \bibinfo {pages} {046} (\bibinfo
  {year} {2016})},\ \Eprint {http://arxiv.org/abs/1512.00440} {arXiv:1512.00440
  [hep-ph]} \BibitemShut {NoStop}%
\bibitem [{\citenamefont {Arhrib}\ \emph {et~al.}(2011)\citenamefont {Arhrib},
  \citenamefont {Benbrik}, \citenamefont {Chabab}, \citenamefont {Moultaka},
  \citenamefont {Peyranere}, \citenamefont {Rahili},\ and\ \citenamefont
  {Ramadan}}]{Arhrib:2011uy}%
  \BibitemOpen
  \bibfield  {author} {\bibinfo {author} {\bibfnamefont {A.}~\bibnamefont
  {Arhrib}}, \bibinfo {author} {\bibfnamefont {R.}~\bibnamefont {Benbrik}},
  \bibinfo {author} {\bibfnamefont {M.}~\bibnamefont {Chabab}}, \bibinfo
  {author} {\bibfnamefont {G.}~\bibnamefont {Moultaka}}, \bibinfo {author}
  {\bibfnamefont {M.~C.}\ \bibnamefont {Peyranere}}, \bibinfo {author}
  {\bibfnamefont {L.}~\bibnamefont {Rahili}}, \ and\ \bibinfo {author}
  {\bibfnamefont {J.}~\bibnamefont {Ramadan}},\ }\href {\doibase
  10.1103/PhysRevD.84.095005} {\bibfield  {journal} {\bibinfo  {journal} {Phys.
  Rev.}\ }\textbf {\bibinfo {volume} {D84}},\ \bibinfo {pages} {095005}
  (\bibinfo {year} {2011})},\ \Eprint {http://arxiv.org/abs/1105.1925}
  {arXiv:1105.1925 [hep-ph]} \BibitemShut {NoStop}%
\bibitem [{\citenamefont {Chun}\ \emph {et~al.}(2012)\citenamefont {Chun},
  \citenamefont {Lee},\ and\ \citenamefont {Sharma}}]{Chun:2012jw}%
  \BibitemOpen
  \bibfield  {author} {\bibinfo {author} {\bibfnamefont {E.~J.}\ \bibnamefont
  {Chun}}, \bibinfo {author} {\bibfnamefont {H.~M.}\ \bibnamefont {Lee}}, \
  and\ \bibinfo {author} {\bibfnamefont {P.}~\bibnamefont {Sharma}},\ }\href
  {\doibase 10.1007/JHEP11(2012)106} {\bibfield  {journal} {\bibinfo  {journal}
  {JHEP}\ }\textbf {\bibinfo {volume} {11}},\ \bibinfo {pages} {106} (\bibinfo
  {year} {2012})},\ \Eprint {http://arxiv.org/abs/1209.1303} {arXiv:1209.1303
  [hep-ph]} \BibitemShut {NoStop}%
\bibitem [{\citenamefont {Bhupal~Dev}\ \emph {et~al.}(2013)\citenamefont
  {Bhupal~Dev}, \citenamefont {Ghosh}, \citenamefont {Okada},\ and\
  \citenamefont {Saha}}]{Dev:2013ff}%
  \BibitemOpen
  \bibfield  {author} {\bibinfo {author} {\bibfnamefont {P.~S.}\ \bibnamefont
  {Bhupal~Dev}}, \bibinfo {author} {\bibfnamefont {D.~K.}\ \bibnamefont
  {Ghosh}}, \bibinfo {author} {\bibfnamefont {N.}~\bibnamefont {Okada}}, \ and\
  \bibinfo {author} {\bibfnamefont {I.}~\bibnamefont {Saha}},\ }\href {\doibase
  10.1007/JHEP03(2013)150, 10.1007/JHEP05(2013)049} {\bibfield  {journal}
  {\bibinfo  {journal} {JHEP}\ }\textbf {\bibinfo {volume} {03}},\ \bibinfo
  {pages} {150} (\bibinfo {year} {2013})},\ \bibinfo {note} {[Erratum:
  JHEP05,049(2013)]},\ \Eprint {http://arxiv.org/abs/1301.3453}
  {arXiv:1301.3453 [hep-ph]} \BibitemShut {NoStop}%
\bibitem [{\citenamefont {Kobakhidze}\ and\ \citenamefont
  {Spencer-Smith}(2013)}]{Kobakhidze:2013pya}%
  \BibitemOpen
  \bibfield  {author} {\bibinfo {author} {\bibfnamefont {A.}~\bibnamefont
  {Kobakhidze}}\ and\ \bibinfo {author} {\bibfnamefont {A.}~\bibnamefont
  {Spencer-Smith}},\ }\href {\doibase 10.1007/JHEP08(2013)036} {\bibfield
  {journal} {\bibinfo  {journal} {JHEP}\ }\textbf {\bibinfo {volume} {08}},\
  \bibinfo {pages} {036} (\bibinfo {year} {2013})},\ \Eprint
  {http://arxiv.org/abs/1305.7283} {arXiv:1305.7283 [hep-ph]} \BibitemShut
  {NoStop}%
\bibitem [{\citenamefont {Chabab}\ \emph {et~al.}(2016)\citenamefont {Chabab},
  \citenamefont {Peyranère},\ and\ \citenamefont {Rahili}}]{Chabab:2015nel}%
  \BibitemOpen
  \bibfield  {author} {\bibinfo {author} {\bibfnamefont {M.}~\bibnamefont
  {Chabab}}, \bibinfo {author} {\bibfnamefont {M.~C.}\ \bibnamefont
  {Peyranère}}, \ and\ \bibinfo {author} {\bibfnamefont {L.}~\bibnamefont
  {Rahili}},\ }\href {\doibase 10.1103/PhysRevD.93.115021} {\bibfield
  {journal} {\bibinfo  {journal} {Phys. Rev.}\ }\textbf {\bibinfo {volume}
  {D93}},\ \bibinfo {pages} {115021} (\bibinfo {year} {2016})},\ \Eprint
  {http://arxiv.org/abs/1512.07280} {arXiv:1512.07280 [hep-ph]} \BibitemShut
  {NoStop}%
\bibitem [{\citenamefont {Haba}\ \emph {et~al.}(2016)\citenamefont {Haba},
  \citenamefont {Ishida}, \citenamefont {Okada},\ and\ \citenamefont
  {Yamaguchi}}]{Haba:2016zbu}%
  \BibitemOpen
  \bibfield  {author} {\bibinfo {author} {\bibfnamefont {N.}~\bibnamefont
  {Haba}}, \bibinfo {author} {\bibfnamefont {H.}~\bibnamefont {Ishida}},
  \bibinfo {author} {\bibfnamefont {N.}~\bibnamefont {Okada}}, \ and\ \bibinfo
  {author} {\bibfnamefont {Y.}~\bibnamefont {Yamaguchi}},\ }\href {\doibase
  10.1140/epjc/s10052-016-4180-z} {\bibfield  {journal} {\bibinfo  {journal}
  {Eur. Phys. J.}\ }\textbf {\bibinfo {volume} {C76}},\ \bibinfo {pages} {333}
  (\bibinfo {year} {2016})},\ \Eprint {http://arxiv.org/abs/1601.05217}
  {arXiv:1601.05217 [hep-ph]} \BibitemShut {NoStop}%
\bibitem [{\citenamefont {Das}\ and\ \citenamefont
  {Santamaria}(2016)}]{Das:2016bir}%
  \BibitemOpen
  \bibfield  {author} {\bibinfo {author} {\bibfnamefont {D.}~\bibnamefont
  {Das}}\ and\ \bibinfo {author} {\bibfnamefont {A.}~\bibnamefont
  {Santamaria}},\ }\href {\doibase 10.1103/PhysRevD.94.015015} {\bibfield
  {journal} {\bibinfo  {journal} {Phys. Rev.}\ }\textbf {\bibinfo {volume}
  {D94}},\ \bibinfo {pages} {015015} (\bibinfo {year} {2016})},\ \Eprint
  {http://arxiv.org/abs/1604.08099} {arXiv:1604.08099 [hep-ph]} \BibitemShut
  {NoStop}%
\bibitem [{\citenamefont {Xu}(2016)}]{Xu:2016klg}%
  \BibitemOpen
  \bibfield  {author} {\bibinfo {author} {\bibfnamefont {X.-J.}\ \bibnamefont
  {Xu}},\ }\href {\doibase 10.1103/PhysRevD.94.115025} {\bibfield  {journal}
  {\bibinfo  {journal} {Phys. Rev.}\ }\textbf {\bibinfo {volume} {D94}},\
  \bibinfo {pages} {115025} (\bibinfo {year} {2016})},\ \Eprint
  {http://arxiv.org/abs/1612.04950} {arXiv:1612.04950 [hep-ph]} \BibitemShut
  {NoStop}%
\bibitem [{\citenamefont {Antusch}\ \emph
  {et~al.}(2018{\natexlab{b}})\citenamefont {Antusch}, \citenamefont {Fischer},
  \citenamefont {Hammad},\ and\ \citenamefont {Scherb}}]{Antusch:2018svb}%
  \BibitemOpen
  \bibfield  {author} {\bibinfo {author} {\bibfnamefont {S.}~\bibnamefont
  {Antusch}}, \bibinfo {author} {\bibfnamefont {O.}~\bibnamefont {Fischer}},
  \bibinfo {author} {\bibfnamefont {A.}~\bibnamefont {Hammad}}, \ and\ \bibinfo
  {author} {\bibfnamefont {C.}~\bibnamefont {Scherb}},\ }\href@noop {} {\
  (\bibinfo {year} {2018}{\natexlab{b}})},\ \Eprint
  {http://arxiv.org/abs/1811.03476} {arXiv:1811.03476 [hep-ph]} \BibitemShut
  {NoStop}%
\bibitem [{\citenamefont {Baldini}\ \emph {et~al.}(2016)\citenamefont {Baldini}
  \emph {et~al.}}]{TheMEG:2016wtm}%
  \BibitemOpen
  \bibfield  {author} {\bibinfo {author} {\bibfnamefont {A.~M.}\ \bibnamefont
  {Baldini}} \emph {et~al.} (\bibinfo {collaboration} {MEG}),\ }\href {\doibase
  10.1140/epjc/s10052-016-4271-x} {\bibfield  {journal} {\bibinfo  {journal}
  {Eur. Phys. J.}\ }\textbf {\bibinfo {volume} {C76}},\ \bibinfo {pages} {434}
  (\bibinfo {year} {2016})},\ \Eprint {http://arxiv.org/abs/1605.05081}
  {arXiv:1605.05081 [hep-ex]} \BibitemShut {NoStop}%
\bibitem [{\citenamefont {Bellgardt}\ \emph {et~al.}(1988)\citenamefont
  {Bellgardt} \emph {et~al.}}]{Bellgardt:1987du}%
  \BibitemOpen
  \bibfield  {author} {\bibinfo {author} {\bibfnamefont {U.}~\bibnamefont
  {Bellgardt}} \emph {et~al.} (\bibinfo {collaboration} {SINDRUM}),\ }\href
  {\doibase 10.1016/0550-3213(88)90462-2} {\bibfield  {journal} {\bibinfo
  {journal} {Nucl. Phys.}\ }\textbf {\bibinfo {volume} {B299}},\ \bibinfo
  {pages} {1} (\bibinfo {year} {1988})}\BibitemShut {NoStop}%
\bibitem [{\citenamefont {Abdallah}\ \emph {et~al.}(2003)\citenamefont
  {Abdallah} \emph {et~al.}}]{Abdallah:2002qj}%
  \BibitemOpen
  \bibfield  {author} {\bibinfo {author} {\bibfnamefont {J.}~\bibnamefont
  {Abdallah}} \emph {et~al.} (\bibinfo {collaboration} {DELPHI}),\ }\href
  {\doibase 10.1016/S0370-2693(02)03125-8} {\bibfield  {journal} {\bibinfo
  {journal} {Phys. Lett.}\ }\textbf {\bibinfo {volume} {B552}},\ \bibinfo
  {pages} {127} (\bibinfo {year} {2003})},\ \Eprint
  {http://arxiv.org/abs/hep-ex/0303026} {arXiv:hep-ex/0303026 [hep-ex]}
  \BibitemShut {NoStop}%
\bibitem [{\citenamefont {Abbiendi}\ \emph {et~al.}(2003)\citenamefont
  {Abbiendi} \emph {et~al.}}]{Abbiendi:2003pr}%
  \BibitemOpen
  \bibfield  {author} {\bibinfo {author} {\bibfnamefont {G.}~\bibnamefont
  {Abbiendi}} \emph {et~al.} (\bibinfo {collaboration} {OPAL}),\ }\href
  {\doibase 10.1016/j.physletb.2003.10.034} {\bibfield  {journal} {\bibinfo
  {journal} {Phys. Lett.}\ }\textbf {\bibinfo {volume} {B577}},\ \bibinfo
  {pages} {93} (\bibinfo {year} {2003})},\ \Eprint
  {http://arxiv.org/abs/hep-ex/0308052} {arXiv:hep-ex/0308052 [hep-ex]}
  \BibitemShut {NoStop}%
\bibitem [{\citenamefont {Achard}\ \emph {et~al.}(2003)\citenamefont {Achard}
  \emph {et~al.}}]{Achard:2003mv}%
  \BibitemOpen
  \bibfield  {author} {\bibinfo {author} {\bibfnamefont {P.}~\bibnamefont
  {Achard}} \emph {et~al.} (\bibinfo {collaboration} {L3}),\ }\href {\doibase
  10.1016/j.physletb.2003.09.082} {\bibfield  {journal} {\bibinfo  {journal}
  {Phys. Lett.}\ }\textbf {\bibinfo {volume} {B576}},\ \bibinfo {pages} {18}
  (\bibinfo {year} {2003})},\ \Eprint {http://arxiv.org/abs/hep-ex/0309076}
  {arXiv:hep-ex/0309076 [hep-ex]} \BibitemShut {NoStop}%
\bibitem [{\citenamefont {Abdallah}\ \emph {et~al.}(2006)\citenamefont
  {Abdallah} \emph {et~al.}}]{Abdallah:2005ph}%
  \BibitemOpen
  \bibfield  {author} {\bibinfo {author} {\bibfnamefont {J.}~\bibnamefont
  {Abdallah}} \emph {et~al.} (\bibinfo {collaboration} {DELPHI}),\ }\href
  {\doibase 10.1140/epjc/s2005-02461-0} {\bibfield  {journal} {\bibinfo
  {journal} {Eur. Phys. J.}\ }\textbf {\bibinfo {volume} {C45}},\ \bibinfo
  {pages} {589} (\bibinfo {year} {2006})},\ \Eprint
  {http://arxiv.org/abs/hep-ex/0512012} {arXiv:hep-ex/0512012 [hep-ex]}
  \BibitemShut {NoStop}%
\bibitem [{\citenamefont {Willmann}\ \emph {et~al.}(1999)\citenamefont
  {Willmann} \emph {et~al.}}]{Willmann:1998gd}%
  \BibitemOpen
  \bibfield  {author} {\bibinfo {author} {\bibfnamefont {L.}~\bibnamefont
  {Willmann}} \emph {et~al.},\ }\href {\doibase 10.1103/PhysRevLett.82.49}
  {\bibfield  {journal} {\bibinfo  {journal} {Phys. Rev. Lett.}\ }\textbf
  {\bibinfo {volume} {82}},\ \bibinfo {pages} {49} (\bibinfo {year} {1999})},\
  \Eprint {http://arxiv.org/abs/hep-ex/9807011} {arXiv:hep-ex/9807011 [hep-ex]}
  \BibitemShut {NoStop}%
\bibitem [{\citenamefont {Amhis}\ \emph {et~al.}(2017)\citenamefont {Amhis}
  \emph {et~al.}}]{Amhis:2016xyh}%
  \BibitemOpen
  \bibfield  {author} {\bibinfo {author} {\bibfnamefont {Y.}~\bibnamefont
  {Amhis}} \emph {et~al.} (\bibinfo {collaboration} {HFLAV}),\ }\href {\doibase
  10.1140/epjc/s10052-017-5058-4} {\bibfield  {journal} {\bibinfo  {journal}
  {Eur. Phys. J.}\ }\textbf {\bibinfo {volume} {C77}},\ \bibinfo {pages} {895}
  (\bibinfo {year} {2017})},\ \Eprint {http://arxiv.org/abs/1612.07233}
  {arXiv:1612.07233 [hep-ex]} \BibitemShut {NoStop}%
\bibitem [{\citenamefont {Alloul}\ \emph {et~al.}(2014)\citenamefont {Alloul},
  \citenamefont {Christensen}, \citenamefont {Degrande}, \citenamefont {Duhr},\
  and\ \citenamefont {Fuks}}]{Alloul:2013bka}%
  \BibitemOpen
  \bibfield  {author} {\bibinfo {author} {\bibfnamefont {A.}~\bibnamefont
  {Alloul}}, \bibinfo {author} {\bibfnamefont {N.~D.}\ \bibnamefont
  {Christensen}}, \bibinfo {author} {\bibfnamefont {C.}~\bibnamefont
  {Degrande}}, \bibinfo {author} {\bibfnamefont {C.}~\bibnamefont {Duhr}}, \
  and\ \bibinfo {author} {\bibfnamefont {B.}~\bibnamefont {Fuks}},\ }\href
  {\doibase 10.1016/j.cpc.2014.04.012} {\bibfield  {journal} {\bibinfo
  {journal} {Comput. Phys. Commun.}\ }\textbf {\bibinfo {volume} {185}},\
  \bibinfo {pages} {2250} (\bibinfo {year} {2014})},\ \Eprint
  {http://arxiv.org/abs/1310.1921} {arXiv:1310.1921 [hep-ph]} \BibitemShut
  {NoStop}%
\bibitem [{\citenamefont {Alwall}\ \emph {et~al.}(2014)\citenamefont {Alwall},
  \citenamefont {Frederix}, \citenamefont {Frixione}, \citenamefont {Hirschi},
  \citenamefont {Maltoni}, \citenamefont {Mattelaer}, \citenamefont {Shao},
  \citenamefont {Stelzer}, \citenamefont {Torrielli},\ and\ \citenamefont
  {Zaro}}]{Alwall:2014hca}%
  \BibitemOpen
  \bibfield  {author} {\bibinfo {author} {\bibfnamefont {J.}~\bibnamefont
  {Alwall}}, \bibinfo {author} {\bibfnamefont {R.}~\bibnamefont {Frederix}},
  \bibinfo {author} {\bibfnamefont {S.}~\bibnamefont {Frixione}}, \bibinfo
  {author} {\bibfnamefont {V.}~\bibnamefont {Hirschi}}, \bibinfo {author}
  {\bibfnamefont {F.}~\bibnamefont {Maltoni}}, \bibinfo {author} {\bibfnamefont
  {O.}~\bibnamefont {Mattelaer}}, \bibinfo {author} {\bibfnamefont {H.~S.}\
  \bibnamefont {Shao}}, \bibinfo {author} {\bibfnamefont {T.}~\bibnamefont
  {Stelzer}}, \bibinfo {author} {\bibfnamefont {P.}~\bibnamefont {Torrielli}},
  \ and\ \bibinfo {author} {\bibfnamefont {M.}~\bibnamefont {Zaro}},\ }\href
  {\doibase 10.1007/JHEP07(2014)079} {\bibfield  {journal} {\bibinfo  {journal}
  {JHEP}\ }\textbf {\bibinfo {volume} {07}},\ \bibinfo {pages} {079} (\bibinfo
  {year} {2014})},\ \Eprint {http://arxiv.org/abs/1405.0301} {arXiv:1405.0301
  [hep-ph]} \BibitemShut {NoStop}%
\bibitem [{\citenamefont {Pumplin}\ \emph {et~al.}(2002)\citenamefont
  {Pumplin}, \citenamefont {Stump}, \citenamefont {Huston}, \citenamefont
  {Lai}, \citenamefont {Nadolsky},\ and\ \citenamefont
  {Tung}}]{Pumplin:2002vw}%
  \BibitemOpen
  \bibfield  {author} {\bibinfo {author} {\bibfnamefont {J.}~\bibnamefont
  {Pumplin}}, \bibinfo {author} {\bibfnamefont {D.~R.}\ \bibnamefont {Stump}},
  \bibinfo {author} {\bibfnamefont {J.}~\bibnamefont {Huston}}, \bibinfo
  {author} {\bibfnamefont {H.~L.}\ \bibnamefont {Lai}}, \bibinfo {author}
  {\bibfnamefont {P.~M.}\ \bibnamefont {Nadolsky}}, \ and\ \bibinfo {author}
  {\bibfnamefont {W.~K.}\ \bibnamefont {Tung}},\ }\href {\doibase
  10.1088/1126-6708/2002/07/012} {\bibfield  {journal} {\bibinfo  {journal}
  {JHEP}\ }\textbf {\bibinfo {volume} {07}},\ \bibinfo {pages} {012} (\bibinfo
  {year} {2002})},\ \Eprint {http://arxiv.org/abs/hep-ph/0201195}
  {arXiv:hep-ph/0201195 [hep-ph]} \BibitemShut {NoStop}%
\bibitem [{\citenamefont {Sjostrand}\ \emph {et~al.}(2006)\citenamefont
  {Sjostrand}, \citenamefont {Mrenna},\ and\ \citenamefont
  {Skands}}]{Sjostrand:2006za}%
  \BibitemOpen
  \bibfield  {author} {\bibinfo {author} {\bibfnamefont {T.}~\bibnamefont
  {Sjostrand}}, \bibinfo {author} {\bibfnamefont {S.}~\bibnamefont {Mrenna}}, \
  and\ \bibinfo {author} {\bibfnamefont {P.~Z.}\ \bibnamefont {Skands}},\
  }\href {\doibase 10.1088/1126-6708/2006/05/026} {\bibfield  {journal}
  {\bibinfo  {journal} {JHEP}\ }\textbf {\bibinfo {volume} {05}},\ \bibinfo
  {pages} {026} (\bibinfo {year} {2006})},\ \Eprint
  {http://arxiv.org/abs/hep-ph/0603175} {arXiv:hep-ph/0603175 [hep-ph]}
  \BibitemShut {NoStop}%
\bibitem [{\citenamefont {de~Favereau}\ \emph {et~al.}(2014)\citenamefont
  {de~Favereau}, \citenamefont {Delaere}, \citenamefont {Demin}, \citenamefont
  {Giammanco}, \citenamefont {Lema\'{i}tre}, \citenamefont {Mertens},\ and\
  \citenamefont {Selvaggi}}]{deFavereau:2013fsa}%
  \BibitemOpen
  \bibfield  {author} {\bibinfo {author} {\bibfnamefont {J.}~\bibnamefont
  {de~Favereau}}, \bibinfo {author} {\bibfnamefont {C.}~\bibnamefont
  {Delaere}}, \bibinfo {author} {\bibfnamefont {P.}~\bibnamefont {Demin}},
  \bibinfo {author} {\bibfnamefont {A.}~\bibnamefont {Giammanco}}, \bibinfo
  {author} {\bibfnamefont {V.}~\bibnamefont {Lema\'{i}tre}}, \bibinfo {author}
  {\bibfnamefont {A.}~\bibnamefont {Mertens}}, \ and\ \bibinfo {author}
  {\bibfnamefont {M.}~\bibnamefont {Selvaggi}} (\bibinfo {collaboration}
  {DELPHES 3}),\ }\href {\doibase 10.1007/JHEP02(2014)057} {\bibfield
  {journal} {\bibinfo  {journal} {JHEP}\ }\textbf {\bibinfo {volume} {02}},\
  \bibinfo {pages} {057} (\bibinfo {year} {2014})},\ \Eprint
  {http://arxiv.org/abs/1307.6346} {arXiv:1307.6346 [hep-ex]} \BibitemShut
  {NoStop}%
\bibitem [{\citenamefont {Cacciari}\ \emph {et~al.}(2008)\citenamefont
  {Cacciari}, \citenamefont {Salam},\ and\ \citenamefont
  {Soyez}}]{Cacciari:2008gp}%
  \BibitemOpen
  \bibfield  {author} {\bibinfo {author} {\bibfnamefont {M.}~\bibnamefont
  {Cacciari}}, \bibinfo {author} {\bibfnamefont {G.~P.}\ \bibnamefont {Salam}},
  \ and\ \bibinfo {author} {\bibfnamefont {G.}~\bibnamefont {Soyez}},\ }\href
  {\doibase 10.1088/1126-6708/2008/04/063} {\bibfield  {journal} {\bibinfo
  {journal} {JHEP}\ }\textbf {\bibinfo {volume} {04}},\ \bibinfo {pages} {063}
  (\bibinfo {year} {2008})},\ \Eprint {http://arxiv.org/abs/0802.1189}
  {arXiv:0802.1189 [hep-ph]} \BibitemShut {NoStop}%
\bibitem [{\citenamefont {Cacciari}\ \emph {et~al.}(2012)\citenamefont
  {Cacciari}, \citenamefont {Salam},\ and\ \citenamefont
  {Soyez}}]{Cacciari:2011ma}%
  \BibitemOpen
  \bibfield  {author} {\bibinfo {author} {\bibfnamefont {M.}~\bibnamefont
  {Cacciari}}, \bibinfo {author} {\bibfnamefont {G.~P.}\ \bibnamefont {Salam}},
  \ and\ \bibinfo {author} {\bibfnamefont {G.}~\bibnamefont {Soyez}},\ }\href
  {\doibase 10.1140/epjc/s10052-012-1896-2} {\bibfield  {journal} {\bibinfo
  {journal} {Eur. Phys. J.}\ }\textbf {\bibinfo {volume} {C72}},\ \bibinfo
  {pages} {1896} (\bibinfo {year} {2012})},\ \Eprint
  {http://arxiv.org/abs/1111.6097} {arXiv:1111.6097 [hep-ph]} \BibitemShut
  {NoStop}%
\bibitem [{\citenamefont {Aad}\ \emph {et~al.}(2011)\citenamefont {Aad} \emph
  {et~al.}}]{Aad:2010fh}%
  \BibitemOpen
  \bibfield  {author} {\bibinfo {author} {\bibfnamefont {G.}~\bibnamefont
  {Aad}} \emph {et~al.} (\bibinfo {collaboration} {ATLAS}),\ }\href {\doibase
  10.1103/PhysRevD.83.112001} {\bibfield  {journal} {\bibinfo  {journal} {Phys.
  Rev.}\ }\textbf {\bibinfo {volume} {D83}},\ \bibinfo {pages} {112001}
  (\bibinfo {year} {2011})},\ \Eprint {http://arxiv.org/abs/1012.0791}
  {arXiv:1012.0791 [hep-ex]} \BibitemShut {NoStop}%
\bibitem [{ATL(2016)}]{ATLAS:2016iqc}%
  \BibitemOpen
  \href@noop {} {\emph {\bibinfo {title} {{Electron efficiency measurements
  with the ATLAS detector using the 2015 LHC proton-proton collision data}}}},\
  \bibinfo {type} {Tech. Rep.}\ \bibinfo {number} {ATLAS-CONF-2016-024}\
  (\bibinfo  {institution} {CERN},\ \bibinfo {address} {Geneva},\ \bibinfo
  {year} {2016})\BibitemShut {NoStop}%
\bibitem [{\citenamefont {Khachatryan}\ \emph {et~al.}(2016)\citenamefont
  {Khachatryan} \emph {et~al.}}]{Khachatryan:2016olu}%
  \BibitemOpen
  \bibfield  {author} {\bibinfo {author} {\bibfnamefont {V.}~\bibnamefont
  {Khachatryan}} \emph {et~al.} (\bibinfo {collaboration} {CMS}),\ }\href
  {\doibase 10.1007/JHEP04(2016)169} {\bibfield  {journal} {\bibinfo  {journal}
  {JHEP}\ }\textbf {\bibinfo {volume} {04}},\ \bibinfo {pages} {169} (\bibinfo
  {year} {2016})},\ \Eprint {http://arxiv.org/abs/1603.02248} {arXiv:1603.02248
  [hep-ex]} \BibitemShut {NoStop}%
\bibitem [{\citenamefont {Cowan}\ \emph {et~al.}(2011)\citenamefont {Cowan},
  \citenamefont {Cranmer}, \citenamefont {Gross},\ and\ \citenamefont
  {Vitells}}]{Cowan:2010js}%
  \BibitemOpen
  \bibfield  {author} {\bibinfo {author} {\bibfnamefont {G.}~\bibnamefont
  {Cowan}}, \bibinfo {author} {\bibfnamefont {K.}~\bibnamefont {Cranmer}},
  \bibinfo {author} {\bibfnamefont {E.}~\bibnamefont {Gross}}, \ and\ \bibinfo
  {author} {\bibfnamefont {O.}~\bibnamefont {Vitells}},\ }\href {\doibase
  10.1140/epjc/s10052-011-1554-0, 10.1140/epjc/s10052-013-2501-z} {\bibfield
  {journal} {\bibinfo  {journal} {Eur. Phys. J.}\ }\textbf {\bibinfo {volume}
  {C71}},\ \bibinfo {pages} {1554} (\bibinfo {year} {2011})},\ \bibinfo {note}
  {[Erratum: Eur. Phys. J.C73,2501(2013)]},\ \Eprint
  {http://arxiv.org/abs/1007.1727} {arXiv:1007.1727 [physics.data-an]}
  \BibitemShut {NoStop}%
\bibitem [{\citenamefont {Acosta}\ \emph {et~al.}(2004)\citenamefont {Acosta}
  \emph {et~al.}}]{Acosta:2004uj}%
  \BibitemOpen
  \bibfield  {author} {\bibinfo {author} {\bibfnamefont {D.}~\bibnamefont
  {Acosta}} \emph {et~al.} (\bibinfo {collaboration} {CDF}),\ }\href {\doibase
  10.1103/PhysRevLett.93.221802} {\bibfield  {journal} {\bibinfo  {journal}
  {Phys. Rev. Lett.}\ }\textbf {\bibinfo {volume} {93}},\ \bibinfo {pages}
  {221802} (\bibinfo {year} {2004})},\ \Eprint
  {http://arxiv.org/abs/hep-ex/0406073} {arXiv:hep-ex/0406073 [hep-ex]}
  \BibitemShut {NoStop}%
\bibitem [{\citenamefont {Abazov}\ \emph {et~al.}(2008)\citenamefont {Abazov}
  \emph {et~al.}}]{Abazov:2008ab}%
  \BibitemOpen
  \bibfield  {author} {\bibinfo {author} {\bibfnamefont {V.~M.}\ \bibnamefont
  {Abazov}} \emph {et~al.} (\bibinfo {collaboration} {D0}),\ }\href {\doibase
  10.1103/PhysRevLett.101.071803} {\bibfield  {journal} {\bibinfo  {journal}
  {Phys. Rev. Lett.}\ }\textbf {\bibinfo {volume} {101}},\ \bibinfo {pages}
  {071803} (\bibinfo {year} {2008})},\ \Eprint {http://arxiv.org/abs/0803.1534}
  {arXiv:0803.1534 [hep-ex]} \BibitemShut {NoStop}%
\end{thebibliography}%

\end{document}